Carlo Scotto

# Radiopropagazione ionosferica HF

*con problemi ed applicazioni informatiche*

# CAPITOLO 1

## Fondamenti di fisica del plasma ionosferico

**Riassunto**

Le caratteristiche del plasma ionosferico vengono introdotte assieme ad alcuni parametri fisici adatti a caratterizzarlo, quali la frequenza di plasma e la distanza di Debye. Dal punto di vista della propagazione delle onde radio, vengono messe in risalto le similitudini e le differenze rispetto ai dielettrici ordinari. Vengono ricavate le equazioni costitutive della materia, presentandole come l'estensione al caso dei magnetoplasmi della teoria della polarizzazione, che viene usata per i dielettrici ordinari. Viene introdotta la condizione di plasma non caldo, accennando la relazione con la dispersione in frequenza.

### 1.1 Caratteristiche del plasma ionosferico

La ionosfera costituisce la parte dell'atmosfera che si trova ad altitudini superiori a 50 km, dove i gas sono più rarefatti e gli elettroni liberi possono esistere per lunghi periodi di tempo, prima che la ricombinazione abbia luogo. L'agente ionizzante è costituito dalla radiazione elettromagnetica (principalmente raggi UV ed X) di origine solare che ionizza le molecole neutre con produzione di elettroni e ioni positivi. Un gas ionizzato, poi, costituisce un *plasma*, sebbene questa definizione necessiti di alcune precisazioni che faremo nel seguito. La materia nella ionosfera è in realtà incorporata in una maggioranza di molecole neutre, come $N_2$, $N$, $NO$, $O_2$, $O$, $H$ ed $H_e$. La densità numerica di elettroni $N_e$ è molto minore della densità $N_n$ delle particelle neutre, vale a dire che il *grado di ionizzazione* è molto piccolo: si parla perciò di *plasma minoritario*. Il rapporto $N_e / N_n$ all'altezza del massimo dello strato F2, raggiunge $10^{-3}$.

Le specie chimiche coinvolte nel processo di ionizzazione ad opera dei fotoni solari, sono diverse a seconda della quota. Ai confini esterni dell'atmosfera, il flusso di radiazione solare è di circa 1370 W/m$^2$, ma la densità degli atomi è bassa, per cui si ha un *tasso di ionizzazione* modesto. Anche alle quote più basse, questo tasso è modesto. Esso, infatti, da un lato tenderebbe ad aumentare perché le molecole sono più dense ma, allo stesso tempo, la radiazione elettromagnetica ionizzante subisce una maggiore attenuazione. Ci aspettiamo quindi, per ogni specie ionica dominante, il verificarsi di un massimo di produzione ad una quota intermedia. Trova così spiegazione l'effetto di stratificazione orizzontale della ionosfera con la formazione di diversi *strati* o *regioni*, quali la regione D, la regione E e la regione F.

La regione D è quella che giace a quota più bassa, essendo compresa tra 50 km e 90 km di altitudine. Ivi il tasso di ricombinazione è molto alto, e, di conseguenza, le densità di elettroni liberi e di ioni sono piccole. La densità di elettroni non è sufficiente a riflettere le onde HF (3-30 MHz), ma ne causa l'attenuazione per assorbimento. La densità di elettroni liberi, che tipicamente si trova a queste quote, è intorno a $10^9$-$10^{10}$ elettroni/m$^3$. La regione D appare all'alba, ha un comportamento quasi sincrono con il sole e scompare praticamente del tutto di notte.

La regione E ha la sua massima densità di elettroni all' altezza di circa 115 km. Essa viene osservata durante il giorno, con $N_e \sim 10^{11}$ m$^{-3}$ evidenziando un forte controllo solare; risulta infatti che $N_e$ dipende soltanto dell'angolo zenitale solare e dal numero di macchie solari.

La regione F è molto più ionizzata. L'altezza del massimo di della densità elettronica in funzione della quota $N_e(h)$ varia nell'intervallo da 200 a 400 km. Le variazioni giornaliere e stagionali della regione F2 sono più complesse di quelle dello regione E. Durante il giorno $N_e(h)$ mostra un flesso aggiuntivo al di sotto del massimo F2, che è conosciuto come *strato F1*. La massima densità elettronica dello strato F è dell'ordine di $10^{12}$ m$^{-3}$ ed è la massima assoluta che si osserva nella ionosfera.

D'altra parte, mentre le regioni ionosferiche vengono identificate attraverso la densità del plasma, la struttura dell'atmosfera viene solitamente descritta dal profilo di temperatura. La Fig. 1.1 mostra uno schema della ionosfera, con un andamento di $N_e(h)$ tipicamente diurno e che ne mostra i vari strati. Contestualmente viene anche mostrata la nomenclatura delle varie regioni atmosferiche, in funzione della temperatura dell'atmosfera neutra. Si noti che, da questo punto di vista, le regioni E ed F giacciono nella termosfera, mentre la regione D nella mesosfera.

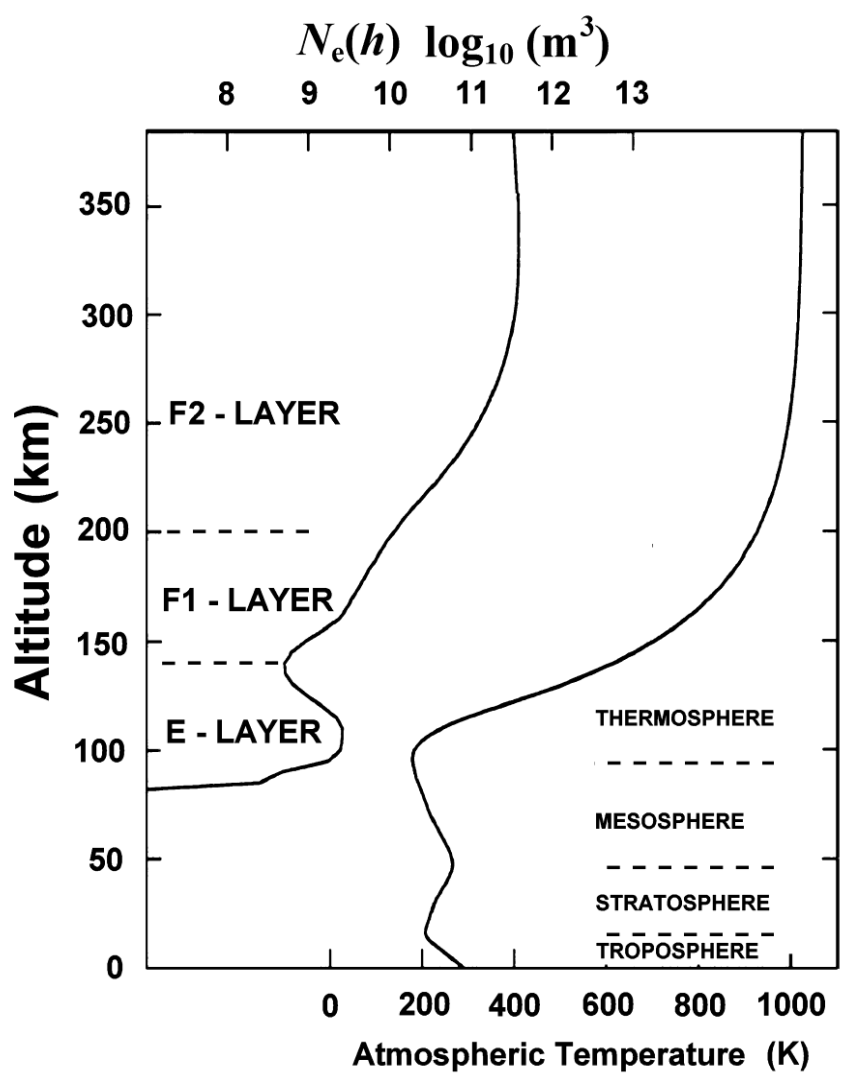

Fig. 1.1. Uno schema della ionosfera, con l'andamento tipico diurno della densità elettronica, che mostra i vari strati. Per confronto è anche indicata la nomenclatura in funzione della temperatura dell'atmosfera neutra.

## 1.2. La temperatura della ionosfera

Nella ionosfera le coppie elettroni-ioni vengono generate per fotoionizzazione. Nel processo di fotoionizzazione, per la loro minore massa, gli elettroni acquistano in media una velocità maggiore, per cui essi tendono ad avere una distribuzione delle velocità Maxwelliane compatibile con una temperatura più elevata rispetto a quella degli ioni.

La temperatura che effettivamente si stabilisce alle diverse quote della ionosfera, separatamente per gli ioni e per gli elettroni, dipende poi dall'equilibrio dinamico che si viene a creare con gli scambi di energia elettroni-particelle neutre, ioni-particelle neutre e ioni-elettrone, scambi che avvengono tramite le collisioni: queste sono la causa della *termalizzazione*, cioè del processo che porta popolazioni inizialmente a temperature differenti al raggiungimento dell'equilibrio termodinamico.
Se indichiamo con $\tau_{ee}$ il tempo caratteristico in cui due popolazioni di elettroni inizialmente a diversa temperatura raggiungono l'equilibrio termico, si può dimostrare che:

$$\tau_{ee} = \left(\nu_{ee}\right)^{-1}. \tag{1.1}$$

Allo stesso modo per gli ioni si avrà:

$$\tau_{ii} = \left(\nu_{ii}\right)^{-1}. \tag{1.2}$$

Nelle (1.1) e (1.2), $\nu_{ee}$ e $\nu_{ii}$ esprimono le *frequenze delle collisioni* elettrone-elettrone e ione-ione, rispettivamente. In queste relazioni interviene il fatto fatto che quando due particelle di ugual massa collidono, ciascuna cede la propria energia all'altra particella in misura completa. Nel caso invece consideriamo la termalizzazione degli ioni con gli elettroni, si deve tenere conto che l'elettrone cede solo una frazione pari a $(m_e/m_i)$ della propria energia allo ione, come può essere dimostrato applicando le conservazioni dell'impulso e dell'energia nell'urto fra le due particelle. La termalizzazione richiede dunque un tempo più lungo; precisamente si può dimostrare che risulta:

$$\tau_{ie} = \frac{m_e}{m_i}\left(\nu_{ie}\right)^{-1}. \tag{1.3}$$

In un plasma gli elettroni e gli ioni possono così esistere per lunghi tempi a temperature diverse.
Se consideriamo le collisioni fra particelle cariche e molecole neutre, i tempi di termalizzazione possono essere ricavati da relazioni simili alla precedente. In questo caso, si dovrà considerare che, detta $M$ la massa delle molecole neutre, si ha $m_e \ll M$ mentre si ha $m_i \sim M$.
In generale si deve poi osservare che, col crescere della densità particellare, cresce la frequenza delle collisioni. Questa, d'altra parte, dipende anche dalla natura delle interazioni fra le particelle coinvolte negli urti. Quando si tratta di urti fra elettroni e ioni, l'efficienza della interazione coulombiana tende a fare in modo che la frequenza delle collisioni sia maggiore che nel caso di urti che coinvolgono molecole neutre. Alla luce di queste considerazioni generali, discutiamo gli andamenti delle temperature ioniche, elettroniche e delle molecole neutre che sono riportate nella Fig. 1.2.
La regione D è caratterizzata da elevata densità di particelle neutre per cui $\nu_{en}$ e $\nu_{in}$ sono elevate. Questo fa in modo che sia elettroni che ioni risultino termalizzati alla temperatura delle molecole neutre.
Nella regione E, a causa della diminuzione della densità di particelle neutre, sia $\nu_{en}$ che $\nu_{in}$ tendono a diminuire. Tuttavia poiché gli ioni hanno maggiori dimensioni rispetto agli elettroni si ha che $\nu_{in} \gg \nu_{en}$. Pertanto, al crescere della quota, $T_n$ e $T_i$ si mantengono quasi le stesse, mentre $T_e$ inizia a deviare verso valori più alti durante il giorno, già alla quota di 130 km. Le misurazioni indicano che $T_n$ e $T_i$ sono le medesime fino a circa 250 km.
Alle altitudini ancora più elevate, corrispondenti alla regione F2 ed al suo topside, l'accoppiamento coulombiano tra ioni e elettroni comincia a prevalere, in seguito alla diminuzione della densità di molecole neutre che, oltre $\nu_{en}$, rende piccolo anche $\nu_{in}$. Alla quota di circa 800 km la temperatura degli elettroni è praticamente la stessa degli ioni.

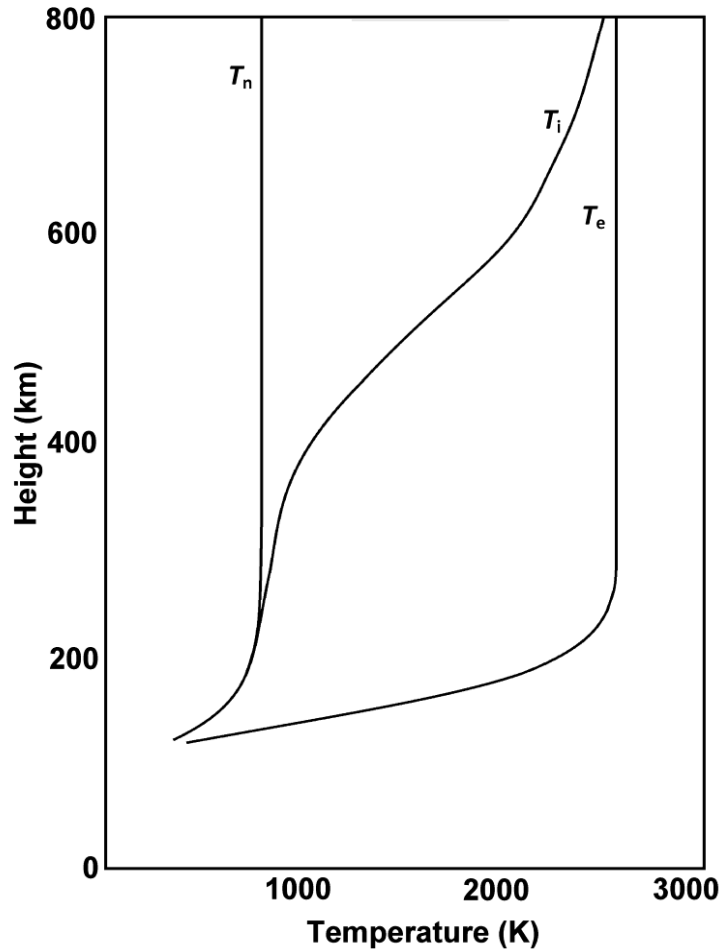

Fig. 1.2. Gli andamenti delle temperature di elettroni $T_e$, ioni $T_i$ e molecole neutre $T_n$ della ionosfera. Generalmente $T_e$ è molto più alta di $T_n$. Gli elettroni e le molecole neutre hanno le stesse temperature solo a quote inferiori a 100 km, dove le collisioni sono importanti. Gli ioni, invece, a causa della loro grande dimensione, transferiscono energia con le molecole neutre, in modo più efficiente degli elettroni; per questa ragione essi hanno la stessa temperatura delle molecole neutre fino a circa 250 km. A quote più elevate, la temperatura si discosta da quella delle particelle neutre, per arrivare alla temperatura degli elettroni, quando le collisioni fra particelle cariche e neutre sono così rare da rendere le interazioni couloumbiane predominanti (rfr1-http://utd500.utdallas.edu/ionosphere.htm).
.

## 1.3 Frequenza di plasma

All'interno di un plasma, consideriamo una area $S$ sulla quale, per qualche motivo, c'è stata una separazione di carica su una distanza $\xi$, come schematizzato nella Fig. 1.3. Ogni elettrone ha massa $m_e$ e carica elettrica $e$. Se c'è una densità di elettroni $N_e$ si può supporre che, a partire dal volume considerato, è stata rimossa una quantità di carica $Q = N_e \cdot e \cdot \xi \cdot S$, che giace ora su entrambe le piastre del condensatore che si è costituito. Per questo motivo, le armature hanno densità superficiale di carica $\sigma = Q / S = N_e \cdot e$, e, di conseguenza, all'interno del condensatore abbiamo un vettore campo elettrico di modulo $E = \sigma / \varepsilon_0 = N_e \cdot e \cdot \xi / \varepsilon_0$.

Questo campo agisce sugli elettroni, in direzione opposta allo spostamento $\xi$ così la legge di Newton può essere scritta come:

$$-\mathbf{E}e = m_e\ddot{\xi}. \tag{1.4}$$

In questa e, usando l'espressione $E = N_e \cdot e \cdot \xi / \varepsilon_0$, abbiamo:

$$\ddot{\xi} + \frac{N_e e^2 \xi}{m_e \varepsilon_0} = 0. \tag{1.5}$$

Si può dunque porre:

$$\omega_p^2 = \frac{N_e e^2}{m_e \varepsilon_0}. \tag{1.6}$$

La grandezza $\omega_p$ si chiama *frequenza di plasma* e rappresenta la frequenza con cui gli elettroni oscillano in un plasma per ristabilire la neutralità, una volta che questa è stata perturbata.

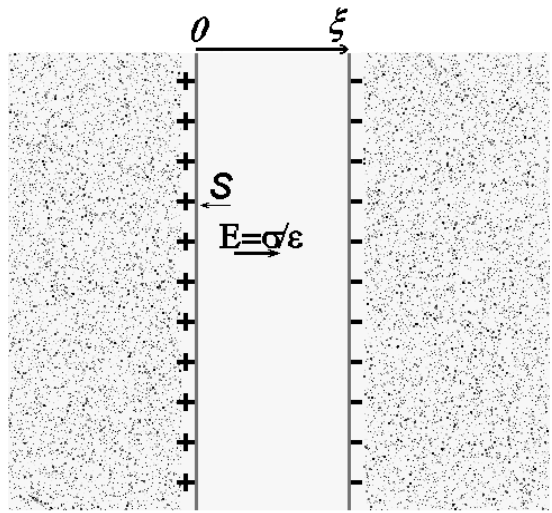

Fig. 1.3. Se in una superficie $S$ all'interno di un plasma c'è uno spostamento di carica, tale che si formano due strati a distanza $\xi$, viene a crearsi un campo elettrico come un condensatore sulle cui armature c'è una densità di carica $\sigma = Q / S = N \cdot e \cdot S \cdot \xi / S$. Questo campo elettrico è di intensità $E = \sigma/\varepsilon_0 = N \cdot e \cdot \xi/\varepsilon_0$. Esso agisce sulla carica elettrica come una forza elastica di intensità $F = N \cdot e^2 \cdot \xi/\varepsilon_0$, con direzione ortogonale alle armature del condensatore e verso tale da ristabilire la neutralità.

## 1.4 La distanza di Debye

Consideriamo un plasma imperturbato ove venga inserita una carica puntuale positiva $Q$. Una nube di elettroni sarà attratta da $Q$ mentre le cariche con lo stesso segno saranno respinte. Così, il campo elettrico generato da $Q$ sarà completamente schermato dal resto del plasma. Questo sarebbe vero se il movimento termico non fosse considerato, come sarebbe corretto allo zero assoluto.
Ora supponiamo di riscaldare il plasma fino ad una temperatura generica $T$, in modo che le particelle acquisiscano il loro moto di agitazione termica. Gli elettroni si disporranno intorno a $Q$ in una nube, all'interno della quale il campo elettrico non sarà completamente schermato dagli elettroni stessi. Al bordo della nuvola, il campo dovuto a $Q$ sarà quasi completamente schermato dagli elettroni.

La *distanza di Debye* $\lambda_D$ è dunque la distanza caratteristica alla quale i portatori di carica mobili schermano i campi elettrici nei plasmi e negli altri conduttori. In altre parole, è la distanza entro la quale si possono verificare separazioni di cariche importanti. Euristicamente $\lambda_D$ può essere ottenuta ponendo l'energia potenziale coloumbiana pari alla energia di agitazione termica che compete a un grado di libertà. D'altra parte, l'energia potenziale coloumbiana può essere calcolata attraverso la massima energia cinetica del moto armonico compiuto dalla carica elettrica per ripristinare la neutralità del plasma, una volta che questa è stata perturbata:

$$\frac{1}{2} m_e \left(\omega_p \lambda_D\right)^2 = \frac{1}{2} K_B T . \tag{1.7}$$

In questa relazione, oltre alle grandezze che abbiamo già introdotto, compare la costante di Boltzmann $K_B = 1.38 \cdot 10^{-23}$ J/K. Per quanto riguarda la temperatura $T$, facciamo qui riferimento a una temperatura da ascriversi genericamente tanto agli elettroni che agli ioni. In realtà, questa temperatura dovrebbe essere posta tale che $T^{-1} = T_i^{-1} + T_e^{-1}$.
Quindi dalla (1.7) si ottiene:

$$\left(\lambda_D\right)^2 = \frac{K_B T}{m \omega_p^2}, \tag{1.8}$$

da cui, usando la (1.6), otteniamo:

$$\left(\lambda_D\right)^2 = \frac{K_B T}{m} \frac{m \varepsilon_0}{N e^2}, \tag{1.9}$$

e quindi:

$$\lambda_D = \sqrt{\frac{K_B T \varepsilon_0}{N e^2}} . \tag{1.10}$$

Nel primo paragrafo è stato introdotto il concetto di plasma assimilandolo a un gas ionizzato. Questo modo di definire i plasmi è senz'altro limitativo e, come abbiamo accennato, necessita di alcune precisazioni. Essi, infatti, posseggono caratteristiche uniche rispetto agli altri stati tradizionali, solido, liquido e gassoso e vengono spesso considerati, a buon titolo, il quarto stato di aggregazione della materia. Analogamente a quanto avviene nelle transizioni solido-liquido e liquido-gas, anche la transizione gas-plasma si ottiene scaldando il sistema. A temperature sufficientemente alte, le molecole o gli atomi del gas si scindono in elettroni e ioni positivi, formando il plasma. Nei plasmi sono quindi presenti cariche elettriche libere di muoversi generando densità di carica e di corrente, che a loro volta generano campi elettrici e magnetici. Il plasma dunque è caratterizzato da cariche libere con comportamenti collettivi.
Vediamo ora che la grandezza $\lambda_D$ è utile anche per valutare se gli elettroni di un gas ionizzato possono essere considerati liberi, per valutare cioè se sono verificate le cosiddette *condizioni di plasma*. A questo scopo confrontiamo l'energia di accoppiamento elettroni - ioni con l'energia del moto di agitazione termica, imponendo che sia quest'ultima a prevalere:

$$\frac{1}{2} K_B T \gg \frac{1}{4\pi\varepsilon_0} \frac{e^2}{r}, \tag{1.11}$$

da cui

$$\frac{1}{2} K_B \frac{\lambda_D^2}{K_B} \frac{N_e e^2}{\varepsilon_0} \gg \frac{1}{4\pi\varepsilon_0} \frac{e^2}{r}. \qquad (1.12)$$

La densità di particelle cariche e la distanze interparticellari sono ovviamente collegate da $N_e^{(1/3)} = r^{-1}$, per cui si ha:

$$\frac{1}{2} K_B \frac{\lambda_D^2}{K_B} \frac{e^2}{\varepsilon_0} \frac{1}{r^3} \gg \frac{1}{4\pi\varepsilon_0} \frac{e^2}{r}, \qquad (1.13)$$

e

$$\frac{\lambda_D^2}{r^2} \gg \frac{1}{2\pi}. \qquad (1.14)$$

Ciò significa che un gas ionizzato si comporta come un plasma se $\lambda_D$ è molto più grande delle distanze interatomiche, cioè, se entro una sfera di Debye ci sono molti elettroni. Ciò corrisponde con l'affermazione che l'interazione couloumbiana con gli ioni vicini è debole rispetto all'agitazione termica. Nella ionosfera questa condizione è sempre verificata (rfr1-1Howard, 2002; Krainov, 1992)

### 1.5 I dielettrici ordinari e la trattazione macroscopica

In un dielettrico ordinario, il campo elettrico esterno agisce sulle molecole provocando delle modificazioni nella loro struttura. Precisamente si formano dei *dipoli elettrici*. Un dipolo elettrico è un sistema composto da due cariche elettriche $q$ uguali e opposte di segno, separate da una distanza $r$ costante nel tempo. Associato al dipolo elettrico si può definire il *momento elettrico*, o *momento di dipolo* per quantificare la polarità del sistema, ovvero la separazione tra le cariche positive e negative. Esso è una grandezza vettoriale definita come:

$$\mathbf{p} = q\mathbf{r} \qquad (1.15)$$

ove $\mathbf{r}$ è il vettore spostamento di una carica rispetto all'altra, orientato da quella negativa a quella positiva.

Il fenomeno con il quale si formano dipoli elettrici, sotto l'azione di un campo elettrico esterno, prende il nome di *polarizzazione*, e il dielettrico che ha subito tale modifica si dice *polarizzato*.

L'effetto della polarizzazione elettrica viene descritto ponendo in relazione la polarizzazione dei dipoli microscopici con una grandezza vettoriale macroscopica che si introduce appositamente e che descrive il comportamento globale del materiale soggetto alla presenza di un campo elettrico esterno. La grandezza vettoriale in questione è il *vettore intensità di polarizzazione*, anche detto *vettore di polarizzazione elettrica* o *di polarizzazione dielettrica*, è solitamente indicato con **P**, e rappresenta il dipolo elettrico per unità di volume posseduto dal materiale. Esso è definito come:

$$\mathbf{P} = Nq\mathbf{r} \qquad (1.16)$$

essendo $q\mathbf{r}$ il valore medio del momento di dipolo elettrico proprio di $N$ coppie di cariche $q$ e $-q$ contenute in un volume $V$.

Si può dimostrare che il vettore **P**, così definito, ha la proprietà di avere

$$\operatorname{div}(\mathbf{P}) = -\rho_{\text{micr}}. \tag{1.17}$$

Inoltre, per il teorema di Gauss espresso in forma differenziale, si ha:

$$\operatorname{div}(\mathbf{E}) = \frac{\rho_{\text{micr}} + \rho_{\text{macr}}}{\varepsilon_0}, \tag{1.18}$$

ove $\rho_{\text{micr}}$ rappresenta la densità di cariche microscopiche.
Viene poi introdotto un vettore *spostamento dielettrico* **D**, definito come:

$$\mathbf{D} = \varepsilon_0 \mathbf{E} + \mathbf{P}, \tag{1.19}$$

che, in base alle (1.17) e (1.18) ha la proprietà:

$$\operatorname{div}(\mathbf{D}) = \rho_{\text{macr}}. \tag{1.20}$$

Si può poi definire una *costante dielettrica relativa* $\varepsilon_r$, attraverso la relazione:

$$\mathbf{D} = \varepsilon_0 \varepsilon_r \mathbf{E}, \tag{1.21}$$

ed una *suscettività dielettrica* $\chi$ attraverso la relazione:

$$\mathbf{P} = \varepsilon_0 \chi \mathbf{E}. \tag{1.22}$$

Sostituendo nella (1.19) le (1.21) e (1.22) si ha:

$$\varepsilon_0 \varepsilon_r \mathbf{E} = \varepsilon_0 \mathbf{E} + \varepsilon_0 \chi \mathbf{E}, \tag{1.23}$$

da cui:

$$\varepsilon_r = 1 + \chi. \tag{1.24}$$

Abbiamo detto che, noto il campo elettrico **E**, **P** è in grado di descrivere il comportamento macroscopico del mezzo. In base alla (1.19), si vede che la stessa descrizione può essere ottenuta anche tramite il vettore **D**. Dalla (1.21) vediamo che **D** può essere ottenuto tramite $\varepsilon_r$, mentre dalla (1.22) vediamo che **P** può essere ottenuto tramite $\chi$. D'altra parte dalla (1.24) vediamo che le conoscenze di $\varepsilon_r$ o di $\chi$ sono equivalenti. In altre parole, si vede che una sola fra le grandezze **D**, **P**, $\varepsilon_r$ e $\chi$ è sufficiente per descrivere il comportamento macroscopico del mezzo, in risposta al campo elettrico applicato **E**.
Vedremo nel paragrafo 1.10 che per onde radio aventi lunghezza d'onda $\lambda$ molto più grande di una certa distanza caratteristica, anche per il plasma si può fare la teoria macroscopica, analoga a quella che abbiamo accennato in questo paragrafo, per i dielettrici ordinari. Vedremo, nello stesso paragrafo 1.10, che quando il plasma è non caldo, se non sono presenti vincoli esterni ad introdurre l'anisotropia, come ad esempio un campo magnetico costante, $\chi$ è uno scalare come quello che abbiamo introdotto con la (1.22) per un dielettrico ordinario.

Nel paragrafo seguente, invece, procederemo col ricavare χ per le onde radio che si propagano nel plasma ionosferico. Per far questo, dovremo modellare l'interazione fra **E** e le particelle cariche nel plasma, per cui sarà necessario fare delle ipotesi sulla natura del mezzo.

**1.6 Suscettività senza alcun campo magnetico**

Per ricavare χ in un mezzo è necessario studiare, a livello microscopico, lo spostamento delle particelle cariche sotto l'azione del campo elettrico applicato. Nel caso di nostro interesse, utilizziamo la seconda legge di Newton:

$$e\mathbf{E} = m\frac{d^2\mathbf{r}}{dt^2}. \tag{1.25}$$

Siccome si assume che il campo elettrico localmente è sinusoidale nella forma $\mathbf{E} = \mathbf{E}_o \cdot \exp(i \cdot \omega \cdot t)$ lo spostamento **r** è anche nella forma $\mathbf{r} = \mathbf{r}_o \cdot \exp(i \cdot \omega \cdot t)$, e dunque $d^2\mathbf{r}/dt^2 = -\omega^2 \cdot \mathbf{r}$. Pertanto, dalla (1.25) abbiamo:

$$e\mathbf{E} = -m\omega^2 \mathbf{r}. \tag{1.26}$$

Da questa, essendo $\mathbf{P} = N \cdot e \cdot \mathbf{r}$, secondo la (1.22), si ottiene:

$$\mathbf{P} = -\varepsilon_0 \frac{Ne^2}{\varepsilon_0 m\omega^2} \mathbf{E} = -\varepsilon_0 \frac{\omega_p^2}{\omega^2} \mathbf{E}, \tag{1.27}$$

cosicché per χ risulta:

$$\chi = -\frac{\omega_p^2}{\omega^2}. \tag{1.28}$$

Nell'ipotesi di plasma in assenza di campo magnetico, vediamo dunque che χ è uno scalare. Esso poi dipende unicamente dalla frequenza e non dal vettore d'onda **k**. Si parla infatti in questo caso *dispersione in frequenza* del mezzo (rfr1-Landau, 1981; Krainov, 1992). Si osservi che questo fenomeno è legato al fatto che abbiamo assunto che un campo elettrico che localmente varia come $\mathbf{E} = \mathbf{E}_o \cdot \exp(i \cdot \omega \cdot t)$ induca uno spostamento **r** della medesima forma: questo è quanto effettivamente succede nel caso definito nel paragrafo 1.10.

**1.7 Onde elettromagnetiche in un dielettrico ordinario**

Scriviamo le equazioni di Maxwell:

$$\begin{cases} \text{div}(\mathbf{D}) = \rho & (1.29a) \\ \text{div}(\mathbf{B}) = 0 & (1.29b) \\ \text{curl}(\mathbf{E}) = -\dfrac{\partial \mathbf{B}}{\partial t} & (1.29c) \\ \text{curl}(\mathbf{H}) = \mathbf{j} + \dfrac{\partial \mathbf{D}}{dt} & (1.29d) \end{cases}$$

È noto che in queste equazioni **E** e **D** sono rispettivamente i vettori campo elettrico e induzione dielettrica che abbiamo già introdotto, mentre **B** è il *vettore induzione magnetica*, **H** è il *vettore intensità del campo magnetico*, $\rho$ la *densità delle cariche macroscopiche*, e **j** la *densità di corrente elettrica*. Per quanto abbiamo detto nei paragrafi 1.5 e 1.6, per il magnetoplasma ionosferico, **P** può essere espresso tramite a **E** con la relazione lineare **P**=$\varepsilon_0 \cdot \chi \cdot$**E**, analogamente a quanto si fa nei dielettrici ordinari. Per le stesse ragioni, secondo la (1.21), possiamo scrivere **D** nella forma **D**=$\varepsilon_0 \varepsilon_r$**E**, dove, secondo la (1.24), $\varepsilon_r = 1+\chi$. Analogamente fra i vettori **B** e **H** vale la relazione **B**=$\mu_0 \cdot \mu_r \cdot$**H** che è la corrispondente per il campo magnetico della (1.21), che è valida per il campo elettrico.
Pertanto le equazioni di Maxwell, divengono:

$$\begin{cases} \text{div}(\mathbf{D}) = 0, & \text{(1.30a)} \\ \text{div}(\mathbf{B}) = 0, & \text{(1.30b)} \\ \text{curl}(\mathbf{E}) = -\dfrac{\partial \mathbf{B}}{\partial t}, & \text{(1.30c)} \\ \text{curl}(\mathbf{B}) = \mu_0 \mathbf{j} + \mu_0 \mu_r \varepsilon_0 \varepsilon_r \dfrac{\partial \mathbf{E}}{dt}. & \text{(1.30d)} \end{cases}$$

Applicando l'operatore rotore a entrambi i membri della terza e quarta equazione, otteniamo rispettivamente:

$$\text{curl}\left[\text{curl}(\mathbf{B})\right] = \frac{\partial \text{curl}(\mathbf{B})}{dt} = -\mu_0 \mu_r \varepsilon_0 \varepsilon_r \frac{\partial^2 \mathbf{B}}{\partial t^2}, \qquad (1.31a)$$

e

$$\text{curl}\left[\text{curl}(\mathbf{E})\right] = \mu_0 \mu_r \varepsilon_0 \varepsilon_r \frac{\partial \mathbf{E}}{dt} = -\mu_0 \mu_r \varepsilon_0 \varepsilon_r \frac{\partial^2 \mathbf{E}}{\partial t^2}. \qquad (1.31b)$$

Ora, se consideriamo l'identità vettoriale seguente: curl[curl(**A**)]=grad[div[A]]-$\nabla^2$**A**, si ha:

$$\text{curl}\left[\text{curl}(\mathbf{B})\right] = \text{grad}\left[\text{div}(\mathbf{B})\right] - \nabla^2 \mathbf{B} \qquad (1.32a)$$

e

$$\text{curl}\left[\text{curl}(\mathbf{E})\right] = \text{grad}\left[\text{div}(\mathbf{E})\right] - \nabla^2 \mathbf{E}, \qquad (1.32b)$$

dalle quali otteniamo le seguenti relazioni:

$$\nabla^2 \mathbf{E} - \mu_0 \mu_r \varepsilon_0 \varepsilon_r \frac{\partial^2 \mathbf{E}}{\partial t^2} = 0, \qquad (1.33a)$$

e

$$\nabla^2 \mathbf{B} - \mu_0 \mu_r \varepsilon_0 \varepsilon_r \frac{\partial^2 \mathbf{B}}{\partial t^2} = 0. \qquad (1.33b)$$

Se consideriamo onde elettromagnetiche che si propagano lungo l'asse $x$, **E** e **B** dipenderanno solo da $x$. Si avrà così:

$$\frac{\partial^2 \mathbf{E}}{\partial x^2} - \mu_0 \mu_r \varepsilon_0 \varepsilon_r \frac{\partial^2 \mathbf{E}}{\partial t^2} = 0, \tag{1.34a}$$

e

$$\frac{\partial^2 \mathbf{B}}{\partial x^2} - \mu_0 \mu_r \varepsilon_0 \varepsilon_r \frac{\partial^2 \mathbf{B}}{\partial t^2} = 0. \tag{1.34a}$$

Per ciascuna componente dei vettori **E** e **B**, varrà una soluzione del tipo $F(x,t)=A \cdot f(x+v \cdot t)+B \cdot f(x-v \cdot t)$. Questo può essere facilmente verificato considerando per esempio la componente $y$ della (1.34a)

$$\frac{\partial^2 E_y(x,t)}{\partial x^2} - \mu_0 \mu_r \varepsilon_0 \varepsilon_r \frac{\partial^2 E_y(x,t)}{\partial t^2} = 0 \tag{1.35}$$

e sostituendovi sia $\partial^2 E_y/\partial x^2 = \partial^2 F(x,t)/\partial x^2$ che $\partial^2 E_y/\partial t^2 = \partial^2 F(x,t)/\partial t$. Risulta:

$$\frac{\partial^2 E_y(x,t)}{\partial x^2} = \frac{\partial^2}{\partial x^2}\left[Af(x+vt) + Bf(x-vt)\right] = Af(x+vt) + Bf(x-vt) \tag{1.36}$$

e

$$\frac{\partial^2 E_y(x,t)}{\partial t^2} = \frac{\partial^2}{\partial t^2}\left[Af(x+vt) + Bf(x-vt)\right] = Av^2 f(x+vt) + Bv^2 f(x-vt), \tag{1.37}$$

che sostituite nella (1.35) dànno:

$$Af(x+vt) + Bf(x-vt) - \mu_0 \mu_r \varepsilon_0 \varepsilon_r \left[Av^2 f(x+vt) + Bv^2 f(x-vt)\right] = 0, \tag{1.39}$$

se poniamo:

$$v = \frac{1}{\sqrt{\mu_0 \mu_r \varepsilon_0 \varepsilon_r}}. \tag{1.40}$$

Quindi, vediamo che, al fine di ottenere un'espressione per la velocità con cui un'onda elettromagnetica viaggia in un mezzo, abbiamo bisogno di conoscere la costante dielettrica $\varepsilon_r$, la quale corrisponde a conoscere $\chi$, che, a sua volta, corrisponde a conoscere il rapporto tra il modulo dei vettori **E** e **P**.

Si noti che in questo paragrafo abbiamo solo supposto che $\chi$ sia costante e sia uno scalare. Abbiamo cioè supposto un dielettrico omogeneo isotropo. Per casi più complicati, come quelli che si incontrano in generale, questo tipo di trattazione non è sufficiente. Per il plasma ionosferico, che è la materia di nostro interesse, abbiamo la necessità di fare una trattazione ben più complessa.

Dobbiamo cioè determinare la relazione vettoriale tra **E** e **P,** esplicitata nelle sue componenti cartesiane, che è nota come *equazioni costitutive della materia*.

**1.8 Equazioni costitutive della materia per un plasma senza campo magnetico**

Al fine di ottenere la relazione tra **P** ed **E**, dobbiamo considerare la forza che agisce su un singolo elettrone. Questa è la somma della forza elettrostatica e·**E** e della forza di Lorentz e·(**v**×**B**). La seconda legge di Newton è quindi:

$$e(\mathbf{E}+\mathbf{v} \times \mathbf{B})=m(d^2\mathbf{x}/dt^2). \tag{1.41}$$

Ora abbiamo, in base alla (1.16): **P**=$N \cdot e \cdot$**x**, da cui $i \cdot \omega \cdot$**P**=$Ne(d\mathbf{x}/dt)$ e quindi $-\omega^2$ **P**=$N \cdot e \cdot (d^2\mathbf{x}/dt^2)$. Pertanto:

$$\frac{d\mathbf{x}}{dt} = \mathbf{v} = \frac{i\omega}{Ne}\mathbf{P} \tag{1.42}$$

e

$$\frac{d^2\mathbf{x}}{dt^2} = -\frac{\omega^2}{Ne}\mathbf{P}. \tag{1.43}$$

Introducendo la (1.42) e la (1.43) nella (1.41), si ha:

$$e\left(\mathbf{E} + \frac{i\omega}{Ne}\mathbf{P} \wedge \mathbf{B}\right) = -\frac{m\omega^2}{Ne}\mathbf{P}. \tag{1.44}$$

che può essere scritta come:

$$\frac{Ne^2}{m\omega^2}\left(\mathbf{E} + \frac{i\omega}{Ne}\mathbf{P} \wedge \mathbf{B}\right) = -\mathbf{P}, \tag{1.45}$$

o

$$\frac{\varepsilon_0 Ne^2}{\varepsilon_0 m\omega^2}\mathbf{E} + i\frac{e}{m\omega}\mathbf{P} \wedge \mathbf{B} = -\mathbf{P}. \tag{1.46}$$

Introducendo in questa equazione l'espressione della frequenza di plasma $\omega_p$ e ponendo $X=\omega_p^2/\omega^2$, abbiamo:

$$\varepsilon_0 X \cdot \mathbf{E} + i\frac{e}{m\omega}\mathbf{P} \wedge \mathbf{B} = -\mathbf{P}. \tag{1.47}$$

Ora possiamo esprimere il termine $e/(m \cdot \omega)$ P×B, che coinvolge un prodotto vettoriale, attraverso il formalismo matriciale, nel modo seguente:

$$\frac{e}{m\omega}\mathbf{P} \wedge \mathbf{B} = \frac{e}{m\omega}\begin{pmatrix} \mathbf{i} & \mathbf{j} & \mathbf{k} \\ P_x & P_y & P_z \\ 0 & B_T & B_L \end{pmatrix} = \frac{e}{m\omega}\begin{pmatrix} P_y B_L - P_z B_T \\ -P_x B_L \\ P_x B_T \end{pmatrix} = \begin{pmatrix} P_y Y_L - P_z Y_T \\ -P_x Y_L \\ P_x Y_T \end{pmatrix}, \quad (1.48)$$

che può essere esplicitata nelle componenti cartesiane, ottenedo il sistema:

$$\begin{cases} \varepsilon_0 X E_x + i\left(P_y Y_L - P_z Y_T\right) = -P_x, & (1.49a) \\ \varepsilon_0 X E_x - i P_x Y_L = -P_y, & (1.49b) \\ \varepsilon_0 X E_z + i P_x Y_T = -P_z. & (1.49c) \end{cases}$$

Questo, solitamente viene scritto nella forma:

$$\begin{cases} \varepsilon_0 X E_x = -P_x - i Y_L P_y + i Y_T P_z, & (1.50a) \\ \varepsilon_0 X E_y = -P_y + i Y_L P_x, & (1.50b) \\ \varepsilon_0 X E_z = -P_z - i Y_T P_x. & (1.50c) \end{cases}$$

Le relazioni sopra sono le relazioni costitutive di un plasma non caldo in un campo magnetico costante (rfr1-Ratcliffe, 1959). . L'ipotesi di plasma non caldo verrà precisata nel seguito.

## 1.9 Equazioni costitutitive di un plasma con collisioni fra elettroni molecole neutre in un campo magnetico

Tenendo conto delle collisioni fra elettrone e molecole neutre, l'equazione delle forze viene modificata come segue:

$$e(\mathbf{E}+\mathbf{v} \wedge \mathbf{B}) - m\nu\,\mathbf{v} = m\,(d^2\mathbf{x}/dt^2). \quad (1.51)$$

Se facciamo in questa equazione operiamo la stessa sostituzione vista nel paragrafo precedente $\mathbf{v}=(i\omega/Ne)\mathbf{P}$ si ottiene:

$$(\mathbf{E} + \mathbf{v} \wedge \mathbf{B}) - m\nu\left(\frac{i\omega}{Ne^2}\right)\mathbf{P} = m\left(\frac{\omega^2}{Ne^2}\right)\mathbf{P}. \quad (1.52)$$

Osserviamo che questa equazione corrisponde alla (1.30), se effettuatuiamo la seguente sostituzione:

$$\mathbf{E} \to \mathbf{E} - m\nu\left(\frac{i\omega}{Ne^2}\right)\mathbf{P}. \quad (1.53)$$

Come conseguenza dalla (1.40) si possono ricavare le stesse espressioni che si ricavano dalla (1.30), che sono le (1.37), purché in queste si operi la stessa sostituzione (1.41). Consideriamo dunque, a titolo esemplificativo, la (1.37a); da questa, applicando la (1.41) otteniamo:

$$\varepsilon_0 X E_x \to \varepsilon_0 X \left[ E_x - m\nu \left( \frac{i\omega}{Ne^2} \right) P_x \right] = \varepsilon_0 X E_x - \varepsilon_0 \frac{Ne^2}{m\varepsilon_0 \omega^2} m\nu \left( \frac{i\omega}{Ne^2} \right) P_x = \varepsilon_0 X E_x - iZP_x \quad (1.54)$$

dove $Z=\nu/\omega$.

Pertanto, le equazioni costitutive del plasma freddo con le collisioni in un campo magnetico costante sono (rfr1-Ratcliffe, 1959):

$$\begin{cases} \varepsilon_0 X E_x = -(P_x - iZP_x) - iY_L P_y + iY_T P_z, & (1.55a) \\ \varepsilon_0 X E_x = -(P_y - iZP_y) + iY_L P_x, & (1.55b) \\ \varepsilon_0 X E_z = -(P_z - iZP_z) - iY_T P_x. & (1.55c) \end{cases}$$

**1.10 Condizioni di plasma non caldo**

Abbiamo visto che il vettore di polarizzione **P** può essere definito come **P**=$N \cdot e \cdot$**r**($t$), ove **r**($t$) rappresenta lo spostamento medio delle cariche elettriche nell'elemento di volume d$V$ considerato. Osserviamo poi che nel paragrafo 1.7 abbiamo usato la relazione d**P**/d$t$=$i \cdot \omega \cdot$**P**. Questa relazione è basata sul fatto che per r($t$) abbiamo assunto **r**($t$)= **r**$_0$($t$)$\cdot$exp($i \cdot \omega$ - $k \cdot x$), ove per $x$ non è stata introdotta nessuna dipendenza da $t$. Abbiamo cioè supposto che lo spostamento delle cariche elettriche avvenga solo in conseguenza del campo elettrico associato all'onda radio che si propaga. Con questa assunzione si ha infatti:

$$\frac{d\mathbf{P}}{dt} = Ne \cdot \frac{d[\mathbf{r}(t)]}{dt} = Ne \cdot \frac{d\left[\mathbf{r}_o e^{i(\omega t - kx)}\right]}{dt} = Nei\omega \mathbf{r}(t) = i\omega \mathbf{P}. \quad (1.56)$$

In realtà, siccome le particelle cariche sono dotate di moto di agitazione termica, abbiamo:

$$x(t) = v_t \cdot t \quad (1.57)$$

e dunque:

$$\frac{d\left[\mathbf{x}_o e^{i[\omega t - kx(t)]}\right]}{dt} = \frac{d\left[\mathbf{x}_o e^{i[\omega - kv_t]t}\right]}{dt}. \quad (1.58)$$

In quest'ultima relazione, se supponiamo:

$$k \cdot v_t \ll \omega, \quad (1.59)$$

otteniamo:

$$\frac{d\left[\mathbf{x}_o e^{i[\omega - kv_t]t}\right]}{dt} = i\omega \mathbf{x}(t). \quad (1.60)$$

Poiché $v_f=\omega/k$ la condizione (1.59), è equivalente a porre:

$v_t \ll v_f$. (1.61)

La relazione $d\mathbf{P}/dt=i\cdot\omega\cdot\mathbf{P}$, risulta perciò valida finché la velocità di fase della onda radio è molto maggiore della velocità di agitazione termica delle particelle cariche.
Moltiplicando entrambi i membri della (1.61) per il periodo $T=2\pi/\omega$, otteniamo:

$l \ll \lambda$ (1.62)

ove $l = v_t \cdot T$ è la distanza percorsa dalla particella carica nel tempo di una oscillazione del campo elettromagnetico.
D'altra parte, è ragionevole pensare che la posizione e l' impulso di una particella in un certo istante, non dipenda dai campi elettrici sperimentati durante tutta la sua traiettoria, ma solo da quelli sperimentati nell'ultimo suo tratto $l$. Si può così pensare che, in assenza di collisioni, le particelle che si trovano entro una sfera di raggio $l$, hanno perso memoria dell'effetto dei campi sperimentati in precedenza, quando si trovavano al di fuori di tale sfera. Esse perciò rispondono al campo elettrico dell'onda incidente in modo coerente fra loro.
Nel caso poi, considerariamo le collisioni, dobbiamo tenere presente che queste ricasualizzano la traiettoria delle particelle. La distanza caratteristica entro la quale le particelle possono reagire al campo elettrico dell'onda elettromagnetica in modo coerente, diventa perciò il libero cammino medio $l_c$. Dovremo perciò porre, in questo caso, $l_c \ll \lambda$.
Precisamente, in caso di plasma collisionale, dunque, si dovrà considerare la distanza più piccola fra fra $l_c$ e $l$, che chiameremo *distanza di correlazione* e indicheremo con $r_{corr}$. Formalmente scriveremo:

$r_{corr} = \min(l, l_c)$. (1.63)

Affinché sia valida la relazione $d\mathbf{P}/dt=i\cdot\omega\cdot\mathbf{P}$, dovremo dunque avere, in luogo della (1.62) la seguente condizione:

$r_{corr} \ll \lambda$.

Vale poi la pena osservare che usualmente, nella ionosfera, per le onde radio alle frequenza HF, si ha $\omega \gg v$. Quindi, essendo $l_c=v_t/v$ e $l=v_t/\omega$ la più piccola fra le due è $l$ che dunque svolge il ruolo di distanza di correlazione.
Vale poi la pena osservare che, in base a quanto detto fino a ora, nei plasmi, la condizione affinché possa essere svolta la teoria macroscopica, è che la distanza di correlazione sia molto più grande della distanza interatomica.
Nei dielettrici ordianari $r_{corr}$ è la distanza interatomica. Affinché, si possa fare la toria macroscopica, deve essere $\lambda \gg r_{corr}$. Quindi, nei dielettrici ordianari, ogni volta che si può fare la teoria macroscopica, avremo una costante dielettrica che dipende unicamente dalla frequenza (rfr1-Landau, 1981; Krainov, 1992)

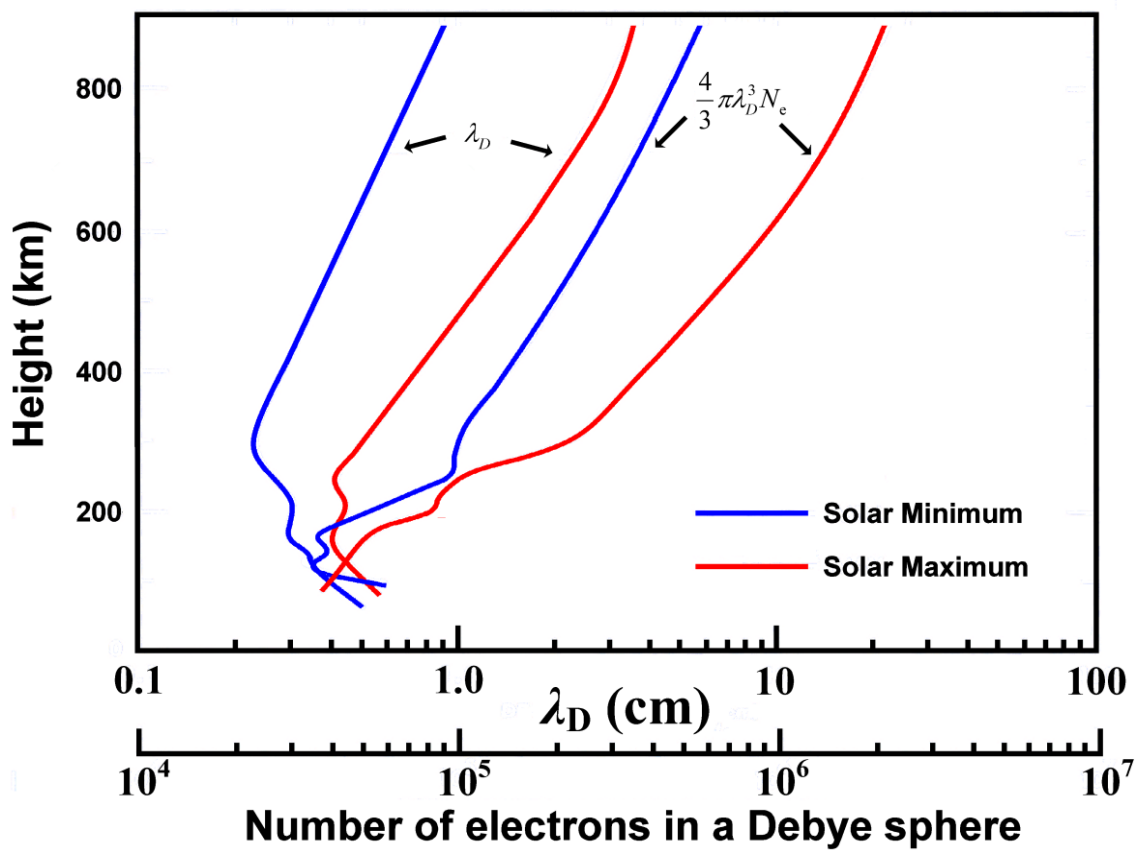

Fig. 1.4. La dimensione della sfera di Debye e il numero di elettroni in essa contenuti.

# Problemi del Capitolo 1

## Problema 1.1

Supponiamo che un fascio di elettroni emessi secondo percorsi paralleli in direzione di un asse $x$, attraversi della materia la cui densità particellare è pari a $n$. Si assuma che ogni particella interagisca con gli elettroni del fascio attraverso la sua *sezione d'urto* $\sigma$ che è l'area data dalla sua proiezione su un piano ortogonale all'asse x stesso. Si determini il *libero cammino medio* $\lambda_m$, cioè la distanza per la quale il flusso cade a $1/e$ del suo valore iniziale.
Suggerimento: si consideri lo strato di materia come composto da tanti strati abbastanza sottili tali che, in ciascun strato, un atomo non possa essere nascosto da un altro, cioè tali che $n \cdot A \cdot dx \cdot \sigma \ll A$.

*Risoluzione*

Consideriamo uno strato di spessore d$x$. Chiameremo $N(x)$ il numero di elettroni che entra in questo strato è $N(x) + dN$ il numero di elettroni che ne escono. Per calcolare d$N$, possiamo pensare che il rapporto fra il numero di particelle che urtano contro gli atomi della materia e il numero totale di particelle del flusso sia pari al rapporto fra la superficie occupata dagli atomi e la superficie totale dello strato. La possibilità di calcolare d$N$ attraverso il rapporto fra dette superfici, è dovuta all'ipotesi che, in ciascun strato, supponiamo che un atomo non possa essere nascosto da un altro.
Nel volume d$V = A \cdot dx$ dello strato saranno contenute d$n = dV \cdot n$ particelle la cui proiezione su un piano ortogonale a x avrà una superficie totale d$S = dn \cdot \sigma = n \cdot A \cdot dx \cdot \sigma$. Quindi risulterà d$N = -N \cdot dS/A = -N(n \cdot A \cdot dx \cdot \sigma / A)$. Per cui abbiamo:

$$\frac{dN}{N} = -\frac{nA\sigma dx}{A} = nA\sigma dx, \tag{P1.1}$$

dalla quale, integrando, risulta:

$$\ln\left[\frac{N(x)}{N_0}\right] = -\int_0^x n\sigma dx = n\sigma x. \tag{P1.2}$$

Da questa espressione possiamo esplicitare $N(x)$ che risulta:

$$N(x) = N_0 e^{-n\sigma x}, \tag{P1.3}$$

perciò il flusso si attenua di $1/e$ quando

$$N\sigma\lambda_m = 1, \tag{P1.4}$$

ossia quando:

$$\frac{1}{N\sigma} = \lambda_m, \tag{P1.5}$$

che costituisce dunque il libero cammino medio.

**Problema 1.2**

Sia dato un sistema costituito da idrogeno puro allo stato gassoso, lasciato isolato per lungo tempo in modo che tutti i processi abbiano raggiunto l'equilibrio. Supponiamo che la temperatura $T$ del sistema sia nota e che sia abbastanza alta da poter assumere per il potenziale chimico $\mu \to -\infty$ (limite classico). L'energia di ionizzazione dell'idrogeno è $\chi$=13.6 eV.
Si trovi una relazione fra il numero di particelle libere e quelle legate agli atomi (rif1-Isavning and Koskinen, 2012).
Suggerimenti:
    si ponga l'energia $E = 0$ quando le velocità di elettroni e protoni sono nulle;
    si ponga l'energia $E = -\chi$, quando le velocità degli atomi neutri è nulla;
    si considerino solo le transizioni fra il livello fondamentale e il primo livello eccitato.

*Risoluzione*

Le distribuzioni statistiche, per fermioni e bosoni, relative agli stati energetici all'equilibrio termico possono essere espresse come:

$$n_k = \frac{g_k}{e^{\frac{E_k - \mu}{K_B T}} \pm 1}. \tag{P1.6}$$

In questa relazione:
il segno positivo va riferito ai fermioni e quello negativo ai bosoni,
$g_k$ esprime la degenarzione dello stato $k$,
$\varepsilon_k$ è l'energia dell'i-imo stato,
$\mu$ è il potenziale chimico,
$K_B$ è la costante di Boltzmann.
Nel limite classico, di alta temperatura, la (P1.6) diventa:

$$n_k = g_k e^{\frac{\mu - E_k}{K_B T}}, \tag{P1.7}$$

per cui il numero di particelle del sistema si otterrà integrando la distribuzione statistica dei livelli energetici (P1.7) su tutto lo spazio delle fasi accessibile al sistema. Risulterà:

$$N = \frac{g}{h^3} \int e^{\frac{\mu - E}{K_B T}} d\Gamma = \frac{g}{h^3} \int e^{\frac{\mu}{K_B T}} e^{-\frac{E}{K_B T}} d\Gamma = \frac{g}{h^3} 4\pi V \int e^{\frac{\mu}{K_B T}} e^{-\frac{p^2}{2m K_B T}} p^2 dp, \tag{P1.8}$$

da cui, passando alla densità, si ha:

$$n = \frac{N}{V} = \frac{g}{h^3} 4\pi e^{\frac{\mu}{K_B T}} \int e^{-\frac{p^2}{2m K_B T}} p^2 dp. \tag{P1.9}$$

Da questa, ricordando che:

$$\int_0^\infty x^2 e^{-\alpha x^2} dx = \frac{1}{4}\sqrt{\frac{\pi}{a}}, \tag{P1.10}$$

si ottiene:

$$n = \frac{N}{V} = \frac{g}{h^3} 4\pi e^{\frac{\mu}{K_B T}} \frac{1}{4}\sqrt{\pi(2mK_B T)^3} = \frac{g}{h^3}(2\pi m K_B T)^{3/2} e^{\frac{\mu}{K_B T}}. \tag{P1.11}$$

La relazione precedente può essere applicata sia agli elettroni che ai protoni liberi, per cui abbiamo, con ovvio significato dei simboli:

$$n_e = \frac{2}{h^3}(2\pi m_e K_B T)^{3/2} e^{\frac{\mu_e}{K_B T}} \tag{P1.12}$$

e

$$n_i = \frac{1}{h^3}(2\pi m_i K_B T)^{3/2} e^{\frac{\mu_i}{K_B T}}. \tag{P1.13}$$

Per quanto riguarda gli atomi neutri, invece, bisogna tenere presente che l'energia corrispondente a velocità nulla va posta uguale a $-\chi$. Di conseguenza nella (P1.10), al posto del termine $\exp[p^2/(2 \cdot m \cdot K_B \cdot T)]$, sarà presente un termine $\exp[p^2/(2 \cdot m \cdot K_B \cdot T) - \chi/(K_B \cdot T)]$. Avremo dunque, in luogo della (P1.13) o (P1.14):

$$n_n = \frac{2}{h^3}(2\pi(m_i + m_e)K_B T)^{3/2} e^{\frac{\mu_n}{K_B T}} e^{\frac{\chi}{K_B T}} \approx \frac{2}{h^3}(2\pi(m_i)K_B T)^{3/2} e^{\frac{\mu_n}{K_B T}} e^{\frac{\chi}{K_B T}}. \tag{P1.14}$$

L'approssimazione eseguita consiste nel trascurare la massa dell'elettrone rispetto a quella del protone, nel valutare la massa dell'atomo di idrogeno $m_i + m_e$. Combinando le (P1.13), (P1.14) e (P1.15), otteniamo:

$$\frac{n_e n_i}{n_n} = \left(\frac{2\pi m_e K_B T}{h^2}\right)^{3/2} e^{\frac{\mu_e}{K_B T}} e^{\frac{\mu_i}{K_B T}} e^{-\frac{\mu_0}{K_B T}} e^{-\frac{\chi}{K_B T}}. \tag{P1.15}$$

Poiché all'equilibrio si ha $\mu_e = \mu_i = \mu_n = 0$, ponendo $n_e = n_i$, si ottiene:

$$\frac{n_e n_i}{n_n} = \left(\frac{2\pi m_e K_B T}{h^2}\right)^{3/2} e^{-\frac{\chi}{K_B T}}, \tag{P1.16}$$

che costituisce l'*equazione di Saha*.

**Problema 1.3**

Si consideri una carica puntiforme $q_T$, in un plasma che ha lunghezza di Debye $\lambda_D$. Si deduca la formula seguente:

$$V(r) = \frac{q_T}{4\pi\varepsilon_0} e^{-(r/\lambda_D)}, \qquad (P1.17)$$

che esprime il potenziale elettrico *V* alla distanza *r* dalla carica considerata (rif1-Isavning and Koskinen, 2012).
.

*Risoluzione*

Riprendiamo la prima equazione di Maxwell, che esprime il teorema di Gauss, in forma differenziale:

$$\text{div}(\mathbf{E}) = \frac{\rho}{\varepsilon_0}, \qquad (P1.18)$$

d'altra parte il campo elettrico è conservativo, per cui risulta:

$$-\text{grad}[V(x,y,z)] = \mathbf{E}, \qquad (P1.19)$$

e, introducendo la (P1.18) nella (P1.19) otteniamo

$$\text{div}(\text{grad}[V(x,y,z)]) = -\frac{\rho}{\varepsilon_0}. \qquad (P1.20)$$

Questa espressione viene solitamente scritta ricorrendo alla notazione vettoriale nel modo seguente:

$$\nabla^2[V(x,y,z)] = -\frac{\rho(x,y,z)}{\varepsilon_0}. \qquad (P1.21)$$

che è nota come *equazione di Poisson*.

Se consideriamo una carica di prova puntiforme *s* posta nel plasma, si può assumere che questo assuma la distribuzione di densità di Boltzmann. Per cui si avrà, per gli ioni, $n_i = n_0 \exp[-e \cdot V(r)/(K_B \cdot T)]$ e per gli elettroni $n_e = n_0 \exp[e \cdot V(r)/(K_B \cdot T)]$. Se inoltre supponiamo che ogni atomo sia ionizzato singolarmente avremo, per la distribuzione di carica:

$$\rho(x) = en_i(x) - en_e(x) + s(x), \qquad (P1.22)$$

da cui

$$\rho(x) = en_0 \left[ e^{-\frac{eV(\mathbf{r})}{K_B \cdot T_i}} - e^{\frac{eV(\mathbf{r})}{K_B \cdot T_e}} \right] + s(x). \qquad (P1.23)$$

Siccome risulta $eV(x) \ll K_B T$, si può usare l'approssimazione $e^\varepsilon = 1 + \varepsilon$, per cui si ha:

$$\rho(x) \approx en_0\left[\left(1-\frac{eV(\mathbf{r})}{K_B \cdot T_i}\right)-\left(1-\frac{eV(\mathbf{r})}{K_B \cdot T_e}\right)\right]+s(x), \tag{P1.24}$$

e quindi:

$$\rho(x) \approx -n_0 e^2\left[\frac{V(\mathbf{r})}{K_B \cdot T_e}+\frac{V(\mathbf{r})}{K_B \cdot T_i}\right]+s(x). \tag{P1.25}$$

L'equazione di Poisson (1.22), con l'inclusione della carica di prova, diventa dunque:

$$\frac{\partial^2 V(x)}{\partial x^2}=\frac{n_0 e^2}{\varepsilon_0}\left[\frac{V(x)}{K_B \cdot T_e}+\frac{V(x)}{K_B \cdot T_i}\right]+s(x), \tag{P1.26}$$

Definiamo dunque la distanza di Debye $\lambda_D$ attraverso la relazione:

$$\frac{1}{\lambda_D^2}=\frac{n_0 e^2}{\varepsilon_0 K_B}\left[\frac{1}{T_e}+\frac{1}{T_i}\right], \tag{P1.27}$$

la quale induce anche l'introduzione di una temperatura efficace $T_{\text{eff}}$ attraverso la realzione:

$$\frac{1}{T_{\text{eff}}}=\frac{1}{T_e}+\frac{1}{T_i}, \tag{P1.28}$$

per cui si ha la seguente espressione:

$$\lambda_D=\sqrt{\frac{\varepsilon_0 K_B T_{\text{eff}}}{n_0 e^2}}. \tag{P1.29}$$

Introducendo la (P1.26) nella (P1.29), questa diventa:

$$\frac{\partial^2 V}{\partial x^2}=\frac{1}{\lambda_D}V+s, \tag{P1.30}$$

la cui soluzione è:

$$V(x)=Ae^{-\frac{x}{\lambda_D}}+Be^{\frac{x}{\lambda_D}}, \tag{P1.31}$$

ove le costanti $A$ e $B$ si determinano imponendo le condizioni al contorno. Queste sono:

$$\lim_{x\to\infty}[V(x)]=0, \tag{P1.32}$$

dalla quale si ottiene $B=0$, e

$$\lim_{x \to 0}[V(x)] = \frac{1}{4\pi\varepsilon_0}\frac{q}{r}, \tag{P1.33}$$

dalla quale si ottiene $A=1/(4\cdot\pi\cdot\varepsilon_0) \cdot q/r$. Introducendo nella (P1.30), i valori delle costanti $A$ e $B$ dianzi determinati otteniamo:

$$V(x) = \frac{1}{4\pi\varepsilon_0}\frac{q}{r}e^{-\frac{r}{\lambda_D}}, \tag{P1.34}$$

che è la relazione proposta nel testo del problema.

**Problema 1.4**

Una regione di spazio è soggetta a un campo magnetico **B** costante. In un sistema di riferimento cartesiano opportunamente scelto, si scrivano le equazioni del moto per ioni ed elettroni, dimostrando che questi si muovono su traiettorie circolari. Si ricavi, separatamente per ioni ed elettroni, le girofrequenze, rispettivamente $\omega_{Bi}$, e $\omega_{Be}$. Si ricavino anche i raggi di tali triettorie, che sono detti raggi *di girazione* o *di Larmor* (rif1-Isavning and Koskinen, 2012).

Risoluzione

Scegliamo un sistema di riferimento avente l'asse $z$ diretto lungo il campo magnetico. Gli altri assi $x$ e $y$ gli scegliamo in direzione generica, purché siano ortogonali ad esso ed ortogonali fra loro. In questo sistema di riferimento la componente lungo $x$ della velocità $v_x$ dà luogo a una forza lungo l'asse $y$ pari a $F_y = -e \cdot B \cdot v_x$, mentre la componente lungo $y$ della velocità dà luogo a una forza diretta lungo $x$ pari a $F_x = e \cdot B \cdot v_y$. La componente lungo $z$ della velocità non dà luogo ad alcuna forza. Applicando la legge di Newton si ha:

$$\begin{cases} \dfrac{dv_x}{dt} = \dfrac{qB}{m}v_y, & \text{(P1.35a)} \\ \dfrac{dv_y}{dt} = -\dfrac{qB}{m}v_x, & \text{(P1.35b)} \\ \dfrac{dv_z}{dt} = 0. & \text{(P1.35c)} \end{cases}$$

Eseguiamo ora la derivata della (P1.32a) e sostituiamo $v_y$ usando la seconda equazione. Otteniamo:

$$\frac{d^2v_x}{dt^2} = \frac{qB}{m}\frac{dv_y}{dt} = -\left(\frac{qB}{m}\right)^2\frac{dv_x}{dt}, \tag{P1.36}$$

dalla quale, ponendo

$$\Omega = \left(\frac{qB}{m}\right), \tag{P1.37}$$

si ottiene:

$$\frac{d^2v_x}{dt^2}+\Omega^2\frac{dv_x}{dt}=0. \tag{P1.38}$$

Similmente, dalla (P1.34b) si ricava:

$$\frac{d^2v_y}{dt^2}+\Omega^2\frac{dv_y}{dt}=0, \tag{P1.39}$$

le cui soluzioni sono:

$$v_x = A_1\cos(\Omega t)+B_1\sin(\Omega t), \tag{P1.39}$$

e

$$v_y = A_2\cos(\Omega t)+B_2\sin(\Omega t). \tag{P1.40}$$

Per come abbiamo scelto il sistema di riferimento, in relazione alle condizioni iniziali, si ha:

1) $v_x(0)=0$ dalla quale si ricava

$$A_1 = 0; \tag{P1.41}$$

2) $v_y(0)=0$ dalla quale si ricava

$$A_2 = 0; \tag{P1.42}$$

3) $v'_x(0)=v_t \cdot B \cdot q/m$ dalla quale si ricava:

$$B_1 = v_t; \tag{P1.43}$$

4) $v'_y(0)=0$ dalla quale si ricava:

$$B_2 = 0; \tag{P1.44}$$

le soluzioni delle (1.39a) e (1.39b) sono dunque:

$$v_x = v_t \sin(\omega t); \tag{P1.45}$$

$$v_y = v_t \cos(\omega t); \tag{P1.46}$$

Ove $v_t=q \cdot B/m$, avendo supposto che la carica $q$ è positiva.

**Problema 1.5**

Si dimostri che il campo elettrico di un'onda elettromagnetica che si propaga in un mezzo lineare con costante dielettrica $\varepsilon$, permettività magnetica $\mu$, conducibilità $\sigma$ e densità di carica $\rho = 0$ può essere determinato attraverso la seguente equazione

$$\nabla^2 \mathbf{E} - \mu\sigma \frac{\partial \mathbf{E}}{\partial t} - \frac{\partial^2 \mathbf{E}}{\partial t^2} = 0. \tag{P1.47}$$

che è detta *equazione del telegrafo*.

*Risoluzione*

Nel mezzo che consideriamo non vi sono cariche macroscopiche $\rho$, dunque la prima equazione di Maxwell div($\mathbf{D}$)=$\rho$ diventa div($\mathbf{D}$)=0. Nella quarta equazione dobbiamo considerare l'effetto della conducibilità non nulla, effetto per il quale è possibile la circolazione di correnti. Nella quarta equazione (curl($\mathbf{H}$)=$\mathbf{j}$+$\partial\mathbf{D}/\partial t$) il termine delle densita di corrente j dovrà perciò essere ritenuto. Le equazioni di Maxwell potranno perciò essere scritte nella forma:

$$\begin{cases} \text{div}(\mathbf{D}) = 0, & \text{(P1.48a)} \\ \text{div}(\mathbf{B}) = 0, & \text{(P1.48b)} \\ \text{curl}(\mathbf{E}) = -\dfrac{\partial \mathbf{B}}{\partial t}, & \text{(P1.48c)} \\ \text{curl}(\mathbf{H}) = j + \dfrac{\partial \mathbf{D}}{\partial t}. & \text{(P1.48d)} \end{cases}$$

Nelle (P1.48) possono essere introdotte le relazioni costitutive della materia che in questo caso assumono la semplice forma $\mathbf{D}=\varepsilon\cdot\mathbf{E}$ e $\mathbf{B}=\mu\cdot\mathbf{H}$. Per quanto riguarda j, essendo nota la conduttività $\sigma$, questo può essere convenientemente scritto in dipendenza di $\mathbf{E}$ nella forma $\mathbf{j}=\sigma\cdot\mathbf{E}$. Si ottengono così:

$$\begin{cases} \text{div}(\mathbf{E}) = 0, & \text{(P1.49a)} \\ \text{div}(\mathbf{B}) = 0, & \text{(P1.49b)} \\ \text{curl}(\mathbf{E}) = -\dfrac{\partial \mathbf{B}}{\partial t}, & \text{(P1.49c)} \\ \text{curl}(\mathbf{B}) = \mu\mathbf{j} + \mu\varepsilon \dfrac{\partial \mathbf{E}}{\partial t}. & \text{(P1.49d)} \end{cases}$$

Considerando le ultime due equazioni, si può procedere in maniera analoga a come di solito si fa quando si intende ricavare l'equazione delle onde nel vuoto o nel caso di un dieltrrico ordinario. Precisamente, applicando il rotore a entrambi i membri delle (P1.48c), e (P1.48d), si ottiene:

$$\begin{cases} \mathrm{curl}\left[\mathrm{curl}(\mathbf{E})\right] = -\frac{\partial}{\partial t}\mathrm{curl}(\mathbf{B}), & \text{(P1.50a)} \\ \mathrm{curl}\left[\mathrm{curl}(\mathbf{B})\right] = \mu \cdot \sigma \cdot \mathrm{curl}(\mathbf{E}) + \mu\varepsilon \cdot \frac{\partial}{\partial t}\mathrm{curl}(\mathbf{E}). & \text{(P1.50b)} \end{cases}$$

Introducendo poi le identità vettoriali:

$$\mathrm{curl}\left[\mathrm{curl}(\mathbf{E})\right] = \mathrm{grad}\left(\mathrm{div}(\mathbf{E})\right) - \nabla^2 \mathbf{E}, \tag{P1.51}$$

$$\mathrm{curl}\left[\mathrm{curl}(\mathbf{B})\right] = \mathrm{grad}\left(\mathrm{div}(\mathbf{B})\right) - \nabla^2 \mathbf{B} \tag{P1.52}$$

e tenendo presente le (P.1.48c) e (P.1.48d), dalle (P.1.49a) e (P.1.49b), si ricavano:

$$\begin{cases} -\nabla^2 \mathbf{E} = -\mu\sigma \frac{\partial \mathbf{E}}{\partial t} - \mu\varepsilon \frac{\partial^2 \mathbf{E}}{\partial t^2}, & \text{(P1.53a)} \\ -\nabla^2 \mathbf{B} = -\mu\sigma \frac{\partial \mathbf{B}}{\partial t} - \mu\varepsilon \frac{\partial^2 \mathbf{B}}{\partial t^2}, & \text{(P1.53b)} \end{cases}$$

e dunque:

$$\begin{cases} \nabla^2 \mathbf{E} - \mu\sigma \frac{\partial \mathbf{E}}{\partial t} - \mu \frac{\partial^2 \mathbf{E}}{\partial t^2} = 0, & \text{(P1.54a)} \\ \nabla^2 \mathbf{B} - \mu \cdot \sigma \cdot \frac{\partial \mathbf{B}}{\partial t} - \mu \cdot \frac{\partial^2 \mathbf{B}}{\partial t^2} = 0. & \text{(P1.54b)} \end{cases}$$

La (P.1.54a) è la risposta all'esercizio proposto. Si vede al contempo che un'equazione analoga è stata ricavata per il campo magnetico associato alla propagazione della onda radio.

**Problema 1.6**

Un sommergibile è in navigazione in mare aperto e comunica usando onde ELF, alla frequenza di 12 Hz. Sapendo che la permebilità magnetica dell'acqua di mare è pari a quella del vuoto, che la costante dielettrica relativa può essere assunta pari a 80 F/m, e che la conduttività solitamente accettata è 4 Ω ·m si determini a quale profondità il segnale cessa di essere ricevuto (rif1-Isavning and Koskinen, 2012).
.

Suggerimento: si applichi l'equazione del telegrafo:

$$\nabla^2 E - \mu\sigma \frac{\partial E}{\partial t} - \mu\varepsilon \frac{\partial^2 E}{\partial t^2} = 0, \tag{P1.47}$$

vista all'esercizio precedente.

*Risoluzione*

Consideriamo un'onda elettromagnetica che si propaga lungo l'asse *x* e scriviamone la componente del campo elettrico secondo la consueta notazione esponenziale $E_x = E_{0x} \exp i(\omega t - kx)$. Risulterà $\partial E_x/\partial t = i\omega E_{0x}$ e $\partial E_x/\partial x = -k^2 E_{0x}$. Dalla (P1.42) risulterà così:

$$k^2 E_x - \mu\sigma i\omega E_x - \mu\varepsilon\omega^2 E_x = 0, \tag{P1.47}$$

da cui:

$$k^2 = \mu\varepsilon\omega^2 + i\mu\sigma\omega, \tag{P1.48}$$

Essendo *k* una grandezza complessa, possiamo usare la sua rappresentazione in forma polare:

$$k = \rho e^{i\alpha}, \tag{P1.49}$$

e quindi

$$k^2 = \rho^2 e^{i2\alpha}. \tag{P1.50}$$

Adesso rappresentando il numero complesso $k^2$, dato dalla (P1.48) in forma polare, risulta:

$$\begin{cases} \rho^4 = \left(\mu\varepsilon\omega^2\right)^2 + \left(\mu\sigma\omega\right)^2, & \text{(P1.51a)} \\ 2\alpha = \arctan\left(\dfrac{\mu\sigma\omega}{\mu\varepsilon\omega^2}\right). & \text{(P1.51b)} \end{cases}$$

Si ottiene così:

$$\rho = \sqrt{\sqrt{\left(\mu\varepsilon\omega^2\right)^2 + \left(\mu\sigma\omega\right)^2}} = \sqrt{\omega\mu\sqrt{\varepsilon\omega^2 + \sigma^2}}, \tag{P1.52}$$

e

$$\alpha = \frac{1}{2}\arctan\left(\frac{\sigma}{\varepsilon\omega}\right). \tag{P1.56}$$

Questi risultati possono ora essere introdotti nella solita espressione esponenziale per le onde allo scopo di valutarne lo smorzamento. Otteniamo:

$$E_x = E_0 e^{i(kz-\omega t)} = E_0 e^{i[\rho(\cos\alpha + i\sin\alpha)z - \omega t]} = E_0 e^{i[\rho(\cos\alpha) - \omega t]} e^{-\rho\sin\alpha z}. \tag{P1.57}$$

Si vede dunque che l'ampiezza è scesa di $1/e$ quando l'onda ha percorso un tratto di lunghezza $\delta$ tale che $\rho \sin\alpha \, \delta = 1$. Si ha dunque:

$$\delta = \frac{1}{\rho \sin\alpha}. \tag{P1.58}$$

Sostituendo i valori numerici, otteniamo:

$$\rho = \sqrt{\omega\mu\sqrt{\varepsilon\omega^2 + \sigma^2}} = \sqrt{2\cdot 3.14\cdot 12\cdot 4\cdot 3.14\cdot 10^{-7}}\sqrt{8.85\cdot 10^{-12}\cdot 80\cdot (2\cdot 3.14\cdot 12)^2 + 5^2} = 0.0266$$

e

$$\alpha = \frac{1}{2}\arctan\left(\frac{\sigma}{\varepsilon\omega}\right) = \frac{1}{2}\arctan\left(\frac{5}{8.85\cdot 10^{-12}\cdot 80\cdot 2\cdot 3.14\cdot 12}\right) = \frac{\pi}{4}$$

per cui abbiamo:

$$\delta = \frac{1}{k\sin\alpha} = \frac{1}{0.0266\cdot \sin(\pi/4)} = 53 \text{ m}.$$

Supponendo dunque che le comunicazioni vadano distrutte quando l'ampiezza del segnale ricevuteo è diminuito di $e^{-1}$, avremo che questo avviene alla profondità di 53 m.

**Problema 1.7**

Per ciascuno dei plasmi seguenti si calcolino la lunghezza di Debye, la frequenza di plasma e la girofrequenza degli elettroni. Si verifichi il sussistere delle condizioni di plasma (rif1-Isavning and Koskinen, 2012).
1) Vento solare alla distanza dal Sole pari a quella della Terra; $T_e \approx 10$ eV, $n_e \approx 10^7$ m$^{-3}$, $B = 5$ nT.
2) Ionosfera diurna alla quota del massimo di densità elettronica nella regione F2 (~ 250 km): $T_e \approx 0.1$ eV, $n_e \approx 5\cdot 10^{11}$ m$^{-3}$, $B = 4.5\cdot 10^4$ nT.
3) Apparato per la fusione $T_e \approx 100$ eV, $n_e \approx 10^{22}$ m$^{-3}$ keV, $B = 1$ T.

*Risoluzione*

Per la lunghezza di Debye vale la seguente relazione, che qui riportiamo per comodità:

$$\lambda_D = \sqrt{\frac{\varepsilon_0 K_B T}{Ne^2}}, \tag{P1.59}$$

Osserviamo poi che nel testo di questo esercizio la temperatura è espressa in eV. Questo significa che viene fornito il valore di $K_B \cdot T$ espresso in eV, mentre, nella formula precedente, noi dobbiamo esprimere il valore di $K_B \cdot T$ in J. Dunque, ricordando la definizione di eV, come l'energia necessaria a far superare a un elettrone una barriera di potenziale di 1 V, risulta $E_{[eV]}\cdot e = E_{[J]}$, per cui dobbiamo sostituire nella (P1.59) $K_B \cdot T$ con $E_{[eV]}\cdot e$. Ricordando infine che la costante dielettrica del vuoto vale $\varepsilon_0 = 8.85\cdot 10^{-12}$ F m$^{-1}$, abbiamo:

$$\lambda_{D[SolarWind]} = \sqrt{\frac{8.85\cdot 10^{-12}(10\cdot 1.6\cdot 10^{-19})}{10^7\cdot 9\cdot 10^{-31}}} = 7.4 \text{ m},$$

$$\lambda_{D[\text{Ionosphere}]} = \sqrt{\frac{8.85 \cdot 10^{-12}(0.1 \cdot 1.6 \cdot 10^{-19})}{5 \cdot 10^{11} \cdot 9 \cdot 10^{-31}}} = 3.3 \cdot 10^{-3} \text{ m}$$

e

$$\lambda_{D[\text{FusionDevice}]} = \sqrt{\frac{8.85 \cdot 10^{-12}(100 \cdot 1.6 \cdot 10^{-19})}{10^{22} \cdot 9 \cdot 10^{-31}}} = 7.4 \cdot 10^{-7} \text{ m}.$$

Per la frequenza di plasma elettronica, vale la (1.6), che qui riportiamo per convenienza:

$$\omega_p = \sqrt{\frac{Ne^2}{m\varepsilon_0}}, \tag{P1.60}$$

per cui, passando ai valori numerici, risulta:

$$\omega_p = \sqrt{\frac{10^7(1.6 \cdot 10^{-19})}{9 \cdot 10^{-31} \cdot 8.85 \cdot 10^{-12}}} = 179 \cdot 10^3 \text{ rad/s},$$

$$\omega_p = \sqrt{\frac{5 \cdot 10^{11}(1.6 \cdot 10^{-19})}{9 \cdot 10^{-31} \cdot 8.85 \cdot 10^{-12}}} = 4 \cdot 10^7 \text{ rad/s},$$

e

$$\omega_p = \sqrt{\frac{10^{22}(1.6 \cdot 10^{-19})}{9 \cdot 10^{-31} \cdot 8.85 \cdot 10^{-12}}} = 2 \cdot 10^{21} \text{ rad/s}.$$

Per la girofrequenza degli elettroni vale la relazione:

$$\omega_B = \frac{eB}{m}, \tag{P1.61}$$

relazione che in questo esercizio assumiamo nota. Per cui passando ai valori numerici, si ha:

$$\omega_{B[\text{SolarWind}]} = \frac{1.6 \cdot 10^{-19} 5 \cdot 10^{-9}}{9 \cdot 10^{-31}} = 888 \text{ rad/s},$$

$$\omega_{B[\text{Ionosphere}]} = \frac{1.6 \cdot 10^{-19} 45000 \cdot 10^{-9}}{9 \cdot 10^{-31}} = 8 \cdot 10^6 \text{ rad/s},$$

e

$$\omega_{B[\text{FusionDevice}]} = \frac{1.6 \cdot 10^{-19} 1}{9 \cdot 10^{-31}} = 1.8 \cdot 10^{11} \text{ rad/s}.$$

Per verificare che le condizioni di plasma siano soddisfatte dobbiamo stabilire se la relazione (1.14):

$$\lambda_D \gg r, \tag{P1.62}$$

dove $r$ è la distanza fra gli elettroni. Questa possono essere stimate attraverso la densità, usando la relazione:

$$r \approx \frac{1}{N^{1/3}}, \tag{P1.63}$$

per cui si ha:

$$r_{\text{SolarWind}} \approx \frac{1}{\left(10^7\right)^{1/3}} = 4 \cdot 10^{-3}\,\text{m},$$

$$r_{\text{SolarWind}} \approx \frac{1}{\left(10^7\right)^{1/3}} = 4 \cdot 10^{-3}\,\text{m},$$

e

$$r_{\text{Ionosphere}} \approx \frac{1}{\left(5 \cdot 10^{11}\right)^{1/3}} = 1.25 \cdot 10^{-4}\,\text{m}.$$

Confrontando questi valori con le corrispondenti distanze di Debye, risultano verificate le seguenti disuguaglianze:

$$r_{\text{SolarWind}} \ll \lambda_{\text{SolarWind}},$$

$$r_{\text{Ionosphere}} \ll \lambda_{\text{Ionosphere}},$$

e

$$r_{\text{PlasmaDevice}} \ll \lambda_{\text{PlasmaDevice}}.$$

Possiamo perciò concludere che nei tre casi proposti le condizioni di plasma sono verificate.

**Problema 1.8**

Un contenitore sferico di raggio 1.5 m è riempito di idrogeno gassoso completamente ionizzato. La densità elettronica nel contenitore è pari a $4 \cdot 10^{23}$ m$^{-3}$. A che temperatura dovrebbe essere l'idrogeno ionizzato affinché siano soddisfatte le condizioni di plasma? Se la temperatura fosse posta a 40 °C, quale dovrebbe essere la densità elettronica affinché siano soddisfatte le condizioni di plasma? (rif1-Isavning and Koskinen, 2012).

*Risoluzione*

Per soddisfare le condizioni di plasma, deve essere verificata la seguente disuguaglianza $r \ll \lambda_D$, dove con $r$ si intende la distanza tipica fra le particelle e $\lambda_D$ è la lunghezza di Debye. Ora risulta:

$$r = \frac{1}{N^{1/3}} = \frac{1}{\left(4 \cdot 10^{23}\right)^{1/3}} = 1.35 \cdot 10^{-8} \, \text{m} \,,$$

mentre $\lambda_D$ può essere ricavata tramite la (1.10), che qui riportiamo per comodità:

$$\lambda_D = \sqrt{\frac{\varepsilon_0 K_B T_{\text{eff}}}{n_0 e^2}} \,. \tag{P1.64}$$

Dovendo essere $r \ll \lambda_D$, possiamo porre:

$$\lambda_D = \sqrt{\frac{\varepsilon_0 K_B \cdot T}{N e^2}} = 10 \frac{1}{N^{1/3}} \,.$$

Possiamo usare questa relazione per calcolare quale temperatura si dovrebbe avere perché la materia sia allo stato di plasma, alla densità data affiché la materia sia allo stato di plasma. Risulta:

$$T = 10^2 \frac{1}{N^{2/3}} \frac{Ne^2}{\varepsilon_0 K_B} = 100 \frac{N^{1/3} e^2}{\varepsilon_0 K_B} = 100 \frac{\left(4 \cdot 10^{23}\right)^{1/3} \left(1.6 \cdot 10^{-19}\right)^2}{8.85 \cdot 10^{-12} \, 1.38 \cdot 10^{-23}} = 1.5 \cdot 10^6 \, \text{K} \,.$$

La stessa relazione può essere usata per determinare quale dovrebbe essere la densità, affinché la materia si comporti come plasma alla temperatura di 40°C. Risulta:

$$N = \left(\frac{T \varepsilon_0 K_B}{100 e^2}\right)^3 = \frac{8.85 \cdot 10^{-12} \left(273 + 40\right) 1.38 \cdot 10^{-23}}{100 \cdot \left(1.6 \cdot 10^{-19}\right)^2} = 9.3 \cdot 10^{18} \, \text{m}^{-3} \,.$$

Calcoliamo anche $\lambda_D$ per i due stati fisici proposti in questo problema. Nel primo caso, con $n_o = 4 \cdot 10^{23}$ m$^{-3}$ e $T = 303$ K si ha:

$$\lambda_D = \sqrt{\frac{\varepsilon_0 K_B \cdot T}{N e^2}} = \frac{8.85 \cdot 10^{-12} \left(1.5 \cdot 10^6\right) \cdot 1.38 \cdot 10^{-23}}{4 \cdot 10^{23} \cdot \left(1.6 \cdot 10^{-19}\right)^2} = 1.33 \cdot 10^{-7} \, \text{m} \,.$$

Nel secondo caso, con $n_o = 9.3 \cdot 10^{18}$ m$^{-3}$ e T=$1.5 \cdot 10^6$ K si ha:

$$\lambda_D = \sqrt{\frac{\varepsilon_0 K_B \cdot T}{N e^2}} = \frac{8.85 \cdot 10^{-12} \left(273 + 40\right) \cdot 1.38 \cdot 10^{-23}}{9.3 \cdot 10^{18} \cdot \left(1.6 \cdot 10^{-19}\right)^2} = 4 \cdot 10^{-7} \, \text{m} \,.$$

**Problema 1.9**

Il grado di ionizzazione è descritto dalla seguente equazione, detta equazione di Saha:

$$\frac{n_i^2}{n_n} = 3 \cdot 10^{27} T^{3/2} e^{-U/T} \tag{P.1.65}$$

ove $n_i$ è la densità numerica della specie ionica considerata, $n_n$ la corrispondente densità numerica degli atomi neutri, $U$ l'energia di ionizzazione (espressa in eV) e T la temperatura (espressa in eV). Nella regione F2 lo ione dominante è O+. Esso ha una energia di ionizzazione di 13.62 eV, e la densità osservata è $10^{11}$ m$^{-3}$. Si calcoli il grado di ionizzazione alla temperatura di 2000 K, 2200 K, 2400 K (rif1-Isavning and Koskinen, 2012).

*Risoluzione*

Esprimiamo le temperature in Joule:

$$T_{1[J]} = T_{1[K]} K_B = 2000 \cdot 1.38 \cdot 10^{-23} = 2.76 \cdot 10^{-20} \, \text{J},$$

$$T_{2[J]} = T_{2[K]} K_B = 2200 \cdot 1.38 \cdot 10^{-23} = 3.04 \cdot 10^{-20} \, \text{J},$$

$$T_{3[J]} = T_{3[K]} K_B = 2400 \cdot 1.38 \cdot 10^{-23} = 3.31 \cdot 10^{-20} \, \text{J}.$$

Da qui passiamo agli eV, ottenendo:

$$T_{1[eV]} = T_{1[J]} e = 2.76 \cdot 10^{-20} \cdot 1.6 \cdot 10^{-19} = 0.44 \, \text{eV},$$

$$T_{2[eV]} = T_{2[J]} e = 3.04 \cdot 10^{-20} \cdot 1.6 \cdot 10^{-19} = 0.49 \, \text{eV},$$

$$T_{3[eV]} = T_{3[J]} e = 3.31 \cdot 10^{-20} \cdot 1.6 \cdot 10^{-19} = 0.53 \, \text{eV}.$$

Quindi, usando la (1.65), otteniamo:

$$[n_i]_{T_1} = \sqrt{8.89 \cdot 10^{-19} \cdot 2000^{3/2} 10^{11} e^{13.62/0.44}} = 1.89 \cdot 10^{-5} \, \text{m}^{-3},$$

$$[n_i]_{T_2} = \sqrt{8.89 \cdot 10^{-19} \cdot 2200^{3/2} 10^{11} e^{13.62/0.49}} = 8.54 \cdot 10^{-5} \, \text{m}^{-3},$$

$$[n_i]_{T_3} = \sqrt{8.89 \cdot 10^{-19} \cdot 2400^{3/2} 10^{11} e^{13.62/0.53}} = 2.28 \cdot 10^{-4} \, \text{m}^{-3}.$$

Si osserva che il grado di ionizzazione aumenta rapidamente con la temperatura.

**Problema 1.10**

Si verifichi che nella ionosfera vale la condizione di plasma non caldo.

*Risoluzione*

Assumiamo per gli elettroni una temperatura $T_e$=1000 K, ed una massa $m_e$=9·10$^{-31}$. Risulta:

$$v_\text{t} = \sqrt{\frac{K_B T}{m}} = \sqrt{\frac{1.38 \cdot 10^{-23} \cdot 1000}{9 \cdot 10^{-31}}} = 5 \cdot 10^5 \,\text{m/s}. \tag{P1.66}$$

Si che la condizione $v_\text{t} \ll c$, di modo che si vede che il plasma nella ionosfera può essere considerato freddo.

**Bibliografia del Capitolo 1**

# CAPITOLO 2

## Indice di rifrazione delle onde radio nella ionosfera

**Riassunto**

A partire dalle equazioni di Maxwell ed usando le equazioni costitutive della materia viene ricavata l'espressione per l'indice di rifrazione per le onde radio nella ionosfera, nota come formula di Appleton-Hartree.

### 2.1 Ipotesi della teoria magnetoionica

Siamo ora interessati a sviluppare una parte molto importante della teoria della propagazione delle onde radio in un magnetoplasma non caldo, nota come *teoria magnetoionica*. Precisamente è scopo di questo capitolo ricavare l'indice di rifrazione per le onde radio che si propagano nella ionosfera.
Come s'è detto, trascurando le collisioni, l'ipotesi che il magnetoplasma non sia caldo si rende necessaria perché dobbiamo supporre che gli elettroni non possano percorrere una distanza $r_{corr}$ maggiore o dell'ordine della lunghezza d'onda $\lambda$ del campo elettromagnetico, nel periodo $T$ di una sua oscillazione. Volumetti di dimensione caratteristica $r_{corr}$, di conseguenza, dimostrano una risposta collettiva. Se questi contengono un grande numero di particelle, sarà possibile scrivere la teoria macroscopica. Le grandezze che interverranno nelle equazioni, dunque, devono essere pensate come mediate su volumi grandi almeno $r_{corr}$. In questo modo possiamo assumere che la risposta del mezzo sia puntuale, ossia la polarizzazione dipende solo dal campo elettrico nel punto considerato secondo una relazione del tipo $\mathbf{P} = \chi \cdot \mathbf{E}$, dove $\chi$ può essere un tensore ma diviene senz'altro uno scalare dipendente unicamente dalla frequenza nel caso di mezzo omogeneo isotropo.

### 2.2 La relazione di polarizzazione

Nelle ipotesi riaasunte nel paragrafo precedente, possiamo cominciare a riprendere le equazioni Maxwell (1.29a), (1.29b), (1.29c) e (1.29d) nella loro forma generale, che abbiamo già introdotto al Paragrafo 1.7, e che qui riscriviamo per comodità:

$$\begin{cases} \mathrm{div}(\mathbf{D}) = \rho & \text{(2.1a)} \\ \mathrm{div}(\mathbf{B}) = 0 & \text{(2.1b)} \\ \mathrm{curl}(\mathbf{E}) = -\dfrac{\partial \mathbf{B}}{\partial t} & \text{(2.1c)} \\ \mathrm{curl}(\mathbf{H}) = \mathbf{j} + \dfrac{\partial \mathbf{D}}{\partial t} & \text{(2.1d)} \end{cases}$$

Noi procederemo tenendo conto della presenza delle cariche elettriche includendo la densità di corrente **j** nella quarta equazione di Maxwell, mentre la costante dielettrica la assumeremo pari a quella del vuoto (**D** = $\varepsilon_0$ **E**) . L'altra possibilità sarebbe stata quella di considerare nulla la corrente elettrica **j** e di includere l'effetto delle cariche elettriche nella costante dielettrica. In questo caso avremmo dovuto scrivere **D** = $\varepsilon_0 \varepsilon_r$ **E.**

Nelle equazioni di Maxwell (2.1a), (2.1b), (2.1c) e (2.1d) introdurremo le seguenti relazioni che sussistono fra **E** e **D**, e fra **H** e **B**:

$$\mathbf{D} = \varepsilon_0 \mathbf{E} \tag{2.2}$$

e

$$\mathbf{B} = \mu_0 \mathbf{H}. \tag{2.3}$$

Anche la permettività magnetica è stata dunque assunta pari a quella del vuoto. Le equazioni di Maxwell (2.1a), (2.1b), (2.1c) e (2.1d) divengono così:

$$\begin{cases} \text{div}(\mathbf{E}) = \rho, & (2.4a) \\ \text{div}(\mathbf{B}) = 0, & (2.4b) \\ \text{curl}(\mathbf{E}) = -\dfrac{\partial \mathbf{B}}{\partial t}, & (2.4c) \\ \text{curl}(\mathbf{B}) = \mu_0 \mathbf{j} + \varepsilon_0 \mu_0 \dfrac{\partial \mathbf{E}}{\partial t}. & (2.4d) \end{cases}$$

I due rotori possono essere sviluppati ottenedo:

$$\text{curl}(\mathbf{E}) \equiv \begin{pmatrix} \mathbf{i} & \mathbf{j} & \mathbf{k} \\ 0 & 0 & -ik_z \\ E_x & E_y & E_z \end{pmatrix} \equiv \begin{pmatrix} ik_z E_y \\ -ik_z E_x \\ 0 \end{pmatrix} \tag{2.5}$$

e

$$\text{curl}(\mathbf{B}) \equiv \begin{pmatrix} \mathbf{i} & \mathbf{j} & \mathbf{k} \\ 0 & 0 & -ik_z \\ B_x & B_y & B_z \end{pmatrix} \equiv \begin{pmatrix} ik_z B_y \\ -ik_z B_x \\ 0 \end{pmatrix}. \tag{2.6}$$

La (2.4c) può dunque essere sviluppata nelle sue componenti, ottenendo:

$$\begin{cases} ik_z E_y = -i\omega B_x, & (2.7a) \\ -ik_z E_x = -i\omega B_y, & (2.7b) \\ 0 = -i\omega B_z, & (2.7c) \end{cases}$$

dalle quali:

$$\begin{cases} -(k_z/\omega)\cdot E_y = B_x, & (2.8a) \\ (k_z/\omega)\cdot E_x = B_y, & (2.8b) \\ 0 = -i\omega B_z. & (2.8c) \end{cases}$$

Procedendo allo stesso modo dalla (2.4d) si possono ottenere le relazioni seguenti:

$$\begin{cases} ik_z B_y = \mu_0 j_x + i\omega\varepsilon_0\mu_0 E_x, & (2.9a) \\ ik_z B_x = \mu_0 j_y + i\omega\varepsilon_0\mu_0 E_y, & (2.9b) \\ 0 = \mu_0 j_z + i\omega\varepsilon_0\mu_0 E_z. & (2.9c) \end{cases}$$

In queste ultime espressioni, è possibile eliminare le componenti di **j** usando le relazioni fra la polarizzazione e le correnti elettriche (**P**=$Ne$**r**, $i\omega$**P**=**j**), relazioni che abbiamo già visto al Paragrafo 1.5. Esse diventano così:

$$\begin{cases} ik_z B_y = i\omega\mu_0 P_x + i\omega\varepsilon_0\mu_0 E_x & (2.10a) \\ -ik_z B_x = i\omega\mu_0 P_y + i\omega\varepsilon_0\mu_0 E_y & (2.10b) \\ 0 = i\omega\mu_0 P_z + i\omega\varepsilon_0\mu_0 E_z & (2.11c) \end{cases}$$

Ora abbiamo dunque una coppia di sistemi di equazioni: le (2.8a), (2.8b) e (2.8c) che derivano da curl(**E**)=-∂**B**/∂$t$, e le (2.10a), (2.10b) e (2.10c), che derivano da curl(**B**)= $\mu_0$**j** +$\varepsilon_0\mu_0$∂**E**/∂$t$. Queste relazioni possono essere messe insieme, usando le (2.8a), (2.8b) e (2.8c) per esprimere nelle (2.10a), (2.10b) e (2.10c) le componenti di **B** in funzione delle componenti di **E**. Si ottiene così:

$$\begin{cases} i(k_z^2/\omega)E_x = i\omega\mu_0 P_x + i\omega\varepsilon_0\mu_0 E_x, & (2.12a) \\ i(k_z^2/\omega)E_y = i\omega\mu_0 P_y + i\omega\varepsilon_0\mu_0 E_y, & (2.12b) \\ 0 = i\omega\mu_0 P_z + i\omega\varepsilon_0\mu_0 E_z, & (2.12c) \end{cases}$$

da cui

$$\begin{cases} (k_z^2/\omega^2)E_x = \mu_0 P_x + \varepsilon_0\mu_0 E_x, & (2.13a) \\ (k_z^2/\omega^2)E_y = \mu_0 P_y + \varepsilon_0\mu_0 E_y, & (2.13b) \\ 0 = P_z + \varepsilon_0 E_z. & (2.13c) \end{cases}$$

In queste relazioni si può far apparire la velocità nella luce nel vuoto $c$ e la velocità di di fase $v_f$ delle onde radio, ricordando che $v_f=\omega/k$ e $c=(\varepsilon_0\mu_0)^{-1/2}$. Esse diventano così:

$$\begin{cases} \left(E_x/v_f^{\;2}\right)-\left(E_x/c^2\right)=\mu_0 P_x, & \text{(2.14a)} \\ \left(E_y/v_f^{\;2}\right)-\left(E_y/c^2\right)=\mu_0 P_y, & \text{(2.14b)} \\ -\varepsilon_0 E_z = P_z. & \text{(2.14c)} \end{cases}$$

In queste può essere utilmente introdotto l'indice di rifrazione di fase attraverso la relazione $n_f = c/v_f$, per cui divengono:

$$\begin{cases} \left[\left(n_f^{\;2}-1\right)/c^2\right]E_x = \mu_0 P_x, & \text{(2.15a)} \\ \left[\left(n_f^{\;2}-1\right)/c^2\right]E_y = \mu_0 P_y, & \text{(2.15b)} \\ -\varepsilon_0 E_z = P_z, & \text{(2.15c)} \end{cases}$$

oppure:

$$\begin{cases} \varepsilon_0\left[n_f^{\;2}-1\right]E_x = P_x, & \text{(2.16a)} \\ \varepsilon_0\left[n_f^{\;2}-1\right]E_y = P_y, & \text{(2.16b)} \\ -\varepsilon_0 E_z = P_z. & \text{(2.16c)} \end{cases}$$

Per ottenere effettivamente l'indice di rifrazione delle onde elettromagnetiche nel magnetoplasma bisogna ricorrere anche alle equazioni costitutive della materia, che abbiamo ricavato nel Paragrafo 1.7 e che andiamo qui a richiamare per comodità. Queste, solitamente vengono scritte nella forma:

$$\begin{cases} \varepsilon_0 X E_x = -P_x - iY_L P_y + iY_T P_z, & \text{(2.17a)} \\ \varepsilon_0 X E_y = -P_y + iY_L P_x, & \text{(2.17b)} \\ \varepsilon_0 X E_z = -P_z - iY_T P_x. & \text{(2.17c)} \end{cases}$$

Dalle equazioni (2.16a) e (2.16b) si può subito vedere che vale la seguente uguaglianza:

$$\frac{E_x}{P_x} = \frac{E_y}{P_y}, \tag{2.18}$$

per cui si vede che è possibile dividere la (2.17a) per $P_x$ e la (2.17b) per $P_y$ ottenendo due equazioni i cui membri di sinistra sono uguali, in forza della (2.18). Poi, se nella (2.17a) e (2.17b) non comparisse $P_z$, uguagliando i membri di sinistra di queste equazioni, sarebbe possibile ricavare il rapporto $P_x/P_y$. Da questo potremmo ricavare $E_x/P_x$ tramite la (2.17a), oppure $E_y/P_y$ tramite la (2.17b); con questo, infine, si potrebbe ricavare $n_f$, tramite la (2.16a) (o 2.16b). Operiamo perciò in questo modo: utilizziamo le relazioni (2.16c) e (2.17c) per sostituire $P_z$ nella (2.17a).
Possiamo introdurre la (2.16c) nella (2.17c), ottenendo:

$$-P_z X = -P_z - iY_T P_x, \tag{2.19}$$

da cui

$$P_z = -\frac{iY_T}{(1-X)} P_x. \tag{2.20}$$

Che può essere inserita nella (2.17a), ottenendo la (2.21a). Per comodità risicriviamo la (2.21a) a sistema con la (2.17b):

$$\begin{cases} \varepsilon_0 X E_x = -P_x - iY_L P_y + \dfrac{Y_T^{\,2}}{(1-X)} P_x, & (2.21a) \\ \varepsilon_0 X E_y = -P_y + iY_L P_x. & (2.21b) \end{cases}$$

Dividiamo la prima equazione per $P_x$ e la seconda per $P_y$, ottenendo:

$$\begin{cases} \varepsilon_0 X \dfrac{E_x}{P_x} = -1 - iY_L \dfrac{P_y}{P_x} + \dfrac{Y_T^{\,2}}{(1-X)}, & (2.22a) \\ \varepsilon_0 X \dfrac{E_y}{P_y} = -1 + iY_L \dfrac{P_x}{P_y}. & (2.22b) \end{cases}$$

Usando la (2.18) possiamo uguagliare i membri di destra, ottenendo:

$$-1 + iY_L \frac{P_x}{P_y} = -1 - iY_L \frac{P_y}{P_x} + \frac{Y_T^{\,2}}{(1-X)}, \tag{2.23}$$

e, introducendo il seguente parametro:

$$\rho = \frac{P_x}{P_y} = \frac{E_x}{E_y}, \tag{2.24}$$

si ricava:

$$+iY_L \rho = -iY_L \frac{1}{\rho} + \frac{Y_T^{\,2}}{(1-X)}, \tag{2.25}$$

e quindi:

$$+iY_L \rho^2 - \frac{Y_T^{\,2}}{(1-X)} \rho + iY_L = 0, \tag{2.26}$$

le cui soluzioni sono:

$$\rho_{1,2} = \frac{1}{i2Y_L} \left[ \frac{Y_T^{\,2}}{(1-X)} \mp \sqrt{\frac{Y_T^{\,4}}{(1-X)^2} + 4Y_L^{\,2}} \right], \tag{2.27}$$

che costituisce la relazione di polarizzazione.

## 2.3 L'equazione di Appleton-Hartree

Dalla (2.16b) (potevamo usare anche la (2.16a)), si ha:

$$n^2 = \frac{1}{\varepsilon_0} \frac{P_y}{E_y} + 1, \tag{2.28}$$

dalla quale si vede che ci serve il rapporto $P_y/E_y$. Questo lo si può ricavare dalla (2.22b), in funzione di $\rho$, che conosciamo. Si ha:

$$\frac{1}{\varepsilon_0 X} \frac{P_y}{E_y} = \frac{1}{-1 + iY_L \frac{P_x}{P_y}}, \tag{2.29}$$

e quindi:

$$\frac{P_y}{E_y} = \frac{\varepsilon_0 X}{-1 + iY_L \rho}. \tag{2.30}$$

In questa ultima relazione introducendo i possibili valori di $\rho$, dati dalla (2.27), si ha:

$$\frac{P_y}{E_y} = \frac{\varepsilon_0 X}{-1 + iY_L \left\{ \frac{1}{i2Y_L} \left[ \frac{Y_T^2}{(1-X)} \pm \sqrt{\frac{Y_T^4}{(1-X)^2} + 4Y_L^2} \right] \right\}}. \tag{2.31}$$

che può essere introdotta nella (2.28) per ottenere:

$$n^2 = \frac{1}{\varepsilon_0} \frac{P_y}{E_y} + 1 = \frac{1}{\varepsilon_0} \frac{\varepsilon_0 X}{-1 + iY_L \left\{ \frac{1}{i2Y_L} \left[ \frac{Y_T^2}{(1-X)} \pm \sqrt{\frac{Y_T^4}{(1-X)^2} + 4Y_L^2} \right] \right\}} + 1 =$$

$$= \frac{X}{-1 + \left\{ \frac{1}{2} \left[ \frac{Y_T^2}{(1-X)} \pm \sqrt{\frac{Y_T^4}{(1-X)^2} + 4Y_L^2} \right] \right\}} + 1. \tag{2.32}$$

e, in definitiva:

$$n^2 = 1 - \frac{X}{1 - \frac{1}{2} \frac{Y_T^2}{(1-X)} \pm \sqrt{\frac{Y_T^4}{4(1-X)^2} + Y_L^2}} \tag{2.33}$$

che è nota come *equazione di Appleton-Hartree* (rfr2-Ratcliffe, 1959; Budden, 1961). La presenza del doppio segno in questa equazione è legata a un fenomeno per certi versi analogo a quello che si verifica in ottica, quando alcune sostanze, generalmente cristalli, dànno luogo al fenomeno della *doppia rifrazione* (o *birifrazione*, o *birifrangenza*), consistente nello sdoppiamento di un raggio incidente in due raggi rifratti. Questi, hanno diverse velocità di propagazione, e, di conseguenza, ad ognuno corrispondono indici di rifrazione diversi.

Nel caso del magnetoplasma ionosferico, rispetto alle onde radio, la birifrangenza è dovuta all'anisotropia indotta da **B**. Il linguaggio, poi, è stato mutuato da quello dell'ottica. Si dirà così che l'onda radio propagantesi nella ionosfera si divide in due *componenti magnetoioniche*, quella *ordinaria* e quella *straordinaria*, a ciascuna delle quali può essere ascritto un rispettivo indice di rifrazione. Con lo stesso linguaggio, parleremo anche di *raggio ordinario* e *raggio straordinario*.

Si può poi dimostrare che, nel caso in cui vengano considerate le collisioni fra elettroni e molecole neutre, la (2.33) diventa:

$$n^2 = 1 - \frac{X}{1 - \frac{1}{2}\frac{Y_T^2}{(1-X-iZ)} \pm \sqrt{\frac{Y_T^4}{4(1-X-iZ)^2} + Y_L^2}} . \qquad (2.34)$$

Ricordiamo anche il significato dei vari termini che compaiono nell'equazione precedente:
$X=\omega_p^2/\omega^2$ (essendo $\omega_p$ la frequenza di plasma e $\omega$ la frequenza angolare dell'onda radio);
$Y_T=Y\cdot\sin\theta$, $Y_L=Y\cdot\cos\theta$ (essendo $\theta$ l'angolo tra il vettore d'onda e il campo magnetico della Terra),
con $Y=\omega_B/\omega$ (essendo $\omega_B$ la girofrequenza, e $B$ l'intensità del campo magnetico della Terra);
$Z = \nu/\omega$ (ove $\nu$ è la frequenza delle collisioni elettrone-molecola neutra);
Gli elementi geometrici della teoria magnetoionica, che intervengono nelle definizioni sono riportati, per maggior chiarezza, per maggior chiarezza, nella Fig. 2.1.

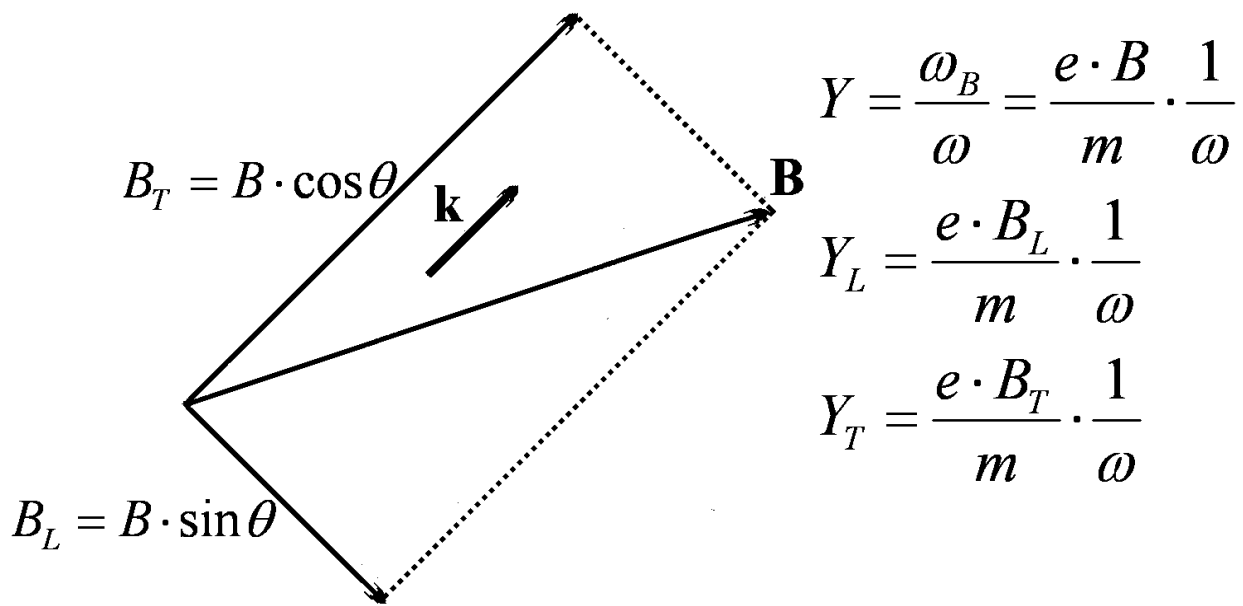

Fig. 2.1. Elementi geometrici della teoria magnetoionica. **k** è il vettore d'onda ed ha la stessa direzione della propagazione dei fronti della onda radio, supposta localmente piana. **B** è il vettore induzione magnetica terrestre.

Infine, vale la pena osservare che a partire dalla (2.34), considerando il caso di assenza di campo magnetico e di collisioni si ottiene la seguente semplice espressione per $\mu_\text{f}$:

$$\mu_\text{f}^2 = 1 - X \,. \tag{2.35}$$

**Bibliografia del Capitolo 2**

# CAPITOLO 3

## Conseguenze dell'equazione di Appleton-Hartree

**Riassunto**

Vengono interpretate le parti reali ed immaginaria dell'indice di rifrazione che si ricava con l'equazione di Appleton-Hartee. Vengono determinate le condizioni di riflessione, sia per la componente ordianaria che per quella straordinaria. Viene posto il problema della corretta scelta del segno nella stessa equazione, e viene introdotta la regola di Booker in modo semplice.

### 3.1 Discussione dell'equazione di Appleton-Hartree

Abbiamo visto, nel capitolo precedente, il risultato fondamentale della teoria magnetoionica, che è l'equazione (2.34):

$$n^2 = 1 - \frac{X}{1 - \frac{1}{2}\frac{Y_T^2}{(1-X-iZ)} \pm \sqrt{\frac{Y_T^4}{4(1-X-iZ)^2} + Y_L^2}}. \tag{3.1}$$

Come abbiamo detto, essa esprime l'indice di rifrazione di fase per le onde elettromagnetiche che si propagano nella ionosfera, nel caso più generico, quando cioè si tiene conto sia della presenza di **B** che delle collisioni fra elettroni e molecole neutre. La presenza del doppio segno esprime la birifrangenza alla quale si è fatto cenno nel capitolo precedente.

### 3.2 Interpretazione dell'indice di rifrazione complesso

Si deve notare che l'indice di rifrazione espresso dalla (3.1) è una grandezza complessa. Per darne una interpretazione dobbiamo scriverlo secondo le sue parti reale $\mu_f$ ed immaginaria $\chi$:

$$n_f = \mu_f - j\,\chi. \tag{3.2}$$

Successivamente dobbiamo ricordare le relazioni con le quali si collegano la velocità di fase $v_f$, il numero d'onda $k$, la frequnza angolare $\omega$ e $\mu_f$. Queste sono:

$$v_f = \frac{\omega}{k}, \tag{3.3}$$

$$\mu_f = \frac{c}{v_f}. \tag{3.4}$$

Si consideri il campo elettrico - ma la stessa cosa vale per il campo magnetico - di un'onda elettromagnetica caratterizzata da un certo $\omega$ in un mezzo avente indice di rifrazione di fase $n_f$. Per una qualsiasi delle componenti di **E** – per fissare le idee consideriamo $E_y$ – riusulta:

$$E_y = A\exp\left[j(\omega t - kx)\right] = A\exp\left[j\omega\left(t - \frac{k}{\omega}x\right)\right] =$$

$$= A\exp\left[j\omega \cdot \frac{1}{c}\left(ct - \frac{c}{v_f}x\right)\right] = A\exp\left[j\omega \cdot \frac{1}{c}(ct - n_f x)\right]. \tag{3.5}$$

Se noi introduciamo la relazione (3.2) nella (3.5), otteniamo:

$$E_y = A\exp j\omega\left[t - \frac{\mu_f}{c}x\right] \cdot \exp\left[-i\omega \cdot \frac{1}{c}(-i\chi)x\right] = \tag{3.6}$$

$$= A\exp j\omega\left[(t - v_f \cdot x)\right] \cdot \exp\left[-\frac{\omega\chi}{c}x\right]. \tag{3.7}$$

Dunque vediamo che quando un mezzo ha un indice di rifrazione complesso, la parte reale è associata alla propagazione dei fronti dell'onda elettromagnetica considerata, mentre la parte immaginaria è associata al suo smorzamento.
Osserviamo poi che nel caso in cui non vi siano collisioni, la (3.1) diventa:

$$n^2 = 1 - \frac{X}{1 - \frac{1}{2}\frac{Y_T^2}{(1-X)} \pm \sqrt{\frac{Y_T^4}{4(1-X)^2} + Y_L^2}}. \tag{3.8}$$

dalla quale deduciamo che in questo caso il quadrato dell'indice di rifrazione è reale, per cui l'indice di rifrazione può essere o reale o puramente immaginario. Ne risulta che non si può avere smorzamneto per l'onda elettromagnetica, perché, come abbiamo appena visto, in questo caso si deve avere un indice di rifrazione in parte reale e in parte immaginario. Un indice di rifrazione puramente immaginario, viceversa, si associa ad onde evanescenti (si veda l'Appendice 1), per le quali non si ha trasporto di energia. Si trae la conclusione, che lo smorzamento delle onde è associato a Z, che è legato alle collisioni. Peraltro, vale la pena far notare che, dal punto di vista termodinamico, le collisioni fra elettroni e molecole neutre costituiscono il meccanismo tramite il quale l'energia ordianata del campo elettromagnetico è convertita nella ionosfera in energia termica.

### 3.3 Condizioni di riflessione in assenza di collisioni

Riveste rilevanza, per le molteplici applicazioni pratiche, determinare quali sono le condizioni di riflessione di un'onda radio che incide la ionosfera verticalmente. Queste condizioni, vengono per ora stabilite in assenza di collisioni e in presenza di campo magnetico. Ci limitiamo a dare due giustificazioni, che non costituiscono delle vere dimostrazioni, del fatto che la condizione di riflessione può essere espressa come $\mu_f = 0$. Un'altra giustificazione verrà data nel capitolo successivo, quando calcoleremo $\mu_g$, con il campo magnetico e senza collisioni. Nel Capitolo 5 mostreremo che le condizioni di riflessione in presenza di collisioni non possono essere espresse in modo altrettanto semplice.

Una prima giustificazione del fatto che $\mu_f = 0$ rappresenta la condizione di riflessione, può essere trovata semplicemente nel fatto che, come abbiamo visto al paragrafo precedente, $\mu_f$ è associato alla propagazione dell'onda. Come conseguenza $\mu_f = 0$ implica che l'onda non si propaga, per cui possiamo pensare di esprimere in questo modo la condizione di riflessione.

Una seconda giustificazione può essere ricavata considerando il caso di assenza di campo magnetico ed assenza di collisioni. In questo caso per l'indice di rifrazione di fase si ha:

$$\mu_f = \frac{c}{v_f} = c\frac{k}{\omega}, \tag{3.9}$$

mentre per quello di gruppo

$$\mu_g = \frac{c}{v_g}. \tag{3.10}$$

Essendo $v_g = d\omega/dk$, la equazione precedente (3.10), diventa:

$$\mu_g = c\frac{dk}{d\omega} = \frac{d}{d\omega}(ck), \tag{3.11}$$

la quale, usando la (3.9), diventa:

$$= \frac{d}{d\omega}(\omega \cdot \mu_f) = \mu_f + \omega\frac{d\mu_f}{d\omega}. \tag{3.12}$$

In questa relazione possiamo ora esplicitare la dipendenza di $n_f$ da ω utilizzando la (2.35), ove abbiamo esplicitato $n_f = 1-X = 1-\omega_p^2/\omega^2$. In questo modo si ottiene:

$$\mu_f = \sqrt{1-\frac{\omega_p^2}{\omega^2}} + \omega\frac{d}{d\omega}\sqrt{\left(1-\frac{\omega_p^2}{\omega^2}\right)} =$$

$$= \sqrt{1-\frac{\omega_p^2}{\omega^2}} + \omega\left(-\frac{1}{2}\right)\frac{(-2)\frac{\omega_p^2}{\omega^3}}{\sqrt{1-\frac{\omega_p^2}{\omega^2}}} =$$

$$= \sqrt{1-\frac{\omega_p^2}{\omega^2}} + \frac{\frac{\omega_p^2}{\omega^2}}{\sqrt{1-\frac{\omega_p^2}{\omega^2}}} =$$

$$= \frac{1-\frac{\omega_p^2}{\omega^2}+\frac{\omega_p^2}{\omega^2}}{\sqrt{1-\frac{\omega_p^2}{\omega^2}}} = \frac{1}{\mu_f}. \tag{3.13}$$

In sintesi otteniamo il seguente risultato importante:

$$\mu_g = \frac{1}{\mu_f}. \qquad (3.14)$$

Si vede così che $\mu_g = \mu_f^{-1}$, per cui, quando abbiamo $\mu_f \to 0$ abbiamo $\mu_g \to \infty$, cosa che, in base alla (3.10), implica $v_g \to 0$. Siccome è noto che $v_g$ è associata al trasporto di energia, questo significa che, per $\mu_f \to 0$, l'energia non si propaga più per cui si verifica la riflessione dell'onda elettromagnetica. Tenendo questo risultato in mente (cioè che $\mu_g \to \infty$, quando $\mu_f \to 0$), valido senza campo magnetico e senza collisioni, possiamo supporre che, anche in presenza di campo magnetico, $\mu_g \to \infty$, quando $\mu_f \to 0$.

### 3.4 Gli zeri dell'equazione di Appleton Hartree

In base a quanto detto nel paragrafo precedente, si capisce che è importante determinare gli zeri dell'equazione di Appleton Hartree, poiché esse esprimono le condizioni di riflessione. Considereremo separatamente i casi di propagazione longitudinale e trasversale.

*Caso di propagazione longitudinale ($Y_T = 0$)*

La relazione (3.1) diventa:

$$\mu_f^2 = 1 - \frac{X}{1 \pm Y}, \qquad (3.15)$$

per cui, ponendo $\mu^2 = 0$ si ha:

$$X = 1 \pm Y. \qquad (3.16)$$

*Caso di propagazione trasversale ($Y_L = 0$)*

a) Segno superiore

Con il segno superiore la relazione (3.1) assume valori diversi a seconda che $(1-X)$ è maggiore o minore di 0, cioè a seconda che $X$ è maggiore o minore di 1.

Se $X < 1$, allora abbiamo

$$\mu^2 = 1 - X, \qquad (3.17)$$

per cui $\mu^2 = 0$ implica

$$X = 1; \qquad (3.18)$$

se $X > 1$, allora abbiamo:

$$\mu^2 = 1 - \frac{X}{1 - \frac{Y^2}{(1-X)}}, \qquad (3.19)$$

per cui $\mu^2 = 0$ implica

$$X = 1 + Y. \tag{3.20}$$

b) Segno inferiore

Se $X<1$, allora abbiamo:

$$\mu^2 = 1 - \frac{X}{1 - \frac{Y^2}{(1-X)}}, \tag{3.21}$$

per cui $\mu^2 = 0$ implica

$$X = 1 - Y; \tag{3.22}$$

se $X > 1$, allora abbiamo:

$$\mu^2 = 1 - X, \tag{3.23}$$

per cui $\mu^2 = 0$ implica:

$$X = 1. \tag{3.24}$$

Risulta naturale interpretare le soluzioni (3.18) e (3.24) come appartenenti allo stesso modo di propagazione, mentre la (3.20) e la (3.22) all'altro. Questo risulterà più chiaro nel seguito, ove studieremo la forma delle funzioni $\mu_f(f)$ che si ottengono dalla (3.1), considerando entrambi i segni che vi compaiono.
Inoltre siccome nelle (3.20) e (3.22) interviene, attraverso $Y$, **B** e l'angolo $\theta$ che questo forma con la direzione di **k**, esse saranno riferite al modo di propagazione che definiremo *straordinario*. Siccome nelle (3.18) e (3.24), invece, non intervengono i valori del campo magnetico, esse saranno riferite al modo di propagazione che definiremo *ordinario*.
In altre parole, quando poniamo $\mu_f = 0$ nella cerchiamo (3.1) e cerchiamo le soluzioni dobbiamo operare nel modo seguente:
  a) se $X<1$ per il modo ordinario si prende il segno + e per lo straordinario il segno -;
  b) se $X>1$ per il modo ordinario si prende il segno − e per lo straordinario il +.
Quanto abbiamo appena affermato è noto come *regola di Booker*.
La necessità di questa regola si mostra evidente quando andiamo a studiare i grafici di $\mu_f^2$ in funzione di $X$, per vari valori di $\theta$, senza di applicare la regola di Booker. La Fig. 3.1 riporta in rosso il valore di $\mu_f$ che indichiamo con $\mu_{f+}$ ricavato dalla (3.1) considerando in questa il segno positivo. Nella stessa figura è riportato in verde il corrsipondente valore $\mu_{f-}$ che si ottiene considerando il segno negativo. Si vede che, sia $\mu_{f+}$ che $\mu_{f-}$ sono discontinui per $X=1$. Questa discontinuità viene eliminata, come si vede dalla Fig. 3.2, applicando la regola di Booker.

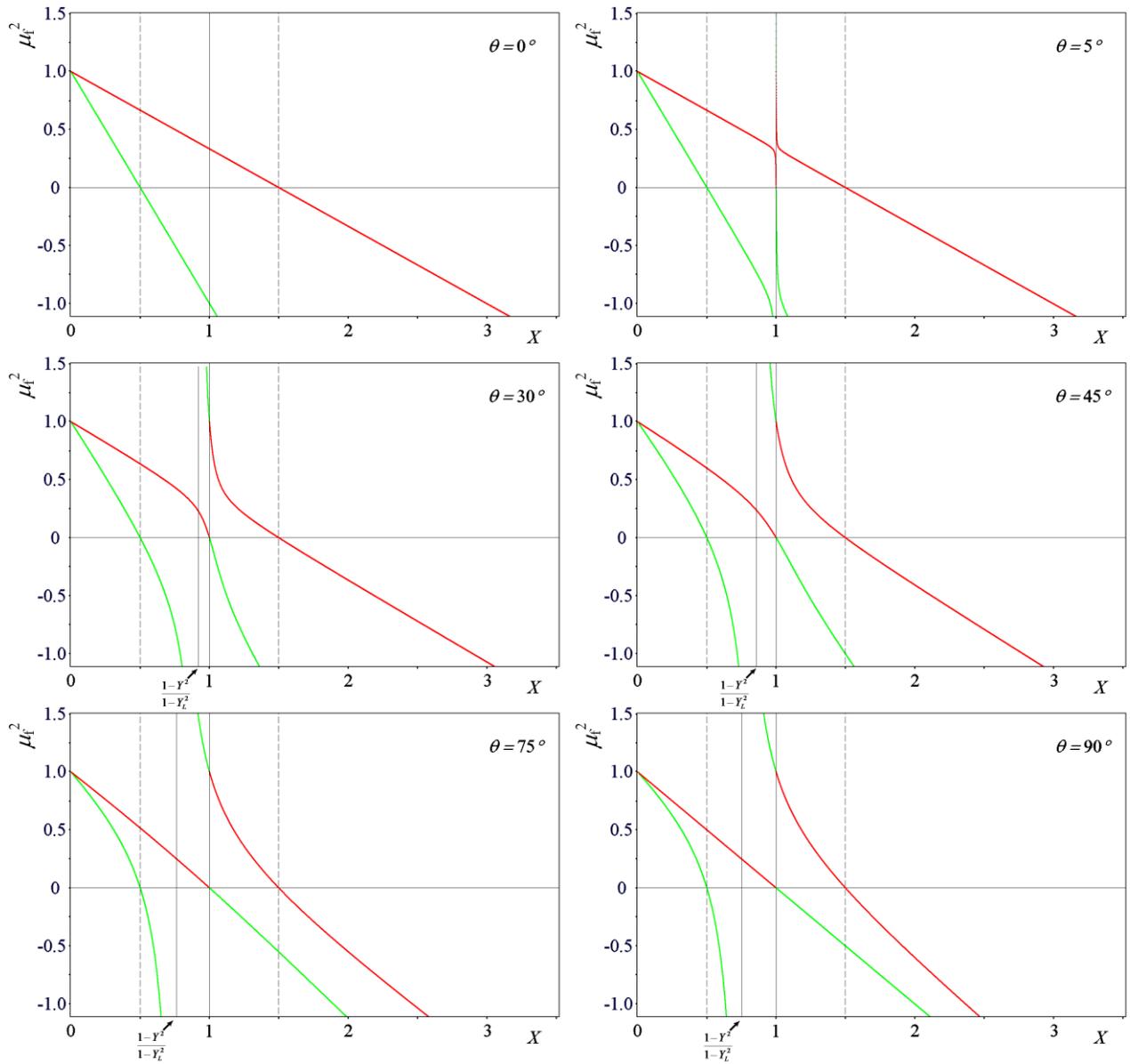

Fig. 3.1. Il quadrato dell'indice di rifrazione in assenza di collisioni in funzione di $X$. In rosso è riportato il grafico per il segno + e in verde quello per il segno -. Si capisce che ragioni di continuità impongono di prendere per lo straordianario la soluzione con il segno negativo se $X<1$, mentre quella col segno positivo se $X>1$ (regola di Booker). La soluzione $X = 1$ va bene per l'ordianario, tranne nel caso di propagazione puramente longitudinale.

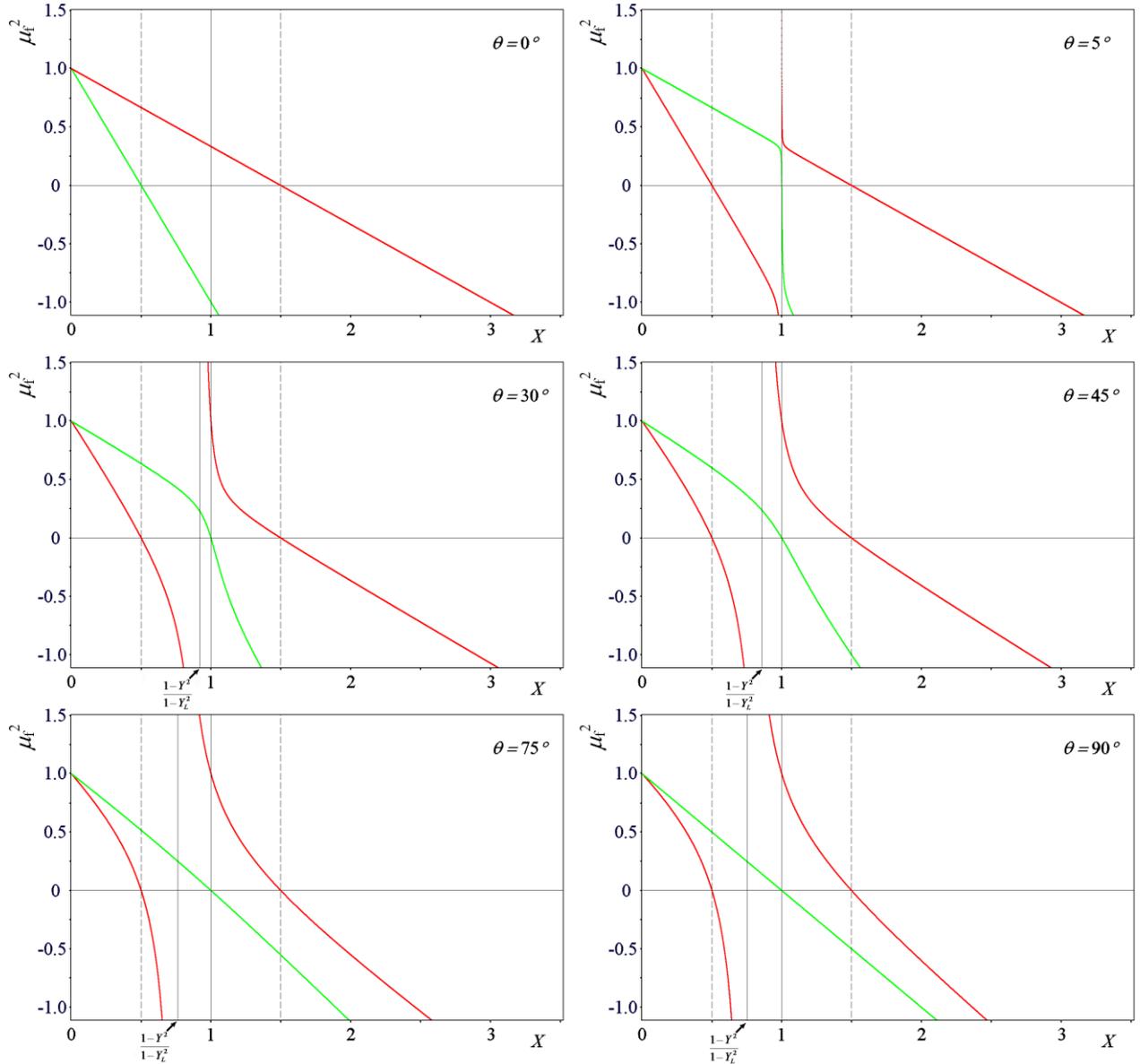

Fig. 3.2. Il quadrato dell'indice di rifrazione in assenza di collisioni in funzione di $X$. In rosso è riportato il grafico per la componete ordinaria e in verde per quella straordinaria. I grafici si ottengono da quelli di Fig. 3.1, applicando la regola di Booker.

Una prima osservazione importante che possiamo fare osservando i grafici della Fig. 3.2 è poi quella che le soluzioni $X=1$ per l'ordinario e $X=1\pm Y$ per lo straordianario non vanno bene solo per la propagazione trasversale, ma per qualsiasi propagazione, eccetto quella puramente longitudinale. Questo, naturalmente, può anche essere mostrato per via algebrica, cosa che in questa sede tralasciamo. Nella Tab. 3.1 vengono riportati i risultati fino a ora ottenuti riguardo le condizioni di riflessione in presenza di **B** e trascurando le collisioni.

Una seconda osservazione riguarda il comportamento di $\mu^2_{f[ext]}$ nell'intervallo $1-Y > X > (1-Y^2)/(1-Y^2_L)$. Si vede che in tale intervallo si ha $\mu^2_{f[ext]} < 0$, per cui abbiamo $\mu_{f[ext]} = -j \cdot \chi$, corrispondente a onde evanescenti, le quali soffrono attenuazione proporzionale a $\chi$ (si veda il Paragrafo 3.2 e l'Appendice 1). In particolare per $X \to (1-Y^2)/(1-Y^2_L)$ si ha $\chi \to \infty$, ricavando una condizione di attenuazione infinita. Ci aspettiamo che la componente straordianaria di una onda radio nella ionosfera, non possa mai propagarsi a un livello dove $X = (1-Y^2)/(1-Y^2_L)$.

Tab. 3.1. Le condizioni di riflessione ottenute come soluzioni, per $\mu_f =0$, dell'equazione di Appleton-Hartree in caso di assenza di collisioni.

|  | Raggio Ordianario | Raggio Straordinario |
|---|---|---|
| Propagazione Transversale | X=1 | X=1±Y |
| Propagazione Longitudinale | X=1+Y | X=1-Y |
| Caso generale | X=1 | X=1±Y |

È poi interessante fare la seguente ulteriore osservazione confrontando $\mu_f^2$ calcolato secondo la (2.35), supponendo cioè che non vi sia campo magnetico e che non vi siano collisioni, e quello calcolato a partire dalla (3.1), che esprime $\mu_f^2$ in assenza di collisioni, ma in presenza di campo magnetico. Precisamente tale confronto viene effettuato considerando il caso in cui la direzione di propagazione della radio onda formi un angolo di 75° con il vettore induzione magnetica terrestre.

Osservando la Fig. 3.3 si vede come il modo di propagazione che corrisponde al segno positivo nella (3.1) per X<1 e negativo per X>1 abbia un indice di rifrazione molto simile a quello che si ricava dalla (2.35). Questo fornisce una ulteriore giustificazione dell'applicazione della regola di Booker e del motivo per il quale il modo di propagazione considerato possa essere definito ordinario.

Si osservi infine, che solo nel caso di propagazione perfettamente longitudinale, al quale si riferiscono i grafici della Fig. 3.4, l'indice di rifrazione del raggio ordinario è lo stesso che si avrebbe in assenza di campo magnetico.

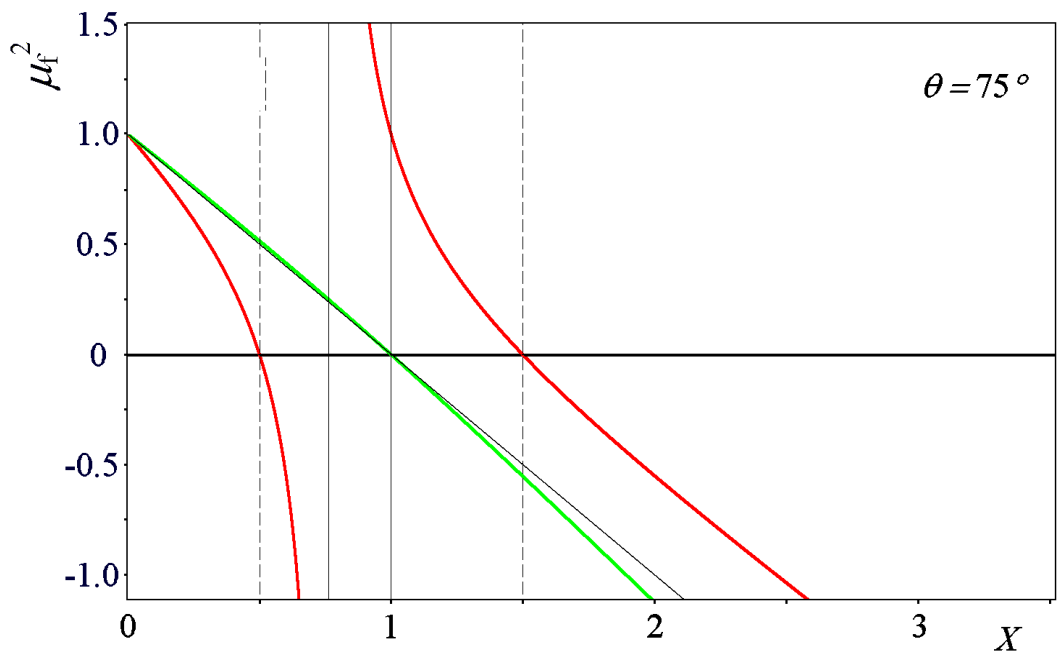

Fig. 3.3. Il quadrato dell'indice di rifrazione in assenza di collisioni e di campo magnetico in funzione di X. In rosso è riportato il grafico per il segno positivo e in verde quello per il segno negativo. In nero è riportato il quadrato dell'indice di rifrazione ricavato in assenza di campo magnetico. Si vede dunque che il modo di propagazione che corrisponde a prendere il segno positivo nella (3.8) per X<1 e negativo per X>1 si propaga quasi come se il campo magnetico non ci fosse, per cui può ragionevolmente essere chiamato "ordinario".

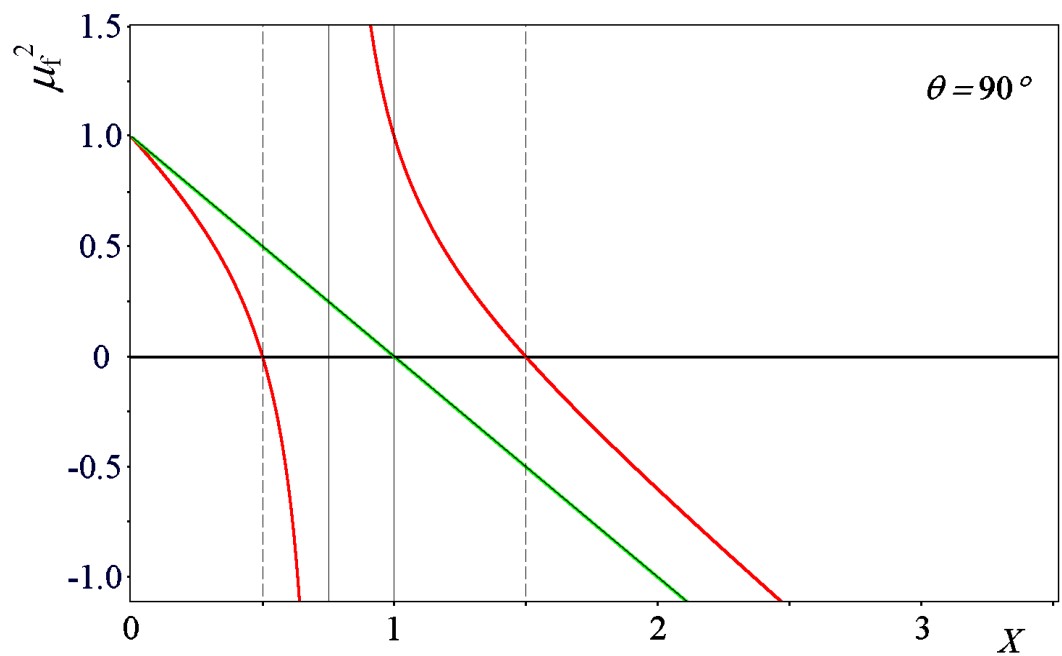

Fig. 3.4. Il quadrato dell'indice di rifrazione in assenza di collisioni e di campo magnetico in funzione di *X*. In rosso è riportato il grafico per il segno + e in verde quello per il segno -. In nero è riportato il quadrato dell'indice di rifrazione ricavato in assenza di campo magnetico. Si vede dunque che in caso di propagazione perfettamente trasversale, il modo ordianario si propaga come se il campo magnetico non ci fosse.

# Applicazioni Informatiche del Capitolo 3

## Applicazione Informatica 3.1

Si consideri il quadrato dell'indice di rifrazione di fase in assenza di collisioni e in presenza di campo magnetico, come esso è espresso dalla relazione di Appleton-Hartree (3.1):

$$\mu_\text{f}^2 = 1 - \frac{X}{1 - \frac{1}{2}\frac{Y_T^2}{(1-X)} \pm \sqrt{\frac{Y_T^4}{4(1-X)^2} + Y_L^2}}. \tag{P3.1}$$

Si riporti su un grafico $\mu^2$ in funzione di $X$, per alcuni particolari valori dell'angolo $\theta$ formato dalla direzione di propagazione col vettore induzione magnetica terrestre **B**.
Si confrontino i grafici che si ottengono con quelli ottenuti tramite la (2.35):

$$\mu_\text{f}^2 = 1 - X, \tag{P3.2}$$

nel caso di assenza di campo magnetico.

**Bibliografia del Capitolo 3**

# CAPITOLO 4

## Indice di rifrazione di gruppo

**Riassunto**

Viene introdotta l'epressione analitica per l'indice di rifrazione di gruppo in presenza di campo magnetico e senza collisioni. Viene introdotto il calcolo numerico dello stesso indice di rifrazione anche in presenza di collisioni. Le condizioni di riflessione vengono discusse da questo punto di vista e viene data la spiegazione del cosidetto raggio z, che si osserva sugli ionogrammi registrati alle latitudini polari.

### 4.1 L'indice di rifrazione di gruppo in presenza di campo magnetico, senza collisioni

Utilizzando la relazione (3.1) è possibile ricavare analiticamente una espressione per l'indice di rifrazione di gruppo $\mu_g$ in presenza di campo magnetico e senza collisioni. Qui non vengono riportati i passaggi matematici che permettono di ricavare $\mu_g$, che furono proposti da rfr4-Shinn and Whale (1952) e che possono essere studiati facendo riferimento al libro di rfr4-Budden (1961). Tali passaggi in questa trattazione vengono proposti come Problema 4.1. L'espressione analitica che può essere ricavata è la seguente:

$$\mu_g = -\frac{1}{2f\mu_f}\left\{\frac{1}{f^2 D}\left[2f_p^2 f - \left(f^2\mu^2 - f^2\right)\frac{Y_L^2 f\left(1-X^2\right)}{\sqrt{\frac{1}{4}Y_T^4 + Y_L^4\left(1-X\right)^2}}\right] + 2f\right\} \qquad (4.1)$$

ove $\mu_g$ è l'indice di rifrazione di fase che si calcola dalla (3.1), $f$ la frequenza dell'onda, $f_p$ la frequenza di plasma, mentre gli altri parametri sono quelli che abbiamo visto al Capitolo 2. Il termine $D$ è:

$$D = \left(1-X\right) - \frac{Y^2 \sin^2\theta}{2} \pm \sqrt{\frac{Y^4 \sin^4\theta}{4} + Y^2\left(1-X\right)^2 \cos\theta} \qquad (4.2)$$

Si osservi che nella espressione (4.1), $\mu_g$ è scritto in forma generica, senza cioè fare riferimento all'indice di rifrazione di gruppo per il raggio straordinario ($\mu_{g[ext]}$) o all'ordinario ($\mu_{g[ord]}$). In realtà, con il segno +, si ottiene $\mu_{g[ord]}$ mentre con il segno – si ottiene $\mu_{g[ext]}$. Naturalmente, anche l'indice di rifrazione di fase $\mu_f$ nella (4.1) è espresso in forma generica. Esso dovrà essere fatto corrispondere, secondo il caso, alla componente ordianaria, indicandolo con $\mu_{f[ord]}$, o a quella straordinaria, indicandolo con $\mu_{f[ext]}$. Per il calcolo di $\mu_{f[ord]}$ e per $\mu_{f[ext]}$, secondo quanto abbiamo detto nel capitolo precedente, vale di nuovo la (3.1) con l'applicazione della regola di Booker.

Graficando queste espressioni in funzione di $X$, si vede che $\mu_{g[ord]} \to \infty$ e $\mu_{g[ext]} \to \infty$, secondo i casi riportati nella Tab. 4.1. Si può notare che questi corrispondono ai casi della Tab. 3.1 che riporta le condizioni per le quali si ha $\mu_{f[ord]} = 0$ e $\mu_{f[ext]} = 0$. È dunque confermato, anche nel caso di propagazione in presenza di campo magnetico, trascurando le collisioni, lo stesso risultato visto al Paragrafo 3.3, in assenza di campo magnetico: $\mu_{g[ord]} \to \infty$ ($\mu_{g[ord]} \to \infty$) quando $\mu_{f[ord]}=0$ ($\mu_{f[ext]}=0$), il che corrisponde al verificarsi della riflessione.

Tab. 4.1. Condizioni per le quali $\mu_g \to \infty$ studiate attraverso la relazione di Shinn and Whale (1952).

|  | Ordianario | Straordinario |
|---|---|---|
| Propagazione Longitudinale | X=1+Y | X=1-Y |
| Caso generale | X=1 | X=1±Y |

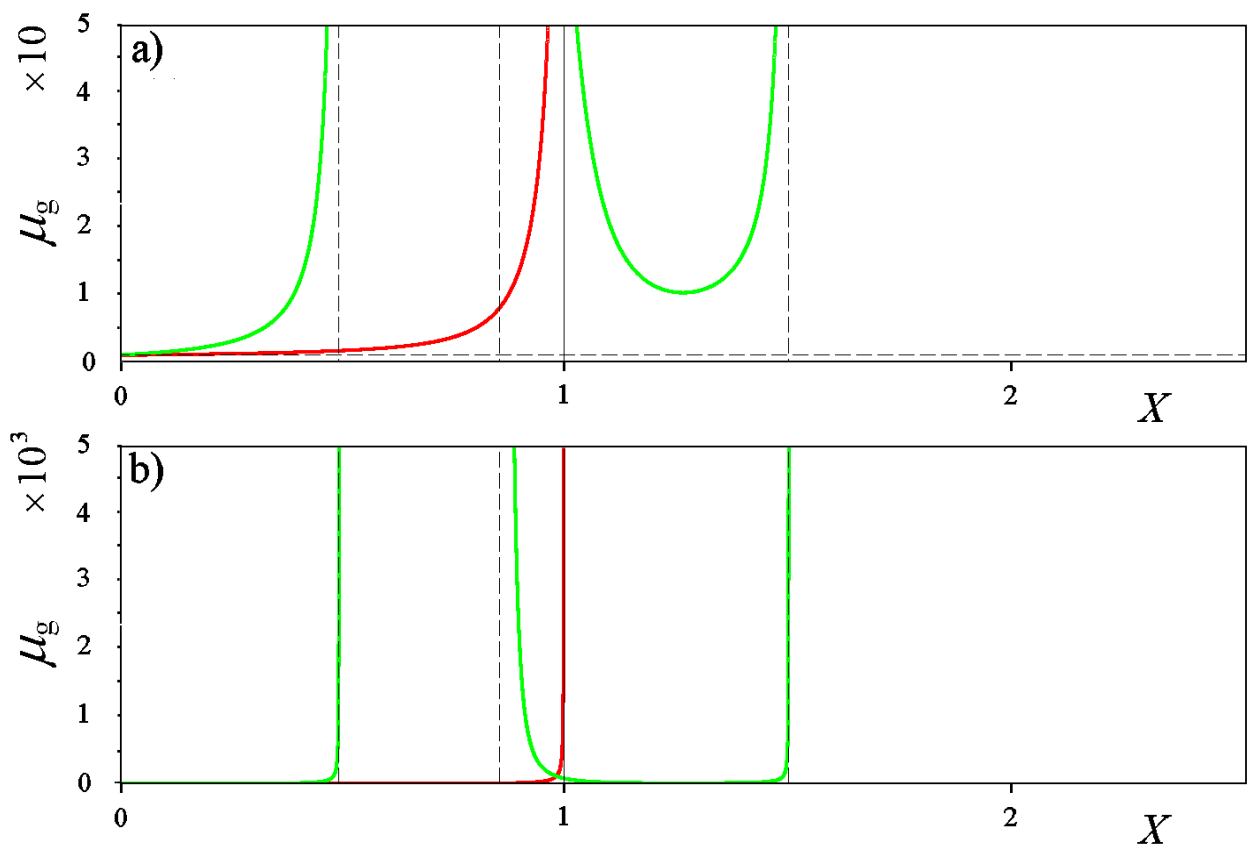

Fig. 4.1. (a)-(b). I grafici di $\mu_{\text{gord}}$ (in rosso) e $\mu_{\text{gext}}$ (in verde) in funzione di $X$, in assenza di collisioni e con campo magnetico ($Y$=0.5). Viene mostrato un caso generale di propagazione, assumendo cioè che **k** formi un angolo $\theta$=45° con **B**. In a) la scala è tale che viene messo in evidenza il fatto che il minimo valore positivo dell'indice di rifrazione è pari a 1. In b) la scala è tale da consentire di mettere in evidenza il comportamento di $\mu_{\text{g[ext]}}$ per $1-Y > X > (1-Y^2)/((1-Y_L^2)$.

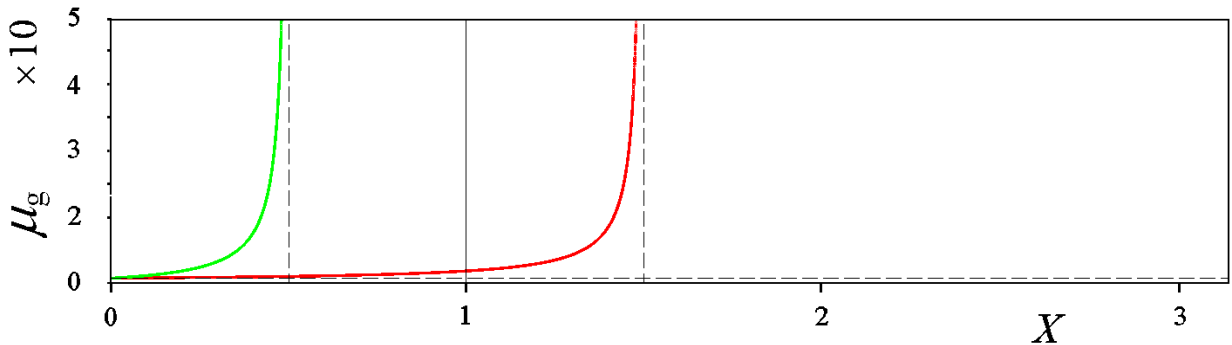

Fig. 4.2. I grafici di $\mu_{g[ord]}$ (in rosso) e $\mu_{g[ext]}$ (in verde) in funzione di $X$, in assenza di collisioni e con campo magnetico ($Y$=0.5). Viene riportato il caso di propagazione perfettamente longitudinale, assumendo cioè che **k** formi un angolo $\theta$=0° con **B**. Si osservano gli asintoti verticali che corrispondono a riflessione in $X$=1-$Y$, per il raggio straordinario, e in $X$=1+$Y$ per quello ordinario.

## 4.2 Ionogrammi

Abbiamo visto al Capitolo 2, che quando inviamo un'onda radio verso la ionosfera, succede che questa si divide nelle sue due componenti magnetoioniche. Di queste componenti, abbiamo poi studiato le condizioni di riflessione trascurando le collisioni e tenendo conto della presenza del campo magnetico della Terra. Nel Paragrafo 3.4, dette condizioni di riflessione sono state determinate studiando i casi ove $\mu_f$ =0. Nel Paragrafo precedente abbiamo visto, che tali casi corrispondono a quelli per i quali si ha $\mu_g$ =0. I risultati principali possono essere riassunti nel modo seguente:

1) in generale, la componente ordianaria di un'onda radio che si propoga nella ionosfera si riflette quando $X$=1, la componente straordinaria quando $X$=1$\pm Y$;
2) in caso di propagazione longitudinale si ha riflessione per il raggio ordinario quando $X$=1+$Y$ e per quello straordinario quando $X$=1-$Y$.

Dal punto di vista dei sondaggi verticali, bisogna tener presente che la propagazione avviene usualmente in condizioni diverse da quella longitudinale, per cui il raggio ordinario viene riflesso quando incontra una zona dove la densità elettronica è tale che

$$X=1 \tag{4.3}$$

cioè:

$$f = f_p. \tag{4.4}$$

Il raggio straordinario, invece, viene riflesso quando incontra una zona dove la densità elettronica è tale che:

$$X = 1 \pm Y \tag{4.5}$$

cioè quando:

$$\frac{f_p^{\ 2}}{f^2} = 1 \pm \frac{f_B}{f}, \tag{4.6}$$

questa equazione può essere risolta coi seguenti semplici passaggi algebrici:

$$f^2 \pm f_B f - f_p^{\ 2} = 0, \tag{4.7}$$

da cui:

$$f_{1,2} = \frac{1}{2}\left\{\pm f_B \pm \sqrt{f_B^{\ 2} + 4 f_p^{\ 2}}\right\} \approx \pm f_p \pm \frac{f_B}{2} \tag{4.8}$$

Nell'ultimo passaggio è pure stata effettuata una approssimazione che tiene conto del fatto che, nelle applicazioni pratiche, la girofrequenza di plasma $f_B$ è molto inferiore alla frequenza di plasma $f_p$. Scartando le soluzioni prive di significato fisico i risultati possibili sono:

$$f_{1,2} = f_p \pm \frac{f_B}{2}. \tag{4.9}$$

Siccome poi l'onda radio con frequenza fissa $f$ viaggia verso strati con densità crescenti $f_p$ cresce, per cui, delle due condizioni espresse dalla relazione precedente, si verifica prima quella con il segno positivo; nella pratica, dunque, il raggio straordinario viene riflesso ove

$$f_x = f_p + \frac{f_B}{2}. \tag{4.10}$$

Sempre per il crescere di $f_p$, mentre l'onda si propaga verso l'alto, succederà che la condizione (4.10), verrà soddisfatta a una quota inferiore rispetto la (4.4). Come conseguenza, in condizioni generali, il raggio straordinario verrà riflesso ad altezze inferiori rispetto a quelle ove avviene la riflessione del raggio ordinario.
Inoltre, per il raggio ordinario, una frequenza superiore alla massima frequenza di plasma nella ionosfera non subisce riflessione, ma penetra la ionosfera. Per il raggio straordinario, invece, la riflessione avviene pure per frequenze superiori a $f_oF2$, purché sia $f < f_oF2 + f_B/2$. Si spiega così la doppia traccia osservata negli ionogrammi.
Nel prossimo paragrafo, invece, ci interesseremo del caso di propagazione perfettamente longitudinale, caso per il quale l'ordinario si riflette ove $X=1+Y$ ( $f = f_p - f_H/2$) e lo straordinario ove $X=1-Y$ ($f = f_p + f_H/2$).

## 4.3 Una prima spiegazione del raggio Z

Quando una ionosonda viene installata a latitudini polari a volte sugli ionogrammi registrati si osserva che la traccia della regione F2, invece che essere doppia come di consueto, è tripla. Il raggio che provoca la traccia aggiuntiva, viene detto raggio z. Un esempio di ionogramma ove è visibile il raggio z è riportato nella Fig. 4.3.

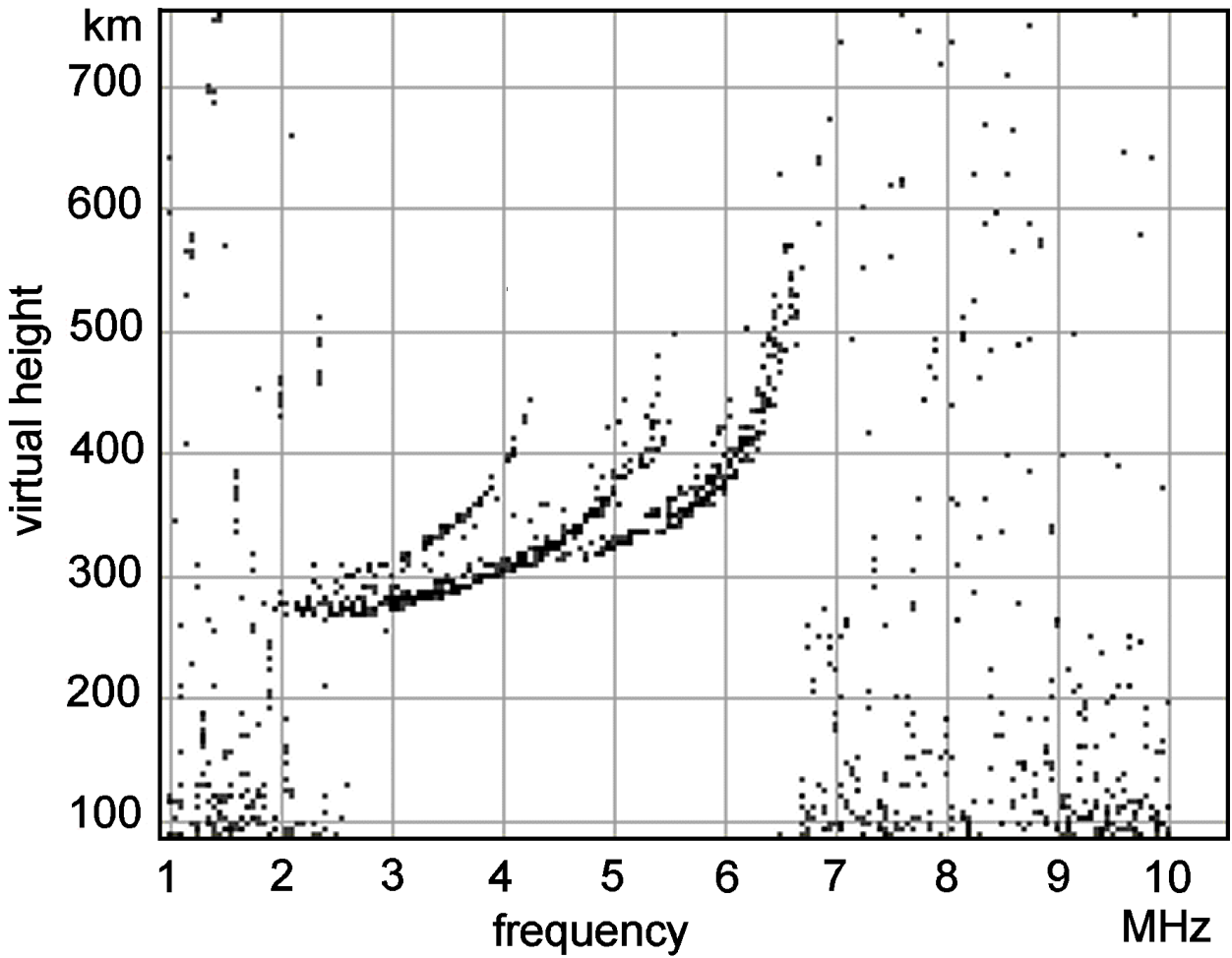

Sounding Station: Baia Terra Nova
Sounding Date: 2004 03 15
Sounding Hour: 07:45

Fig. 4.3. Uno ionogramma registrato a latitudini polari, nel quale è visibile il raggio z.

Per spiegare questo fenomeno si deve tenere presente che l'angolo di radiazione dell'antenna di una ionosonda è molto ampio, poiché alle frequenze HF, non è possible ottenere elevate direttività. L'inclinazione di **B**, d'altro canto, è elevata, per cui ci sarà qualche raggio che si propaga in modo longitudinale, in modo cioè esattamente parallelo a **B**. Per questo raggio, si ha la riflessione per l'ordianario a $X=1+Y$ ($f = f_P - f_H/2$) e per lo straordianario a $X=1-Y$ ($f = f_P + f_H/2$). È perciò come se vedessimo due ionogrammi sovrapposti: lo ionogramma dovuto ai raggi che si propagano in condizioni generali di propagazione e lo ionogramma dovuto alla propagazione perfettamente longitudinale. Nella Fig. 4.4 è riportata sinteticamente la geometria della propagazione per un radio sondaggio ionosferico effettuato nelle regioni polari, e nella Fig. 4.5 il relativo schema coi due ionogrammi sovrapposti, che spiegano la presenza della tripla traccia.

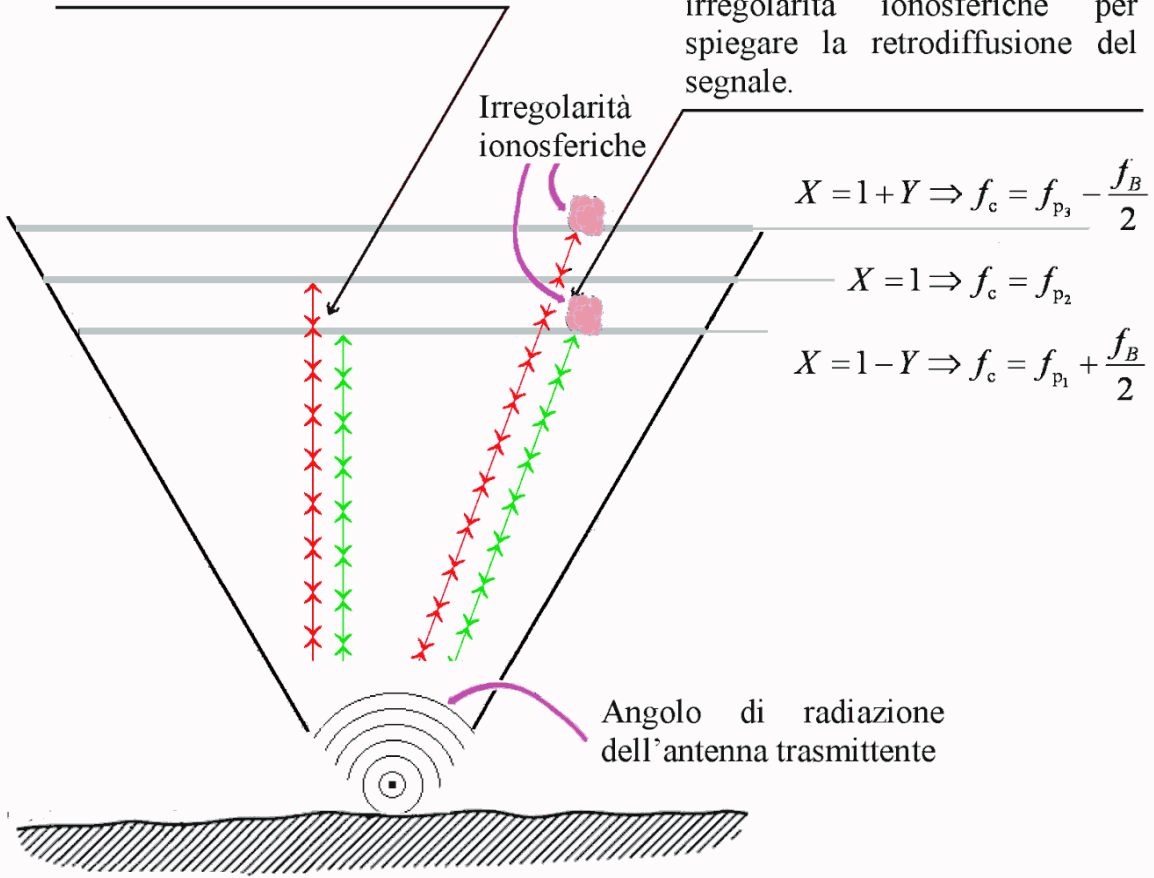

Fig. 4.4. - Radiosondaggio ionosferico verticale alle latitudini polari: geometria e spiegazione del raggio z.

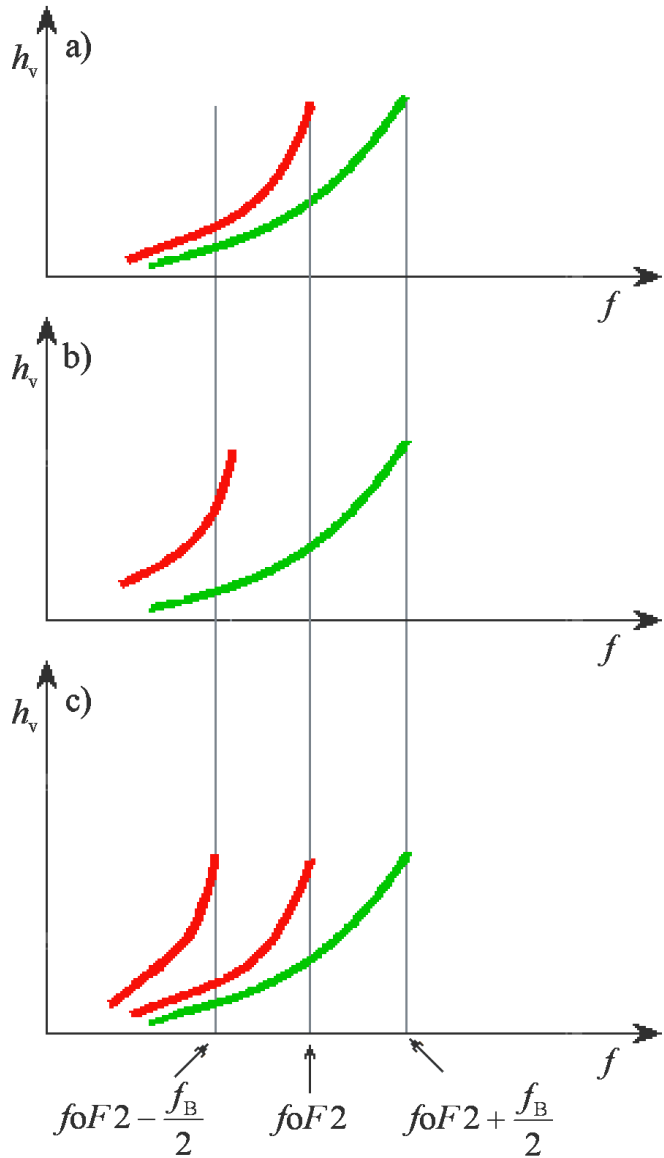

propagazione in condizioni generali

ord: $X = 1 \rightarrow f_o = f_oF2$

ext: $X = 1-Y \rightarrow f_x = f_oF2 + \dfrac{f_B}{2}$

propagazione longitudinale

ord: $X = 1+Y \rightarrow f_z = f_oF2 - \dfrac{f_B}{2}$

ext: $X = 1-Y \rightarrow f_x = f_oF2 + \dfrac{f_B}{2}$

propagazione longitudinale insieme a propagazione in condizioni generali

$$\begin{cases} f_x = f_oF2 - \dfrac{f_B}{2} \\ f_o = f_oF2 \\ f_z = f_oF2 + \dfrac{f_B}{2} \end{cases}$$

Fig. 4.5. Gli ionogrammi che presentano il raggio Z (c) possono essere pensati come costituiti da due ionogrammi sovrapposti: uno ionogramma come quello che si osserva in condizioni di propagazione normale (a) e uno ionogramma che si avrebbe in caso di propagazione perfettamente longitudinale (c).

Questa spiegazione, sebbene affascinante nella sua semplicità, presenta il problema che i raggi ordinari che si propagano in modo perfettamente longitudinale, in grado di penetrare la quota $X=1$, sarebbero compresi in un cono di radiazione con un angolo infinitesimo. Essi dunque sarebbero in grado di trasportare una energia virtualmente nulla o comunque modestissima. Questo non è in accordo con le osservazioni sperimentali di tracce dovute al raggio z che hanno intensità paragonabile a quella delle tracce dovute ai consueti modi di propagazione ordinario e straordinario.

**4.4 Il raggio z secondo la teoria dell'accoppiameto fra raggio ordinario e raggio straordinario**

Secondo la teoria che venne inizialmente accreditata (rfr4-Eckersley, 1950; rfr4-Rydbeck, 1950, 1951), alla quota $X=1$, parte dell'energia del modo ordinario viene trasferita attraverso fenomeni di accoppiamento, a quello straordinario, per cui questa energia si propaga fino alla quota alla quale si verifica $X=1+Y$, ove viene riflessa. L'onda diviene quindi una onda straordinaria che viaggia verso il basso fino a raggiungere la regione di accoppiamento ($X=1$) da sopra. Siccome l'indice di rifrazione di gruppo per lo straordinario diviene $-iA$ con $A \to \infty$ al livello dove $X = (1- Y^2)/(1-Y_L^2)$ ($X<1$), l'onda, a questo stesso livello non potrebbe più propagarsi, divenendo evanescente. Tuttavia, prima di raggiungere questo livello, essa passa di nuovo attraverso la regione dove $X=1$ e dove il processo di accoppiamento agisce in modo inverso, cosicché viene generata un'onda ordinaria in grado di raggiungere il suolo.

**4.5 La teoria del raggio z**

rfr4-Ellis (1953, 1956) mise in evidenza che il meccanismo di accoppiamento non è in grado di spiegare il raggio z che talvolta si osserva anche alle medie latitudini. Infatti, alle medie latitudini, a causa dell'ampiezza dell'angolo $\theta$, affinché il meccanismo di accoppiamento sia efficacie, si dovrebbe avere una frequenza delle collisioni elettrone-molecola neutra troppo elevata per essere realistica.

Il meccanismo per il quale secondo Ellis si può osservare il raggio z, può essere compreso sempre facendo riferimento alla Fig. 4.4, ove si illustra l'usuale splitting fra la componente ordinaria, che va a riflettersi alla quota $X=1$, e quella straordinaria che va a riflettersi alla quota $X=1-Y$. Ellis in realtà descrive una situazione radiopropagativa lievemente più realistica, ove viene tenuto conto che le onde radio non si propagano in linea retta. Le basi del suo ragionamento, e le conclusioni non sono tuattavia diverse da quelle che qui vengono illustrate.

Abbiamo visto, gia nel Capitolo 3, che, se la frequnza delle collisioni $\nu$ è nulla, per $\theta=0$, il raggio ordinario non si riflette alla quota $X=0$, ma alla quota ove $X=1+Y$. Nel Capitolo 4, con le Fig. 4.1 e 4.2, questo fatto può essere visto dal punto di vista di $\mu_{g[ord]}$ osservando che, per $\theta=0$, $\mu_{g[ord]}$ non diverge per $X=0$, ma per $X=1+Y$. Possiamo dimostrare, cosa che faremo nel paragrafo seguente, che un piccolo valore di $\nu > 0$ fa in modo che il comportamento di $\mu_{g[ord]}$, che si osserva nelle Fig. 4.1 e 4.2 per $\theta=0$, si osservi pure finché $\theta < \delta$, essendo $\delta$ un angolo piccolo.

Ne risulta che i raggi che hanno $\theta < \delta$ non sono riflessi al livello $X=1$, come si ha per incidenza verticale, ma continuano a viaggiare fino a che, non raggiungono una quota superiore dove si ha $X=1+Y$, dove cioè $\mu_{g[ord]} \to \infty$ e dove vengono riflessi. Affinché questo meccanismo sia efficace, è comunque necessaria la presenza di irregolarità nella distribuzione di elettroni presso il livello $X=1+Y$, irregolarità che causano una rilevante retrodiffusiuone di energia.

Per quanto abbiamo detto, viene a crearsi nella ionosfera, un "buco" di propagazione, che è stato scoperto da rfr4-Ellis (1956). Presentiamo nel paragrafo seguente uno studio su di $\mu_{g[ord]}$ che ci permette da un lato di semplificare la trattazione e dall'altro ci permette di stimare le dimensioni di tale "buco".

**4.6 L'effetto delle collisioni sull'indice di rifrazione di gruppo: una spiegazione per l'osservazione del raggio z alle latitudini polari**

Come abbiamo detto, l'indice di rifrazione di gruppo può venire calcolato nel caso vi siamo collisioni attraverso la relazione (2.34) che scriviamo nella forma:

$$\left[\mu_f(\omega) - j\chi(\omega)\right]^2 = 1 - \frac{X}{1 - jZ - \frac{Y_T^2}{2\cdot(1-X-jZ)} \pm \sqrt{\frac{Y_T^4}{4\cdot(1-X-jZ)^2} + Y_L^2}}, \qquad (4.10)$$

per mettere in evidenza il fatto che da questa siamo interessati a ricavare la parte reale $\mu_f(\omega)$. Attraverso uno dei software in commercio, capaci di svolgere il calcolo simbolico, è possibile ricavare $\mu_f(\omega)$, ottenendo una relazione estremamente complicata, di difficile interpretazione, che è inutile riportare. Da tale espressione per $\mu_f(\omega)$ è possibile ricavare numericamente l'indice di rifrazione di gruppo attraverso la (3.12) che, adesso, esprimiamo con le differenze finite:

$$\mu_g(\omega) = \mu_f(\omega) + \frac{\mu_f(\omega + \Delta\omega) - \mu_f(\omega)}{\Delta\omega}. \tag{4.11}$$

Con questa relazione possiamo calcolare $\mu_{g[ord]}(\omega)$ per cui, nella Fig. 4.6(a), ne riportiamo gli andamenti per vari valori dell'angolo $\theta > 2°$. Si osserva che $\mu_g(\omega)$ diverge per $X=1$. Questo è in linea con quanto è noto nel caso acollisionale e con quanto si osserva sperimentalmente: fin tanto che la propagazione non è perfettamente longitudinale, il raggio ordinario si riflette per $X=1$. Nella Fig. Fig. 4.6(b) riportiamo il grafico dell'indice di rifrazione ordianario per $\theta \leq 0.5°$. Si osserva che per $\theta \leq 0.5°$, $\mu_{gord}(\omega)$ non diverge in prossimità di $X=1$. Questo corrisponde a dire che includendo l'effetto delle collisioni, si osserva che la divergenza $\mu_g(\omega)$, in $X=1$ viene a mancare non solo per $\theta=0$ (come avviene nel caso a-collisionale), ma per qualsiasi valore $\theta \leq 0.5°$.

Questo approccio ha il vantaggio di dimostrare che un meccanismo come quello descritto al Paragrafo 4.5, può essere invocato per spiegare la propagazione del raggio ordinario oltre la quota $X=1$, e la sua riflessione a $X=1+Y$ per tutti i raggi che giacciono in un cono di radiazione di ampiezza piccola ($\theta \leq 0.5°$) ma non nulla. Le dimensioni di questo cono, per inciso, sono compatibili con quelle individuate sperimentalmente da Ellis (1956).

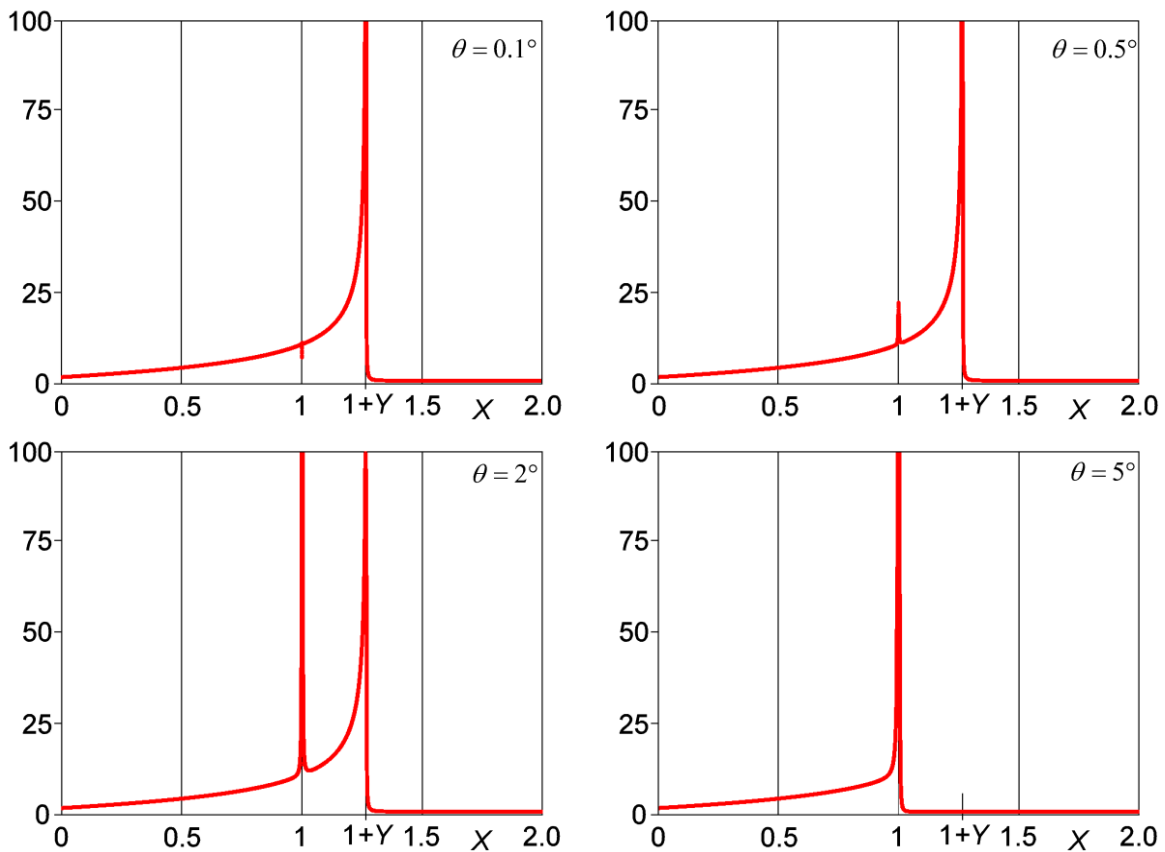

Fig. 5.6. Gli andamenti di $\mu_g(X)$ per vari valori dell'angolo $\theta$, per $Y=0.26$, $Z=5\cdot10^{-4}$, che è compatibile con la frequenza delle collisioni nella regione F2. Si osserva che $\mu_g$ non diverge per $X=1$, per piccoli valori di $\theta$. La condizione limite è $\theta=2°$.

## Applicazioni informatiche del capitolo 4

**Problema 4.1**

Nel Capitolo 4 è stata ricavata l'equazione (2.33) di Appleton-Hartree, che esprime il quadrato della parte reale dell' indice di rifrazione di fase $\mu_f$ per le onde elettromagnetiche che si propagano in un plasma freddo, quale la ionosfera, in presenza di campo magnetico, trascurando le collisioni. La presenza del doppio segno in tale equazione è da dovuta al fenomeno dello splitting magnetoionico, per cui in realtà avremo sia $\mu_{f[ord]}$ che $\mu_{f[ext]}$, con ovvio significato di simboli. Sia $\mu_{f[ord]}$ che $\mu_{f[ext]}$, si ricavano dalla (2.33), tenendo conto del doppio segno e applicando la regola di Booker vista al paragrafo 3.4.
Partendo da questa stessa equazione, si ricavino i corrispondenti indici di rifrazione di gruppo $\mu_{g[ord]}$ e $\mu_{g[ext]}$.

*Suggerimenti*

1) Si consideri l'equazione (2.33) e la si esprima in funzione di $Y$ e di $\theta$. Si mostri che essa assume la forma seguente:

$$\mu^2 = 1 - \frac{X(1-X)}{D}, \qquad (P4.1)$$

dove:

$$D = (1-X) - \frac{Y^2 \sin^2 \theta}{2} \pm \sqrt{\frac{Y^4 \sin^4 \theta}{4} + Y^2(1-X)\cos\theta}. \qquad (P4.2)$$

2) Si parta dall'espressione (P4.1) per ottenere una relazione che collega $\mu_f$ a $f^2 D$.
3) Dalla relazione ricavata al punto 2), si ottenga una relazione che collega $\mu_g$ a $f^2 D$ e a $\partial(f^2 D)/\partial f$.
4) Per portare a termine il compito proposto si dovrà esplicitare $\partial(f^2 D)/\partial f$ ricavando, dalla relazione trovata al punto 2), $\mu_g$. Per far questo si potrà utilizzare la relazione trovata al punto 2), esprimendo $f^2 D$ in funzione di $\mu_f$.

*Risoluzione*

Riprendiamo l'equazione (2.33) di Appleton-Hartree trascurando le collisioni, che è:

$$\mu_f^2 = 1 - \frac{X}{1 - \frac{Y_T^2}{2(1-X)} \pm \sqrt{\frac{Y_T^4}{4(1-X)^2} + Y_L^2}} ; \qquad (P4.3)$$

essa può essere espressa in funzione di $Y$ e $\theta$, introducendo le relazioni $Y_T = Y \cdot \sin\theta$ e $Y_L = Y \cdot \cos\theta$. Otteniamo così:

$$\mu_\text{f}^2 = 1 - \frac{X}{1 - \frac{Y^2 \sin^2\theta}{2(1-X)} \pm \sqrt{\frac{Y^4 \sin^4\theta}{4(1-X)^2} + Y^2 \cos^2\theta}} \,; \tag{P4.4}$$

oppure

$$\mu_\text{f}^2 = 1 - \frac{X(1-X)}{(1-X) - \frac{Y^2 \sin^2\theta}{2} \pm \sqrt{\frac{Y^4 \sin^4\theta}{4} + Y^2(1-X)^2 \cos^2\theta}} \,; \tag{P4.5}$$

che corrisponde a:

$$\mu_\text{f}^2 = 1 - \frac{X(1-X)}{D}, \tag{P4.6}$$

dove:

$$D = (1-X) - \frac{Y^2 \sin^2\theta}{2} \pm \sqrt{\frac{Y^4 \sin^4\theta}{4} + Y^2(1-X)^2 \cos\theta} \,. \tag{P4.7}$$

Partendo dalla (P4.6), e ricordando che $X = f_\text{p}^2/f^2$, abbiamo:

$$\mu_\text{f}^2 = 1 - \frac{\frac{f_p^2}{f^2}\left(1 - \frac{f_p^2}{f^2}\right)}{D}, \tag{P4.8}$$

da cui

$$f^2 \mu_\text{f}^2 - f^2 = -\frac{f_p^2 \left(\frac{f^2 - f_p^2}{f^2}\right)}{D}, \tag{P4.9}$$

e poi:

$$\left(f^2 \mu^2 - f^2\right) f^2 D = f_p^2 \left(f_p^2 - f^2\right); \tag{P4.10}$$

Questa espressione fornisce una relazione fra $\mu_\text{f}$ e $f^2 D$, come suggerito al punto 2).
Differenziando questa si ricava una relazione che collega $\mu_\text{g}$ con $f^2 D$ e con $\partial(f^2 D)/\partial t$. Questo è quello che facciamo nei prossimi passaggi, che portano alla (P4.12).

$$\left(2f\mu^2 + 2f^2\mu\mu' - 2f\right)f^2 D + \left(f^2\mu^2 - f^2\right)\frac{\partial}{\partial f}\left(f^2 D\right) = -2 f_p^2 f \,, \tag{P4.11}$$

e

$$\left(2f\mu\left(\overbrace{\mu+f\mu'}^{\mu_g}\right)-2f\right)f^2D+\left(f^2\mu^2-f^2\right)\frac{\partial}{\partial f}\left(f^2D\right)=-2f_p^2 f. \tag{P4.12}$$

La relazione precedente è quella suggerita al punto 3).

Per poter ricavare $\mu_g$, è necessario procedere al calcolo di $\partial(f^2D)/\partial t$, calcolo per il quale è necessario esplicitare $f^2D$. Secondo quanto suggerito al punto 4), questo può essere fatto a partire dalla (P4.10) ottenendo:

$$f^2D = f^2\left(1-\frac{f_p^2}{f^2}\right)-f^2\frac{\frac{f_B^2}{f^2}\sin^2\theta}{2}\pm f^2\sqrt{\frac{\frac{f_B^4}{f^4}\sin^4\theta}{4}+\frac{f_B^2}{f^2}(1-X)^2\cos\theta}, \tag{P4.13}$$

oppure

$$f^2D = f^2\left(1-\frac{f_p^2}{f^2}\right)-f^2\frac{\frac{f_B^2}{f^2}\sin^2\theta}{2}\pm f^2\sqrt{\frac{\frac{f_B^4}{f^4}\sin^4\theta}{4}+\frac{f_B^2}{f^2}\left(1-2\frac{f_p^2}{f^2}+\frac{f_p^4}{f^4}\right)\cos\theta}, \tag{P4.15}$$

e

$$f^2D = \left(f^2-f_p^2\right)-\frac{f_B^2\sin^2\theta}{2}\pm\sqrt{\frac{f_B^4\sin^4\theta}{4}+f_B^2\left(f^2-2f_p^2+\frac{f_p^4}{f^2}\right)\cos^2\theta}. \tag{P4.16}$$

Dalla quale possiamo calcolare la derivata:

$$\frac{\partial}{\partial f}\left(f^2D\right)=2f\pm\frac{2f_B^2\cos^2\theta\cdot f-\frac{2f_p^4}{f^3}f_B^2\cos^2\theta}{2\sqrt{\frac{f_B^4\sin^4\theta}{4}+f_B^2\left(f^2-2f_p^2+\frac{f_p^4}{f^2}\right)\cos^2\theta}}=$$

$$=\frac{f_B^2\cos^2\theta\left(f-\frac{f_p^4}{f^3}\right)}{\sqrt{\frac{f_B^4\sin^4\theta}{4}+f_B^2 f^2\left(1-2\frac{f_p^2}{f^2}+\frac{f_p^4}{f^4}\right)\cos^2\theta}}. \tag{P4.17}$$

Itroducendo le relazioni $f_T=f_B\sin(\theta)$, e $f_L=f_B\cos(\theta)$ insieme a $Y_T=f_T/f$ e $Y_L=f_L/f$, questa diventa:

$$\frac{\partial}{\partial f}\left(f^{2}D\right)=\frac{\dfrac{f_{L}^{2}}{f^{2}}f^{3}\left(1-\dfrac{f_{p}^{4}}{f^{4}}\right)}{f^{2}\sqrt{\dfrac{1}{4}\dfrac{f_{T}^{4}}{f^{4}}+f_{L}^{2}f^{2}\dfrac{1}{f^{4}}\left(1-2\dfrac{f_{p}^{2}}{f^{2}}+\dfrac{f_{p}^{4}}{f^{4}}\right)}}=$$

$$=\frac{Y_{L}^{2}f\left(1-X^{2}\right)}{\sqrt{\dfrac{1}{4}Y_{T}^{4}+Y_{L}^{4}\left(1-X\right)^{2}}}\,. \tag{P4.18}$$

Introducendo la (P4.18) nella (P4.12) si ottiene:

$$\left(2f\mu\mu_{g}-2f\right)f^{2}D+\left(f^{2}\mu^{2}-f^{2}\right)\frac{Y_{L}^{2}f\left(1-X^{2}\right)}{\sqrt{\dfrac{1}{4}Y_{T}^{4}+Y_{L}^{4}\left(1-X\right)^{2}}}=-2f_{p}^{2}f \tag{P4.19}$$

ed infine:

$$\mu_{g}=-\frac{1}{2f\mu}\left\{\frac{1}{f^{2}D}\left[2f_{p}^{2}f-\left(f^{2}\mu^{2}-f^{2}\right)\frac{Y_{L}^{2}f\left(1-X^{2}\right)}{\sqrt{\dfrac{1}{4}Y_{T}^{4}+Y_{L}^{4}\left(1-X\right)^{2}}}\right]+2f\right\}. \tag{P4.20}$$

che costituisce la relazione cercata.

**Applicazione informatica 4.1**

Nel problema precedente abbiamo visto la relazione che esprime l'indice di rifrazione di gruppo per un'onda radio in un plasma senza collisioni in presenza di un campo magnetico.
Si assuma $Y=0.5$ e $\theta=45°$. Si disegni il grafico di $\mu_{g[\text{ord}]}$ e $\mu_{g[\text{ext}]}$ in funzione di $X$. Si confrontino i risultati con quelli riportati nella Fig. 4.1. Si metta in evidenza il fenomeno di evanescenza della componente straordinaria, che si osserva per $X=(1-Y^2)/((1-Y_L^2)$. Si mettano in evidenza le riflessioni che si hanno per l'ordinario, per $X=1$ e per lo straordinario per $X=1\pm Y$.

**Applicazione informatica 4.2**

Si consideri la stessa relazione che esprime l'indice di rifrazione di gruppo per un'onda radio in un plasma senza collisioni che abbiamo visto nel Problema 4.1.
Si assuma adesso $Y=0.5$ e $\theta=0°$, che corrisponde al caso di propagazione longitudinale. Si disegni il grafico degli indici di rifrazione di gruppo ordinario e $\mu_{\text{ord}}$ e straordinario $\mu_{\text{ext}}$ in funzione di $X$. Si confrontino i risultati con quelli riportati nella Fig. 4.2. Si mettano in evidenza le riflessioni che si hanno per l'ordinario, per $X=1-Y$ e per lo straordinario per $X=1+Y$.

**Applicazione informatica 4.3**

Si consideri la stessa relazione che esprime l'indice di rifrazione di gruppo per un'onda radio in un plasma senza collisioni, relazione che abbiamo visto nel Problema 4.1.
Si assuma adesso $Y=2$ e $\theta=23°$. Si disegnino i grafici degli indici di rifrazione di gruppo ordinario e $\mu_{[\text{ord}]}$ e straordinario $\mu_{[\text{ext}]}$ in funzione di $X$. Si discutano i risultati.

**Applicazione informatica 4.4**

*Premessa*

Simulare uno ionogramma significa simulare il processo con il quale la ionosonda invia brevi impulsi radio verso l'alto e misura il tempo che intercorre fra la trasmissione e la ricezione della loro eco ionosferica. Per ricavare uno ionogramma artificiale, si dovrà dunque calcolare l'integrale:

$$t(f) = \int_0^{h_r} \frac{dh}{v_g} = \frac{1}{C} \int_0^{h_r} \mu_g \, dh, \tag{P4.21}$$

che corrisponde al calcolo di:

$$h' = h_b + \int_{h_b}^{h_r} \mu_g \, dh. \tag{P4.22}$$

In questo integrale è stato posto $h'=c \cdot t(f)$ e si è indicato con $h_b$ l'altezza della base della ionosfera, altezza sotto la quale l'indice di rifrazione delle radio onde può essere considerato unitario. Come abbiamo visto al paragrafo (3.3), nel caso in cui sia l'effetto del campo magnetico della Terra che l'effetto delle collisioni fra elettroni e molecole neutre siano trascurati, per $\mu_g$ e $\mu_f$ sussistono le seguenti relazioni:

$$\mu_g = \frac{1}{\mu_f} = \frac{1}{\sqrt{1-X}} = \frac{1}{\sqrt{1-\frac{\omega_p^2}{\omega^2}}} = \frac{1}{\sqrt{1-\frac{N_e(h)e^2}{4\pi^2\varepsilon_0 m f^2}}}, \tag{P4.23}$$

ove $\omega_p$ è la frequenza angolare di plasma, $\omega$ è la frequenza angolare della radio onda, $N_e(h)$ è la densità elettronica in funzione dell'altezza vera, $e$ ed $m$ sono rispettivamente la carica e la massa dell'elettrone ed $\varepsilon_0$ è la permittività dielettrica del vuoto.
Sostituendo la (P4.23) nella (P4.22) otteniamo:

$$h'(f) = h_b + \int_{h_b}^{h_r} \frac{1}{\sqrt{1-\frac{N_e(h)e^2}{4\pi^2\varepsilon_0 m f^2}}} \, dh. \tag{P4.24}$$

Questa ultima relazione costituisce l'integrale con il quale, dato un certo $N_e(h)$, può essere calcolato il corrispondente ionogramma simulato $h'(f)$. Praticamente, assumendo che $\mu_g$ sia costante su piccoli intervalli di $h$, l'integrale (P4.24) può essere rimpiazzato dalla seguente sommatoria su $i$:

$$h' = h_b + \sum_{i=1}^{p} \mu_g(h_i)\delta h. \tag{P4.25}$$

*Obbiettivo proposto*

Si consideri un profilo di densità elettronica, avente forma parabolica, con il massimo di densità elettronica $N_m$ posto alla quota di 225 km. Si assuma che al di sotto di $h_b$=150 km e al di sopra di 300 la densità elettronica sia nulla. Un tale profilo di densità elettronica si ottiene assumendo:

$$N(h) = \begin{cases} 0, & \text{if } h < h_b, \\ ah^2 + bh + c, & \text{if } 300 \text{ km} \geq h \geq h_b, \\ 0, & \text{if } h > 300 \text{ km}, \end{cases}$$

con

$a = -4 \cdot N_m / (150 \cdot 2 \cdot 300)$ m$^{-3}$ km$^{-2}$,
$b = 4 \cdot N_m / 150$ m$^{-3}$ km,
$c = -3 \cdot N_m$ m$^{-3}$.

Si assuma inoltre $N_m$=10$^{12}$ m$^{-3}$.
Si calcoli numericamente il seguente integrale:

$$h'(f_s) = h_b + \int_{h_b}^{h_r} n_g\left[f_s, \omega_p(h)\right] dh, \tag{P4.26}$$

che è in sostanza lo stesso riportato nella (P4.24). Si riporti $h'(f_s)$ su un grafico, scegliendo il range di variazione di $f_s$ nel campo delle HF.

*Suggerimenti*

Si proceda secondo lo schema suggerito nel seguito.
1. Si introducano le seguenti costanti nel programma: carica dell'elettrone $e = 1.6 \cdot 10^{-19}$ C, massa dell'elettrone $m_e$=9·10$^{-31}$ kg, costante dielettrica del vuoto: $\varepsilon_0$=8.85 10$^{-12}$ F m$^{-1}$.
2. Si calcoli la massima frequenza di plasma di questa ionosfera modellata:

$$f_{max} = \frac{1}{2\pi}\sqrt{\frac{e^2 N_m}{m\varepsilon_0}}. \tag{P4.27}$$

3. Si selezioni una frequenza per l'onda radio che incide verticalmente la ionosfera, minore di $f_{max}$.
4. Si supponga che l'onda radio si trovi ad una certa altezza $h$.
5. Si calcoli $N_e(h)$.
6. Si calcoli la $\omega_p(h)$.

7. Le condizioni di riflessione sono verificate? Se si vai al punto 12, altrimenti prosegui.
8. Si calcoli l'indice di rifrazione di gruppo; a questo scopo si usi la relazione (P4.26).
9. Si calcoli l'incremento di altezza virtuale.
10. Si incrementi l'altezza reale
11. Si ritorni al punto 5.
12. Si riporti su un grafico l'altezza virtuale. Si pongano a zero l'altezza reale e quella virtuale.
13. Si aumenti la frequenza di sondaggio $f_s$. Questa è maggiore o uguale a $f_{max}$? Se si, si termini il programma, altrimenti si vada al punto 5.

Altri suggerimenti utili sono i seguenti:
a) Al punto 7 si prosegua il calcolo finché $X<0.999$; in questo modo si evita la divergenza di mug che si ha per $X=1$, divergenza che manderebbe in errore il programma.
b) Nel calcolare l'integrale (P4.25), attraverso la sommatoria (P4.26), di assuma $\delta h=0.1$ km.
c) Eseguendo il calcolo nel modo che abbiamo qui proposto, la traccia dello ionogramma presenterà delle irregolarità che sono dovute a instabilità di calcolo. Si trascuri questo fenomeno, che verrà affrontato nell'esercizio successivo.

## Applicazione informatica 4.5

Abbiamo visto nel Applicazione Informatica 4.4 precedente, come la simulazione di un sondaggio ionosferico verticale corrisponda corrisponda al calcolo numerico di un integrale, calcolo da ripetersi per ogni valore della frequenza di sondaggio. Dal punto di vista numerico si trattava di effettuare la sommatoria (P4.25), che è la somma di molti cammini ottici virtuali $\delta \cdot \mu_g(h_i)$; abbiamo anche potuto capire che tale sommatoria deve essere interrotta prima che si raggiunga la quota $h_r$ che è quella ove $X=1$ e ove $\mu_g$ diverge. Nel Applicazione Informatica 4.4 è stato proposto, a titolo di esempio, di eseguire la sommatoria finché $X<0.999$ con $\delta=0.1$. In questo caso si osservano delle irregolarità nella traccia dello ionogramma dovute alla divergenza di $\mu_g$.

Per capire come si possono controllare questi fenomeni, si deve considerare una radio onda che viaggia verso strati della ionosfera con valori di $N_e(h)$ crescenti. Sappiamo che $\mu_g$ tende a infinito al tendere di $X = \omega_p^2/\omega^2$ a 1. Dunque, allo scopo di evitare divergenze nel calcolo, abbiamo la necessità di arrestare la sommatoria (P4.25) prima che $\mu_g \to \infty$. Si tratta perciò di porre per $X$ una soglia $\tau$ inferiore a 1, sopra la quale il calcolo deve essere arrestato. In altre parole $p$ che appare nella (P4.25) è numericamente definito dalle relazioni $X(h_p) < \tau$ e $X(h_{p+1}) \geq \tau$. Ovviamente, per avere una valutazione realistica dell'altezza reale di riflessione è necessario che questa soglia debba essere quanto più possibile vicina a 1.

Si consideri per esempio, il seguente semplice strato ionosferico:

$$\begin{cases} N_e(h) = 0 & \text{if } |h-h_0| \geq D \\ N_e(h) = N_0 \left[1 - \left(\frac{h-h_0}{D}\right)^2\right] & \text{if } |h-h_0| < D \end{cases} \quad \text{(P4.28a)}$$
$$\text{(P4.28b)}$$

Dove $N_0$ è la densità elettronica massima, che si trova alla quota $h_0$ e $D$ è il semispessore della parabola. La Fig. P4.1 mostra lo ionogramma ottenuto calcolando numericamente la sommatoria (P4.25) con $h_b = 150$ km e $\delta h = 0.5$ km, assumendo lo strato di forma parabolica secondo la (P4.28b) con $N_0 = 1.3 \cdot 10^{12}$ electroni/$m^3$, $h_0 = 250$ km, e $D = 100$ km, ed eseguendo il calcolo fino a che $X < 0.90$, $X < 0.99$ e $X < 0.999$, rispettivamente.

Si vede dalla stessa figura come, aumentando $\tau$, la cuspide dello ionogramma calcolato diventa via via più realistica. Comunque la Fig. P4.1 illustra anche come sopra un certo valore l'aumento di $\tau$ causi significativa instabilità di calcolo dovuta alla divergenza di $\mu_g$ in prossimità della quota di riflessione.

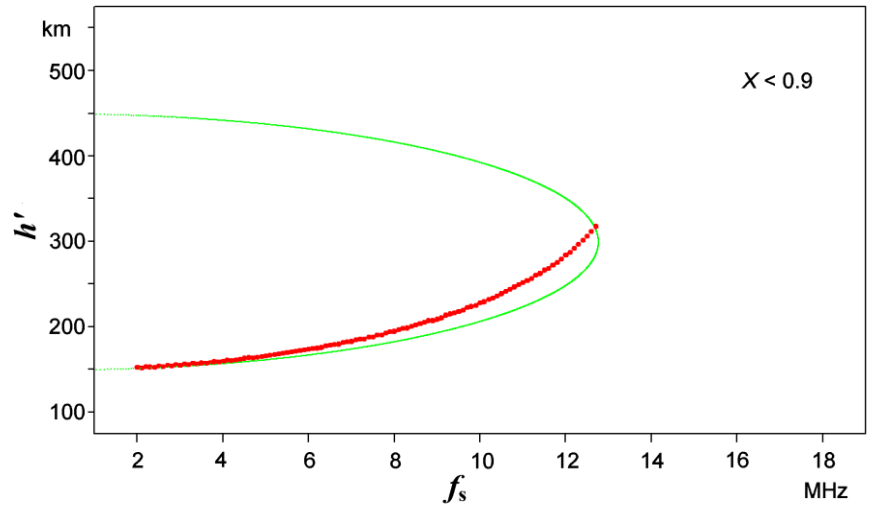

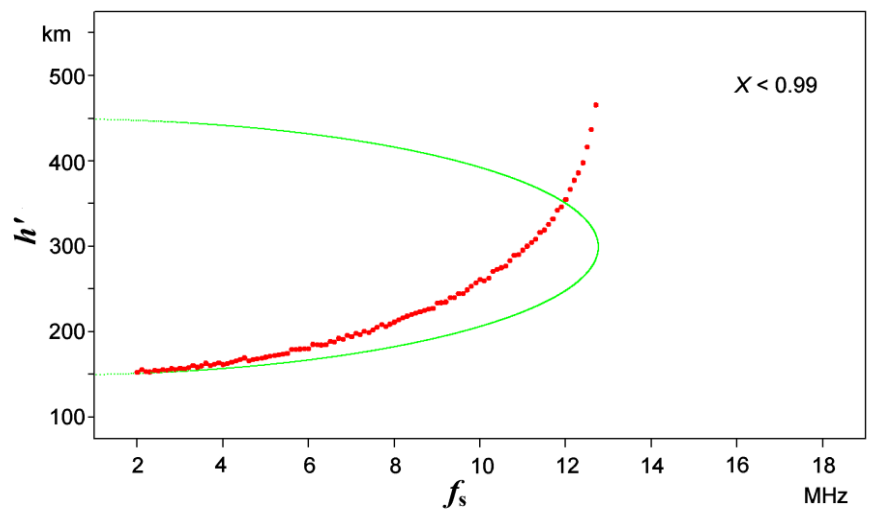

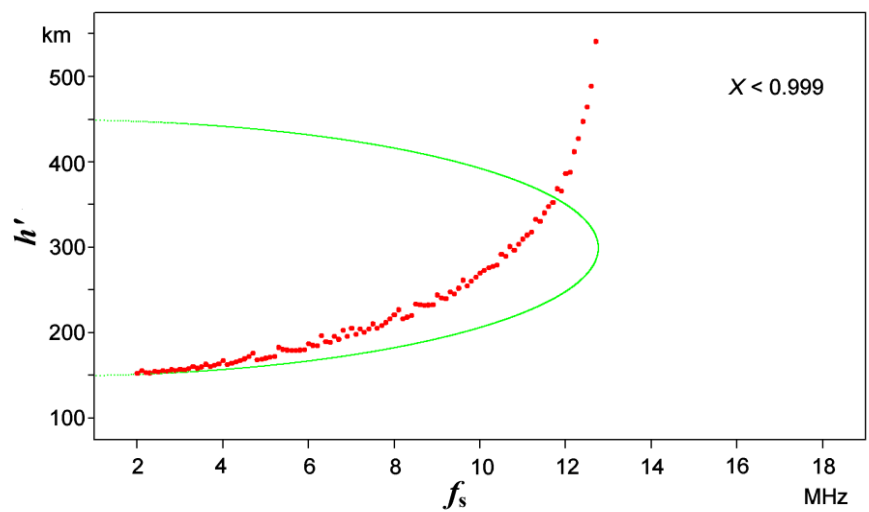

Fig. P4.1. Ionogramma ottenuto calcolando numericamente la sommatoria (P4.25) con $h_b = 150$ km $\delta h = 0.5$ km, considerando lo strato parabolico (P4.28) con $N_0 = 1.3 \cdot 10^{12}$ electroni/m$^3$, $h_0 = 250$ km, e $D = 100$ km. Il calcolo è stato eseguito per $X < \tau$ con $\tau = 0.90$, $0.99$ e $0.999$ rispettivamente. Queste figure sono analoghe a quelle pubblicate da rfr4-Scotto et. al. (2009).

*Compito proposto*

Si esegua la sommatoria (P4.25). L' incremento di altezza $\delta h$ vera però, non si consideri più fissato, ma decrescente come l'indice di rifrazione di gruppo $\mu_g$, secondo la seguente relazione:

$$\delta h_i = \frac{s}{\mu_g(h_i)}. \tag{P4.29}$$

In questo modo, gli incrementi di cammino ottico virtuale sono costanti e sono pari ad $s$, cosa che garantisce la stabilità dello ionogramma calcolato. Il valore di $s$ deve essere consistente con la risoluzione in altezza virtuale dello ionogramma che vogliamo ottenere. Si assuma nel calcolo $s = 1$ km. Si esegua il calcolo per $X < \tau$ con $\tau = 0.90$, 0.99 e 0.999 rispettivamente.

Si osservi che, aumentando $\tau$, la cuspide dello ionogramma simulato, corrispondente al picco di densità elettronica dello strato, diviene sempre più pronunciata, e dunque realistica. Allo stesso tempo è possibile vedere che i fenomeni di instabilità legati alla divergenza dell'indice di rifrazione che erano evidenziati nella Fig. P4.1, non sono più presenti.

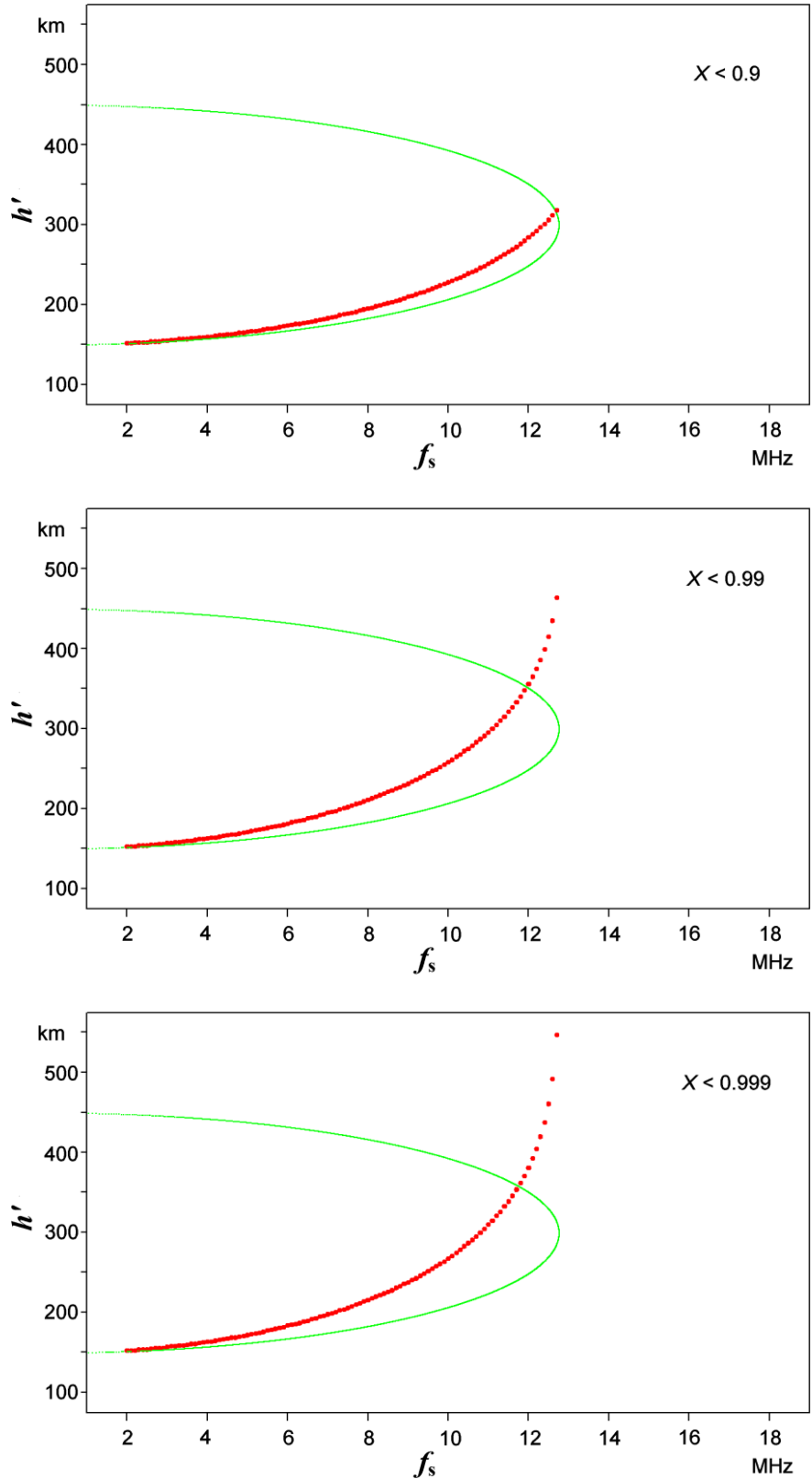

Fig. P4.2. Ionogramma ottenuto calcolando numericamente la sommatoria (P4.25) con $h_b = 150$ km, considerando lo strato parabolico (4.28b) con $N_0 = 1.3 \cdot 10^{12}$ m$^{-3}$, $h_0 = 250$ km e $D = 100$ km. Il calcolo è stato eseguito per $X < \tau$ con $\tau = 0.90$, $0.99$ e $0.999$ rispettivamente. Il passo di integrazione non è stato tenuto fisso (come si era fatto nei risultati presentati nella Fig. P4.1), ma è stato calcolato attraverso la (4.29). Queste figure sono analoghe a quelle pubblicate da Scotto et. al. (2009).

## Applicazione informatica 4.6

*Premessa*

Abbiamo visto nel problema 4.1, che $\mu_{g[ord,ext]}$, in assenza di collisioni e considerando la presenza del campo magnetico, può essere espresso analiticamente attraverso la (P4.20). Invece, per gli analoghi, $\mu_{g[ord,ext]}$ considerando le collisioni, non esiste una corripondente espressione analitica. In questo caso, procedere come si è visto al Paragrafo 4.6, ricavando però, non solo $\mu_{g[ord]}$ ma anche $\mu_{g[ext]}$.
Si osservi che allo stesso modo sarebbe stato possibile ricavare anche le parti immaginarie $\chi_{[ord]}$ e $\chi_{[ext]}$, di $n_{[ord,ext]}$. Tuttavia, come abbiamo detto al Paragrafo 3.2, queste sono associate all'assorbimento delle radio onde, e in questo problema non sono di nostro interesse.

*Obbiettivo proposto*

Si costruisca un programma con il quale si calcoli l'indice di rifrazione di gruppo per diversi valori dell'angolo $\theta$, diversi valori del parametro $Z$, e diversi valori del rapporto $\upsilon/\omega_c$. Si riportinio i valori su un grafico.

## Applicazione informatica 4.7

Si consideri un profilo di densità elettronica avente forma parabolica, analogo a quello proposto nell'applicazione informatica 4.3. Si ponga l'altezza della base della ionosfera pari a $h_1 = 110$ km, altezza al di sotto della quale $N_e(h)$ si assume nulla. Si assuma che il massimo di $N_e(h)$ sia $N_o = 2 \cdot 10^{12}$ m$^{-3}$ e che si trovi alla quota $h_m = 380$ km.

Si riporti su un grafico la grandezza:

$$h'(f_s) = \int_0^{h_R} \mu_{g[ord]}\left[f_s, \omega_p(h)\right] dh, \tag{P4.30}$$

che costituisce la simulazione della traccia ordinaria di uno ionogramma, eseguendo l'integrale numericamente, come indicato nel precedente Applicazione Informatica 4.4. Si esegua il calcolo facendo per l'indice di rifrazione $\mu_g$ diverse assunzioni:

a) assenza di campo magnetico e di collisioni;
b) presenza di campo magnetico B di intensità tale da indurre una girofrequenza elettronica pari a 1.2 MHz;
c) presenza di campo magnetico **B** di intensità tale da indurre una girofrequenza elettronica di 1.2 MHz; si consideri nell'indice di rifrazione anche l'effetto delle collisioni elettrone-molecola neutra; si assuma che la frequenza di queste collisioni vari con la quota $h$ secondo il modello semplificato $\nu = 10^{-0.0125 \cdot h \,[\text{km}] + 5.5}$ [s$^{-1}$].

Quando si fanno le assunzioni b) e c), si considerino diversi casi, ciascuno corrispondente a un diverso valore dell'angolo $\theta$ fra il vettore d'onda **k** e **B**, supponendo che questo sia diretto verticalmente, come avviene in un radio sondaggio ionosferico.

*Suggerimenti*

a) In assenza di campo magnetico e di collisioni $\mu_{g[ord]}$ è espresso dalla (2.35).
b) In presenza di campo magnetico e di collisioni $\mu_{g[ord]}$ è espresso dalla (P4.20).
c) Per l'indice di rifrazione con campo magnetico e con collsioni, non si riesce ad ottenere una formulazione analitica per l'indice di rifrazione di gruppo. Questo può essere ottenuto numericamente procedendo come nell'Applicazione Informatica 4.5.

**Applicazione informatica 4.8**

Si consideri lo stesso profilo di densità elettronica di forma parabolica dell'Applicazione Informatica 4.7 precedente.
Si riporti su un grafico la grandezza:

$$h'(f_s) = \int_0^{h_R} \mu_{g[ext]}\left[f_s, \omega_p(h)\right] dh, \tag{P4.31}$$

che costituisce la simulazione della traccia straordianaria di uno ionogramma. Si eseguano i calcoli facendo per l'indice di rifrazione $\mu_{g[ext]}$ diverse assunzioni, in modo simile a quanto richiesto nel Problema 4.8. In questo caso non si può fare l'assunzione di assenza di campo magnetico, in quanto la traccia straordianaria non si avrebbe se non vi fosse il campo magnetico.
I casi da considerare in questo caso saranno:

a) Presenza di campo magnetico B di intensità tale da indurre una girofrequenza elettronica pari a 1.2 MHz e assenza di collisioni.
b) Presenza di campo magnetico B di intensità tale da indurre una girofrequenza elettronica di 1.2 MHz. Si consideri nell'indice di rifrazione anche l'effetto delle collisioni elettrone-molecola neutra. Si assuma che la frequenza di queste collisioni vari secondo il modello semplificato $\nu = 10^{-0.0125 \cdot h\,[\text{km}]+5.5}$ [s$^{-1}$].

Si considerino diversi casi, ciascuno corrispondente a un diverso valore dell'angolo $\theta$ fra il vettore d'onda e B e il vettore d'onda **k**, supponendo che questo sia diretto verticalmente, come avviene in un radio sondaggio ionosferico.

*Suggerimenti*

a) In presenza di campo magnetico e di collisioni $\mu_{g[ext,ord]}$ è espresso dalla (P4.20).
b) Per l'indice di rifrazione con campo magnetico e con collsioni, non si riesce ad ottenere una formulazione analitica per l'indice di rifrazione di gruppo. Questo può essere ottenuto numericamente procedendo come nel problema P4.6.

**Bibliografia del Capitolo 4**

# CAPITOLO 5

## Condizioni di riflessione in presenza di collisioni

**Riassunto**

Viene studiate la riflessione attraverso le condizioni di Fresnel. Si trovano le condizioni di riflessione, sia per il raggio ordinario che per quello straordinario. Si vede che in questo ambito ha interesse valutare i campi di validità delle approssimazioni quasi longitudinale e trasversale. Questi campi di validità vengono studiati per una ionosfera modellata.

### 5.1 Condizioni di riflessione in presenza di collisioni

Riguardo ciascuna delle due componenti magnetoioniche di una radio onda che si propaga nella ionosfera, abbiamo visto, nei paragrafi 3.3 e 4.1, che, senza considerare le collisioni, si ha la divergenza di $\mu_g$ quando $\mu_f=0$. Abbiamo visto dunque che queste possono alternativamente essere considerate le condizioni di riflessione.

Lo studio delle divergenze di $\mu_g$ o la valutazione dei casi in cui si ha $\mu_f=0$, è sufficiente nel caso in cui non si considerino collisioni. In presenza di collisioni, invece, non si ha mai $\mu_f=0$ (si veda l'Applicazione Informatica 5.4), tuttavia la riflessione avviene. Lo studio di $\mu_{g[collisions]}$, ci permette di determinarne le divergenze, individuando dei casi nei quali la riflessione ionosferica che si osserva risulta spiegata.

Lo studio di $\mu_{g[collisions]}$, tuttavia non è esaustivo. Infatti, le considerazioni che abbiamo fatto fino a questo punto, sono basate sul prendere in esame un singolo raggio che incide la ionosfera. Una trattazione più completa deve essere fatta secondo le *equazioni di Fresnel* (o *condizioni di Fresnel*), condizioni che vennero originariamente ricavate per descrivere il comportamento dei raggi luminosi al passaggio fra due mezzi di diverso indice di rifrazione. Secondo questa teoria, ogni volta che un'onda elettromagnetica passa fra due mezzi aventi indice di rifrazione $\mu_{f[1]}$ e $\mu_{f[2]}$ ha luogo la riflessione di una debole onda avente ampiezza $\delta\mu_f / 2 \cdot \mu_f$, essendo $\delta\mu_f = \mu_{f[2]} - \mu_{f[1]}$ e $\mu_f = 1/2 \cdot (\mu_{f[2]} + \mu_{f[1]})$, con $\delta\mu_f \ll \mu_f$. Uno schema di questa riflessione elementare è riportato in Fig. 5.1.

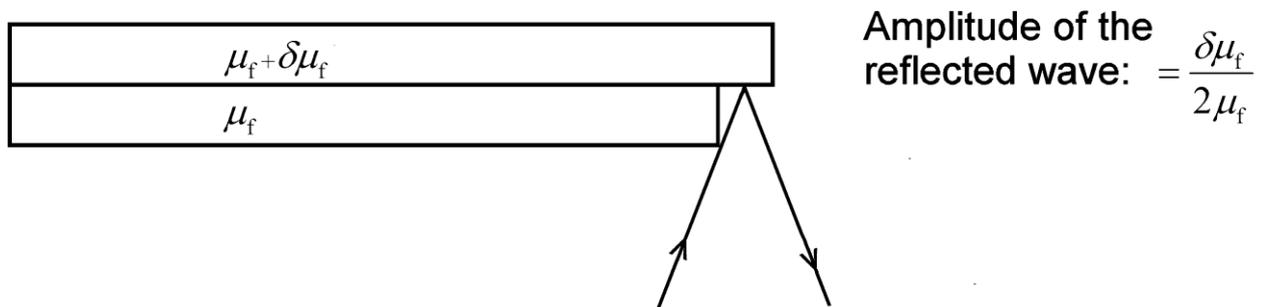

Fig. 5.1. Ampiezza delle onde riflesse fra due strati i cui indici di rifrazione di fase variano di $\delta\mu_f$, secondo Fresnel.

Allo scopo di modellare anche la riflessione ionosferica secondo le equazioni di Fresnel, supponiamo che l'onda radio, mentre procede verso l'alto, vada a incontrare delle zone ove l'indice di rifrazione medio è $\mu_f$ e dove, per esso, vi sia un gradiente. Supponiamo che questo gradiente sia fatto da una serie di uguali passi infinitesimali, ciascuno di spessore δ$z$, ove $\mu_f$ cambia di una quantità δ$\mu_f$, come riportato nella Fig. 5.2(a). Per determinare l'ampiezza della riflessione dovuta a un gradiente di densità elettronica così fatto, sarà necessario sommare, tenendo giusto conto dell'ampiezza e della fase, tutte le onde riflesse dai singoli strati.

Mentre l'ampiezza delle onde riflesse da ogni strato è uguale, la fase è diversa. Come si vede nella Fig. 5.3, il numero $N$ di onde di lunghezza λ compreso in un tratto δ$z$ è $N=\delta z \backslash \lambda$. Siccome, nel caso della riflessione da parte di uno strato infinitesimo, il tratto δ$z$ viene percorso sia in un senso che nell'altro, si ha:

$$N = \frac{\delta z}{\lambda} 2. \tag{5.1}$$

Volendo poi collegare la lunghezza d'onda λ nel mezzo considerato, avente indice di rifrazione $\mu_f$, alla lunghezza d'onda $\lambda_0$ nel vuoto, bisogna ricordare che per il periodo $T$ dell'onda valgono le relazioni: $T = \lambda_0/c = \lambda/v$, ove $c$ è la velocità della luce nel vuoto (ove $\mu_f = 1$), e $v$ quella nel mezzo. Da queste, ricordando che $v = c/\mu_f$:

$$\lambda_0 = \lambda \cdot \mu, \tag{5.2}$$

come riportato sinteticamente ancora nella Fig. 5.3.

Considerando le (5.1) e (5.2), risulta che ciascuno strato introduce una differenza di fase:

$$\phi = 2\pi N = 4\pi \frac{\delta z}{\lambda}. \tag{5.3}$$

Per eseguire la somma delle onde riflesse da ciascuno straterello, riportate nella Fig. 5.2(a), si ricorre al diagramma di ampiezza e fase. Come è noto, la somma di più vettori si esegue col metodo del poligono, come indicato nella Fig. 5.2(b), ove vengono disegnati i primi cinque d$\mathbf{a}_1$, d$\mathbf{a}_2$...d$\mathbf{a}_5$ di $N$ vettori. Questi sono tutti di uguale modulo d$a$, e su essi bisogna eseguire la somma:

$$\mathbf{R} = \sum_{i=1}^{N} d\mathbf{a}_i, \tag{5.4}$$

che ci permette di determinare la risultante **R**.

Nel limite in cui δ$\mu$ e δ$z$ tendono a zero, il poligono diventa una circonferenza. D'altra parte, se ha luogo un certo assorbimento, le onde diventano di ampiezza più debole man mano che viene raggiunto un livello successivo e il cerchio diventa la spirale riportata nella Fig. 5.2(c). L'ampiezza dell'onda risultante riflessa sarà allora rappresentata dal raggio $OR$ indicato nella Fig. 5.2(d). Per stimare questo si dovrà in primo luogo calcolare la lunghezza dell'arco:

$$OST = \int_0^{\pi} \frac{da}{d\phi} \cdot d\phi, \tag{5.4}$$

ove si ha:

$$\frac{da}{d\phi} = \left\{ \frac{\delta\mu}{2\mu} \right\} \div \left\{ 2\pi \frac{\mu}{\lambda} 2\delta z \right\} = \frac{1}{8\pi} \frac{\lambda}{\mu} \frac{d\mu}{dz}, \tag{5.5}$$

per cui eseguendo l'integrale:

$$OST = \frac{1}{8} \frac{\lambda_0}{\mu^2} \frac{d\mu}{dz} \qquad (5.6)$$

e, di conseguenza:

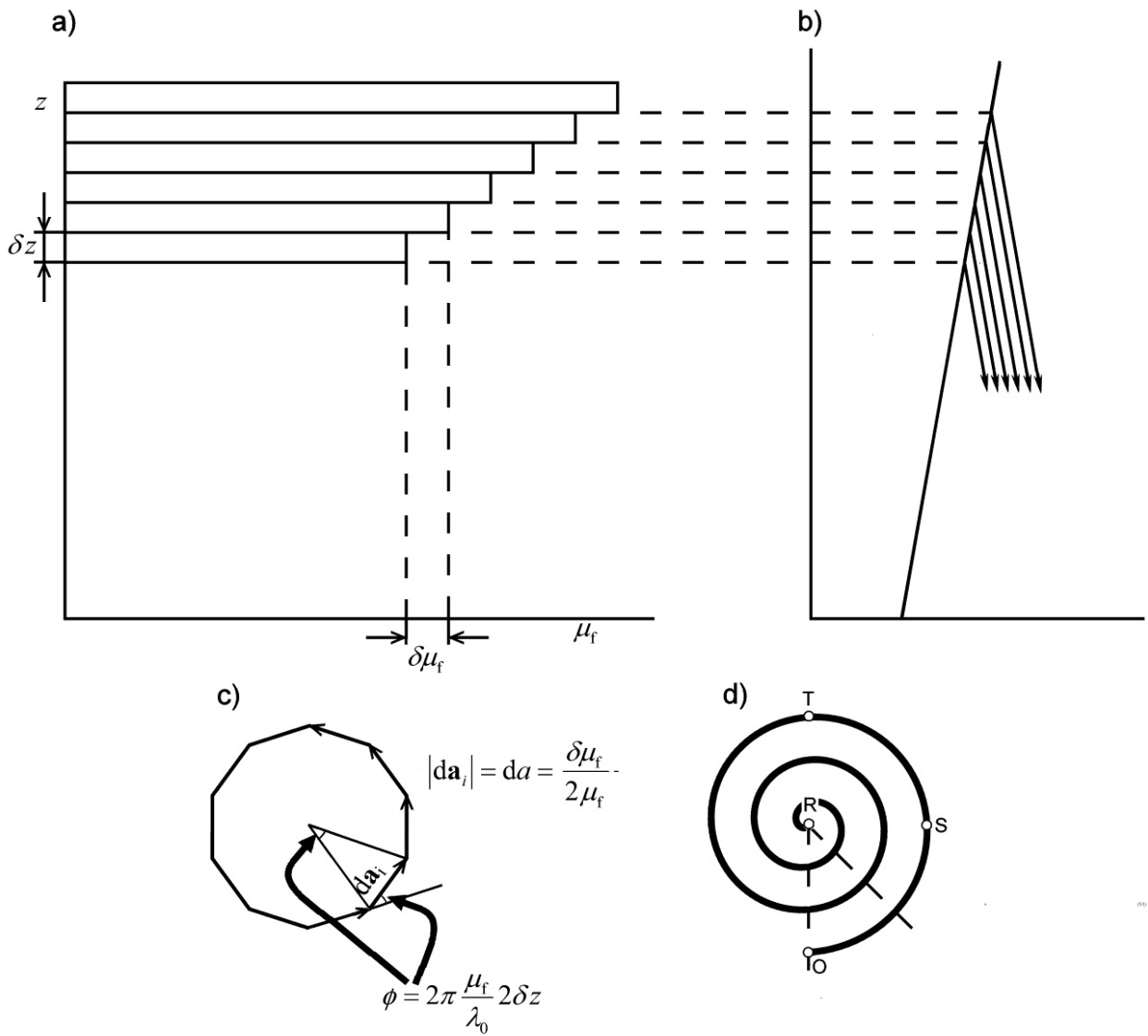

Fig 5.2. (a)-(d) Riflessione di un mezzo ove l'indice di rifrazione di fare $\mu_f$ varia con l'altezza.

$$\overline{OR} = \frac{\overline{OST}}{\pi} = \frac{1}{8\pi} \frac{\lambda_0}{\mu^2} \frac{d\mu}{dz}. \qquad (5.7)$$

Abbiamo così dimostrato che si ha riflessione se:

$$\frac{1}{8\pi}\frac{\lambda_0}{\mu_f^2}\frac{d\mu_f}{dz}\sim 1. \tag{5.8}$$

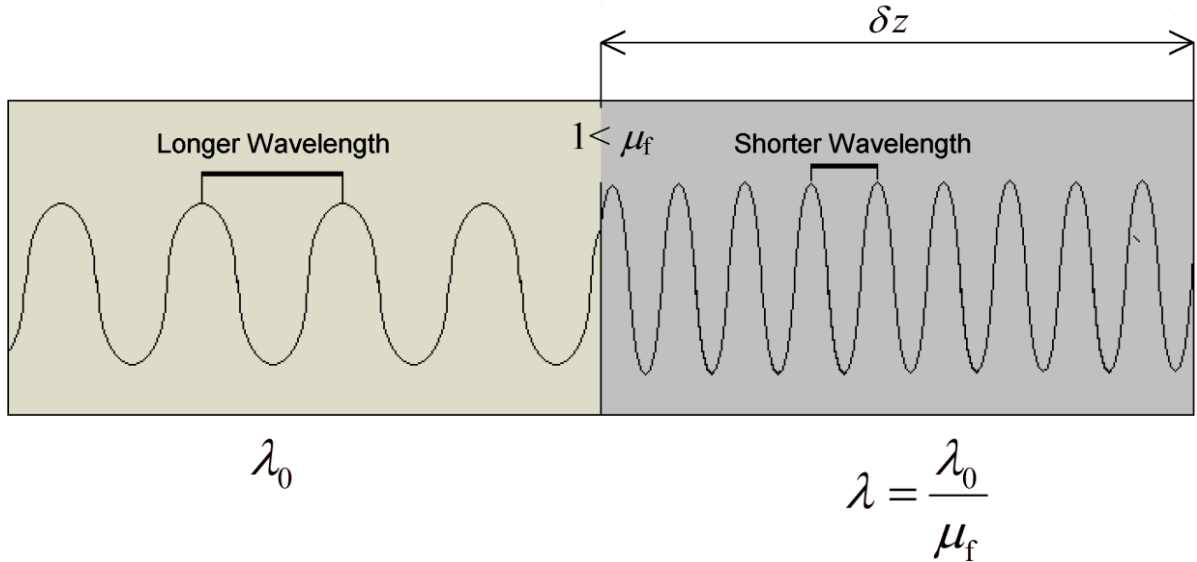

Fig. 5.3. Un'onda elettromagnetica che nel vuoto si propaga con lunghezza d'onda $\lambda_0$, in un mezzo avente indice di rifrazione $\mu_f$ si propaga con lunghezza d'onda $\lambda=\lambda_0/\mu_f$.

## 5.2 Condizioni di riflessione in presenza di collisioni e assenza di campo magnetico

Cerchiamo di valutare, sempre considerando una trattazione con raggi multipli, quali sono le condizioni di riflessione, per un plasma collisionale in assenza di campo magnetico. Si rende dunque necessario, facendo riferimento alla (5.8), procedere con una valutazione separata di $\mu_f$ e di $d\mu_f/dz$.

a) valutazione di $\mu_f$

Partendo dalla (2.34), si vede che, in assenza di campo magnetico, l'indice di rifrazione per le onde radio assume la seguente forma:

$$n_f^2 = 1 - \frac{X}{1-jZ} = 1 - \frac{X\cdot(1+jZ)}{1+Z^2}. \tag{5.9}$$

Se esprimiamo $n_f$, come si è soliti fare, in dipendenza delle sue parti reali ed immaginaria ($n_f=\mu_f-i\chi$), otteniamo:

$$\begin{cases} \mu_f^2 - \chi^2 = 1 - \dfrac{X}{1+Z^2} \approx 1 - X, & (5.10a) \\ 2\mu_f\chi = \dfrac{XZ}{1+Z^2} \approx XZ, & (5.10b) \end{cases}$$

dalle quali, per $X=1$ si ricava:

$$\chi = \mu_\text{f} \approx \sqrt{\frac{1}{2}Z} \,. \tag{5.11}$$

L' approssimazione è valida se $Z^2$ può essere trascurato nei confronti di 1, cioè se:

$$Z < 0.1. \tag{5.12}$$

Abbiamo così visto, partendo da una espressione come la (5.10), che, se $Z<0.1$, $\mu_\text{f}$ è piccolo per $X=1$ e vale $\mu_\text{f} \cong 0.5\, Z^{1/2}$.

b) Valutazione di $d\mu_\text{f}/dz$

Nella (5.8) dobbiamo pure valutare la derivata $d\mu/dz$. Si osservi che la dipendenza di $\mu_\text{f}$ da $z$, può essere espressa come $\mu_\text{f} = \mu_\text{f}(X(\omega_\text{p}(N_\text{e}(z))))$, per cui risulta:

$$\frac{d\mu_\text{f}}{dz} = \frac{d\mu_\text{f}}{dX}\frac{dX}{dz}. \tag{5.14}$$

Per quanto riguarda $dX/dz$, si ha:

$$\frac{dX}{dz} = \frac{d}{dz}\frac{\omega_p^2}{\omega^2} = \frac{e^2}{m\varepsilon_0\omega^2}\frac{dN}{dz}. \tag{5.15}$$

Per quanto riguarda $d\mu_\text{f}/dX$ si può dimostrare che il suo valore è non piccolo al tendere di $X$ a 1, e può essere assunto:

$$\frac{d\mu_\text{f}}{dX} \sim 1 \tag{5.16}$$

per $Z \sim 0.1$. Al diminuire di $Z$, inoltre, $d\mu_\text{f}/dX$ cresce. Questa dimostrazione può essere fatta sia analiticamente che numericamente ed è proposta nel Applicazine Informatica 5.3.

c) Valutazione complessiva della (5.8)

Dobbiamo valutare il termine di sinistra della (5.8). In forza delle (5.14), (5.15) e (5.16) questo può essere scritto come:

$$\frac{1}{8\pi}\frac{\lambda_0}{\mu_\text{f}^2}\frac{e^2}{m\varepsilon_0\omega^2}\frac{dN}{dz} \sim 1\,, \tag{5.17}$$

per $X=1$. Da questa

$$\frac{dN}{dz} \sim \frac{m\varepsilon_0\omega^2}{e^2}\frac{\mu_\text{f}^2}{\lambda_0}8\pi \tag{5.18}$$

Consideriamo dapprima una onda radio alla frequenza $f_s=1$ MHz. Questa ha $\lambda=c/f_s=(3\cdot 10^8/1\cdot 10^6)=300$ m. Possiamo assumere che per tale onda radio la condizione $X=1$ (cioè $f_s =$

$f_p$ ) si verifichi alla quota di 100 km, nella regione E. A questa quota si ha che ν=$10^5$ s$^{-1}$ per cui Z= ν/ω = ν/(2·π· $f_s$)= $10^5$/(6.28·1· $10^6$)=0.016. Siccome Z <0.1, possiamo usare la (5.12) per valutare $\mu_f$. Si ha così $\mu_f \cong 0.5\, Z^{1/2} = 0.5 \cdot 0.016^{1/2}$=0.063. Dunque numericamente la (5.10) si valuta come:

$$\frac{dN}{dz} \sim 8 \cdot 3.14 \cdot \frac{10^{-30} 8.85 \cdot 10^{-12} \left(6.28 \cdot 10^6\right)^2}{(1.6 \cdot 10^{-19})^2} \frac{0.063^2}{300} = 2.04 \cdot 10^6 \, \mathrm{m^{-3}/m}.$$

Che è un gradiente facilmente riscontrabile nella regione E, ove un'onda di 1 MHz si può riflettere. Consideriamo ora una onda radio alla frequenza $f_s$ =10 MHz. Questa ha λ=c/$f_s$=(3·$10^8$/10·$10^6$)=30 m. Possiamo assumere che per tale onda radio la condizione X=1 (cioè $f_s = f_p$ ) si verifichi alla quota di 250 km, nella regione F2. A questa quota si ha che ν=$10^{2.7}$ s$^{-1}$ per cui Z=ν/ω=ν/(2·π· $f_s$)= $10^{2.67}$/(6.28·10· $10^6$)=8· $10^{-6}$. Essendo Z <0.1, possiamo usare la (5.12) per valutare $\mu_f$. Si ha così $\mu_f \cong 0.5\, Z^{1/2} = 0.5 \cdot (8 \cdot 10^{-6})^{1/2}$=1.4· $10^{-3}$. Dunque numericamente la (5.10) si valuta come:

$$\frac{dN}{dz} \sim 8 \cdot 3.14 \cdot \frac{10^{-30} \cdot 8.85 \cdot 10^{-12} \left(6.28 \cdot 10^6\right)^2}{(1.6 \cdot 10^{-19})^2} \frac{0.063^2}{30} = 2.05 \cdot 10^6 \, \mathrm{m^{-3}/m}$$

Che è un gradiente facilmente riscontrabile nella regione F2.
Abbiamo così visto che, in assenza di **B**, la (5.8) è facilmente verificata per X=1, se Z<0.1. In questo caso, di conseguenza, ha luogo la riflessione ionosferica delle onde HF. Si osservi che questo è dovuto al fatto che per X=1, se Z<0.1 si ha $\mu_f^2$<<1. Come vedremo nei paragrafi successivi, questo risultato ha interesse più vasto che il semplice caso irrealistico di **B**=0.

### 5.3 Condizione di riflessione per la componente ordinaria in approssimazione quasi trasversale

Consideriamo l'indice di rifrazione in presenza di collisioni nel caso del raggio ordinario. Opereremo supponendo che per i termini dentro il radicando valga la relazione $Y_T^2/[2\cdot(1\text{-}X\text{-}j\cdot Z)] \gg Y_L^2$, detta *approssimazione quasi trasversale* (QT). Sarà poi evidente che, operando in questa approssimazione, si ha $\omega_t < \nu$, condizione per la quale la regola di Booker impone di considerare per la componente ordinaria il segno positivo nella (2.34) sia per X<1 che per X>1. Da tale relazione, dunque, possiamo scrivere:

$$n^2 = 1 - \frac{X}{1 - jZ - \dfrac{Y_T^2}{2\cdot(1-X-jZ)} + \sqrt{\dfrac{Y_T^4}{4\cdot(1-X-jZ)^2} + Y_L^2}}, \qquad (5.12)$$

nella quale si possono riorganizzare i termini nel modo seguente:

$$n^2 = 1 - \frac{X}{1 - jZ - \dfrac{Y_T^2}{2\cdot(1-X-jZ)} \cdot \left(1 - \sqrt{1 + Y_L^2 \dfrac{4\cdot(1-X-jZ)^2}{Y_T^4}}\right)}. \qquad (5.13)$$

Questa formula può poi essere approssimata usando la relazione $(1+\varepsilon)^{1/2} \approx 1+\varepsilon/2$, per cui si ha:

$$\approx 1 - \cfrac{X}{1 - jZ - \cfrac{Y_T^2}{2 \cdot (1 - X - jZ)} \cdot \left[1 - \left(1 - \cfrac{1}{2} Y_L^2 \cfrac{4 \cdot (1 - X - jZ)^2}{Y_T^4}\right)\right]} =$$

$$= 1 - \frac{X}{1 - jZ + (1 - X - jZ)\dfrac{Y_L^2}{Y_T^2}}. \tag{5.14}$$

In questa relazione possono poi essere introdutte le relazioni $Y_L = Y \cdot \cos(\theta)$ e $Y_T = Y \cdot \sin(\theta)$, (ove $\theta$ è l'angolo fra il campo magnetico e la direzione di propagazione dell'onda), viste al Capitolo 2, ottenendo:

$$= 1 - \frac{X}{1 - jZ + (1 - X - jZ)\dfrac{\cos^2 \theta}{\sin^2 \theta}}. \tag{5.15}$$

Ora per $X=1$, questa diventa:

$$n^2 = 1 - \frac{1}{1 - jZ \cdot (1 + \dfrac{\cos^2 \theta}{\sin^2 \theta})}. \tag{5.16}$$

In analogia a quanto avviene per il caso senza campo magnetico, avremmo che se,

$$Z \cdot (1 + \frac{\cos^2 \theta}{\sin^2 \theta}) < 0.1, \tag{5.17}$$

allora per $\mu$ vale la relazione:

$$\mu = \sqrt{\frac{1}{2} Z \cdot (1 + \frac{\cos^2 \theta}{\sin^2 \theta})}. \tag{5.18}$$

Introducendo poi la relazione trigonometrica:

$$1 + \frac{\cos^2 \theta}{\sin^2 \theta} = \frac{1}{\sin^2 \theta}, \tag{5.19}$$

la (5.17) può più convenientemente essere scritta come:

$$Z < 0.1 \cdot \sin^2 \theta, \tag{5.20}$$

mentre la (5.18) diventa:

$$\mu_f = \frac{1}{\sin \theta} \sqrt{\frac{1}{2} Z}. \tag{5.21}$$

Nelle condizioni in cui ci siamo posti, espresse dall' approssimazione QT e dalla (5.20), la (5.21) ci garantisce che $\mu_f^2 \ll 1$, il che, analogamente a quanto visto al paragrafo precedente, implica la riflessione dell'onda radio.

**5.4 Condizione di riflessione per la componente ordinaria in approssimazione quasi longitudinale**

Se fra i termini che si trovano dentro il radicando sussiste la relazione $Y_T^2/[2\cdot(1-X-j\cdot Z)] \ll Y_L^2$, allora siamo in condizioni di *approssimazione quasi longitudinale* (QL), e l'indice di rifrazione assume la forma:

$$n^2 = 1 - \frac{X}{1 - jZ \pm Y_L}. \tag{5.22}$$

Ponendo:

$$X_{QL} = \frac{X}{1 \pm Y_L}, \tag{5.23}$$

e

$$Z_{QL} = \frac{Z}{1 \pm Y_L}, \tag{5.24}$$

possiamo riscrivere la (5.22) in termini delle nuove grandezze $Z_{QL}$, $X_{QL}$. Per far questo ricaviamo:

$$Z_{QL}(1 \pm Y_L) = Z \tag{5.25}$$

e

$$X_{QL}(1 \pm Y_L) = X, \tag{5.26}$$

le quali, sostituite nella (5.22) dànno:

$$n^2 = 1 - \frac{X_{QL}(1 \pm Y_L)}{1 - jZ_{QL}(1 \pm Y_L) \pm Y_L}, \tag{5.27}$$

e dunque:

$$n^2 = 1 - \frac{X_{QL}(1 \pm Y_L)}{1 - jZ_{QL} \mp jZ_{QL}Y_L \pm Y_L} =$$
$$= 1 - \frac{X_{QL}(1 \pm Y_L)}{-jZ_{QL}(1 \pm Y_L) + 1 \pm Y_L} =$$
$$= 1 - \frac{X_{QL}}{-jZ_{QL} + 1}. \tag{5.28}$$

È semplice osservare che la (5.10) e la (5.28) hanno la medesima forma. Dalla (5.28) possono dunque essere ottenuti risultati simili a quelli che abbiamo ottenuto dalla (5.10); dunque analogamente a quanto avviene nel caso QT ordinario, si ha che la parte reale dell'indice di rifrazione è per $X_{QL}=1$:

$$\mu = \sqrt{\frac{1}{2} Z_{QL}}, \tag{5.29}$$

se $Z_{QL} < 0.1$.

Cioè (usando $X=1 \pm Y_L$):

$$\mu = \sqrt{\frac{1}{2} \frac{Z}{1 \pm Y_L}}, \tag{5.30}$$

se

$$Z_{QL} = \frac{Z}{1 \pm Y_L} < 0.1, \tag{5.30}$$

cioè se

$$Z < 0.1 \cdot (1 \pm Y_L). \tag{5.31}$$

Abbiamo così dimostrato che se $Z<0.1\cdot(1\pm Y_L)$, in approssimazione QL, per $X=1\pm Y_L$ l'indice di rifrazione risulta $\mu_f=[0.5\cdot Z/(1\pm Y_L)]^{1/2}$. Da questo si $\mu_f^2 \ll 1$, il che, analogamente a quanto visto al paragrafo precedente, implica la riflessione dell'onda radio.

**5.5 Zone di validità delle approssimazioni QT e QL**

Riassumendo, abbiamo visto che, in presenza di collisioni si ha $\mu_f^2 \ll 1$, con conseguente riflessione, in approssimazione QT per $X=1$, se $Z < 0.1 \cdot \sin^2\theta$, in approssimazione QL se $Z < 0.1 \cdot (1 \pm Y_L)$ per $X=1\pm Y_L$. Risulta così importante studiare le condizioni sotto le quali sono applicabili le approssimazini QT o QL.
A questo scopo bisogna considerare di nuovo l'equazione (2.34) e, partendo da questa, si deve valutare in quali casi è applicabile l'approssimazione QT e in quali casi quella QL. Riprendendo quanto detto nei paragrafi precedenti, dovremo osservare il termine sotto radice:

$$\frac{Y_T^4}{4 \cdot (1-X-jZ)^2} + Y_L^2, \tag{5.32}$$

dal quale si vede che affinché sia valida l' approssimazione QT dovrà essere:

$$\frac{Y_T^4}{4 \cdot \left[(1-X)^2 + Z^2\right]} \gg Y_L^2, \tag{5.33}$$

viceversa affinché sia valida l'approssimazione QL dovrà essere:

$$\frac{Y_T^4}{4\cdot\left[(1-X)^2+Z^2\right]} \ll Y_L^2. \tag{5.34}$$

*Approssimazione QT*

Si applica la (5.33), dalla quale:

$$\frac{Y_T^4}{4Y_L^2} \gg \left[(1-X)^2+Z^2\right]. \tag{5.35}$$

Ricordando le definizioni di $Y_T$ e $Y_L$ già richiamate al paragrafo 5.3, si ha:

$$\frac{\omega_B^4 \sin(\theta)^4}{\omega^4}\frac{1}{4}\frac{1}{\frac{\omega_B^2\cos(\theta)^2}{\omega^2}} \gg \left[(1-X)^2+Z^2\right], \tag{5.36}$$

e dunque

$$\frac{\omega_B^2}{\omega^2}\frac{1}{4}\frac{\sin(\theta)^4}{\cos(\theta)^2} \gg \left[(1-X)^2+Z^2\right]. \tag{5.37}$$

Usualmente si pone

$$f(\theta) = \frac{1}{2}\frac{\sin(\theta)^2}{\cos(\theta)}, \tag{5.38}$$

di modo che la (5.37) viene scritta come:

$$\frac{\omega_B^2}{\omega^2}\left[f(\theta)\right]^2 \gg \left[(1-X)^2+Z^2\right]. \tag{5.39}$$

Per $X=1$ essa è verificata se:

$$\frac{\omega_B^2}{\omega^2}\left[f(\theta)\right]^2 \gg Z^2, \tag{5.40}$$

la quale, dopo aver definito,

$$\omega_c = \omega_B\left[f(\theta)\right] \tag{5.41}$$

diventa:

$$\frac{\omega_c^2}{\omega^2} \gg \frac{v^2}{\omega^2}. \tag{5.42}$$

Questa relazione può essere espressa attraverso $Z$ e in questo caso corrisponde a:

$$\frac{1}{3}\frac{\omega_c}{\omega} > Z. \tag{5.43}$$

In altre parole, la condizione affinché sia valida l'approssimazione QT per $X=1$, può essere espressa come:

$$\omega_c^2 \gg v^2, \tag{5.44}$$

oppure:

$$\omega_c \frac{1}{3} > v. \tag{5.45}$$

Nel caso contrario avremo l'approssimazione QL, che sussiste se:

$$3 \cdot \omega_c < v. \tag{5.46}$$

Quest'ultima, se espressa tramite $Z=v/\omega$, corrisponde a:

$$3\frac{\omega_c}{\omega} < Z. \tag{5.47}$$

Abbiamo dunque visto in quali casi, per $X=1$, vale l'approssimazione QT. In questi casi se è verificata la (5.20), ($Z< 0.1 \sin^2\theta$) $\mu_f$ può essere valutato attraverso la (5.21) [$\mu_f= (\sin\theta)^{-1}(0.5\ Z)^{1/2}$]. Si ha in questo caso $\mu_f^2 \ll 1$ e abbiamo la riflessione ionosferica.

*Approssimazione QL*

Se la condizione QT sussiste per $X=1$, ci saranno comunque altri valori di $X$, per i quali l'approssimazione da tenere in conto è quella QL. In particolare saremo interessati a sapere se l'approssimazione è valida per $X=1\pm Y_T$, poiché, in questo caso, potremmo valutare $\mu_f$ con la (5.31). Similmente a quanto abbiamo fatto sopra per la condizione QT, affinché si verifichi la condizione QL, si deve imporre:

$$\frac{\omega_B^2}{\omega^2}\left[f(\theta)\right]^2 \ll \left[(1-X)^2 + Z^2\right]. \tag{5.48}$$

Una condizione sufficiente, ma non necessaria, per poter applicare la approssimazione QL, sarà dunque:

$$\frac{\omega_B^2}{\omega^2}\left[f(\theta)\right]^2 \ll \left[(1-X)^2\right]. \tag{5.49}$$

Ci domandiamo ora se questa condizione sussiste per $X=1\pm Y_L$. Dovrà essere:

$$Y^2\left[f(\theta)\right]^2 \ll Y_L^2, \tag{5.50}$$

da cui:

$$Y^2\left[f(\theta)\right]^2 \ll Y^2\cos(\theta)^2 \tag{5.51}$$

e

$$\left[\frac{1}{2}\frac{\sin^2(\theta)}{\cos\theta}\right]^2 \ll \cos^2(\theta). \tag{5.52}$$

Volendo passare a una semplice disuguaglianza si ha:

$$\left[\frac{1}{2}\frac{\sin^2(\theta)}{\cos\theta}\right] < 3\cos(\theta), \tag{5.53}$$

da cui:

$$\left[\frac{\sin(\theta)}{\cos\theta}\right]^2 \ll \frac{3}{2}, \tag{5.54}$$

ed infine:

$$\theta < \arctan\sqrt{3/2} = 40°. \tag{5.55}$$

**5.6 Sintesi dei risultati raggiunti**

I risultati fino a qui raggiunti possono essere sintetizzati qui di seguito:

1) In assenza di campo magnetico, se $Z<0.1$, $\mu_f$ può essere stimato come $\mu_f = ½·Z^{0.5}$, e questo dà luogo alla riflessione ionosferica.
2) Considerando il campo magnetico, se per $X=1$ vale l'approssimazione QT, si ha $\mu_f(X=1) = (½·Z)$, se $Z<0.1 \sin^2\theta$. Questo dà luogo alla riflessione ionosferica, a meno di d$N$/d$h$ modesti, quali quelli che si possono trovare intorno ai massimi di $N_e(h)$.
3) Considerando il campo magnetico, se per $X=1\pm Y_L$ vale l'approssimazione QL, si ha $\mu_f(X=1\pm Y_T) = ½·Z/(1\pm Y_T)$, se $Z<0.1 (1\pm Y_T)$.
4) In relazione a quanto detto al punto 2), per $X=1$ vale l'approssimazione QT se la frequenza della collisioni è abbastanza bassa, cioè $\nu < 1/3\ \omega_c$.

5) In relazione a quanto detto al punto 3), per $X=1\pm Y_L$ vale l'approssimazione QL se se $\theta<40°$.

Nei casi 1), 2) e 3) si ha $\mu_f^2 \ll 1$ che, a meno di d$N$/d$h$ modesti, quali quelli che si possono trovare intorno ai massimi di $N_e(h)$, garantisce la riflessione ionosferica.

# Applicazioni informatiche del capitolo 5

## Applicazione informatica 5.1

Consideriamo un magnetoplasma non caldo, quale quello ionosferico ove il campo magnetico è costante, la densità elettronica e la frequenza delle collisioni sono stratificate in modo parallelo al piano orizzontale, e hanno gradiente nella direzione $z$, che è supposta verticale, con l'orientazione positiva verso l'alto.

Si consideri un'onda radio di frequenza $f_s$, che viaggi verso l'alto, verso zone a densità elettronica $N_e(z)$ via via crescente. Per quanto abbiamo detto, se trascuriamo le collisioni tra elettroni e molecole neutre, l'indice di rifrazione diventerà nullo ad una certa quota, dove, secondo la teoria a raggi singoli, l'onda sarà completamente riflessa.

Si consideri ora uno strato di elettroni, che ha densità eccessiva per permettere la propagazione dell'onda radio considerata: una densità tale da indurre una frequenza di plasma $f_p > f_s$. Esisteranno dunque due livelli, indicati nella Fig. P5.1 con LL' e MM', fra i quali l'indice di rifrazione complesso $n=\mu-i\chi$, sarà puramente immaginario. Sarà cioè $n^2$ negativo, $\mu=0$ e $\chi>0$. Dunque, se la distanza fra i due livelli LL' e MM', non è troppo grande può capitare che l'onda evanescente può non estinguersi completamente, con una parte dell'energia che raggiungerà il livello MM', e continuerà a propagarsi alle quote superiori.

Si consideri un profilo di densità elettronico costituito da un sottile strato di forma parabolica, quale quello illustrato nella Fig. P5.1. Esso è rappresentabile nella modo seguente:

$$N(h) = \begin{cases} 0 & \text{if } h < 101.5 \text{ km} \\ ah^2 + bh + c & \text{if } 102 \text{ km} \geq h \geq 101.5 \\ 0 & \text{if } h > 102 \text{ km} \end{cases}$$

con

$a = -9 \cdot 10^9$ m$^{-3}$ km$^{-2}$,

$b = 2 \cdot 10^{12}$ m$^{-3}$ km,

$c = -1.12$ m$^{-3}$.

Questo profilo di densità elettronico è stato inoltre costruito assumendo un valore della densità elettronica massima pari a $N_m = 0.993 \cdot 10^{12}$ m$^{-3}$, valore compatibile con una frequenza di plasma massima pari a $f_{p[max]} = 9$ MHz.

Scegliendo alcune frequenze inferiori a $f_{p[max]}$, si calcoli la percentuale di energia che, attraverso l'onda evanescente, può continuare a propagarsi nella ionosfera sovrastante lo strato considerato.

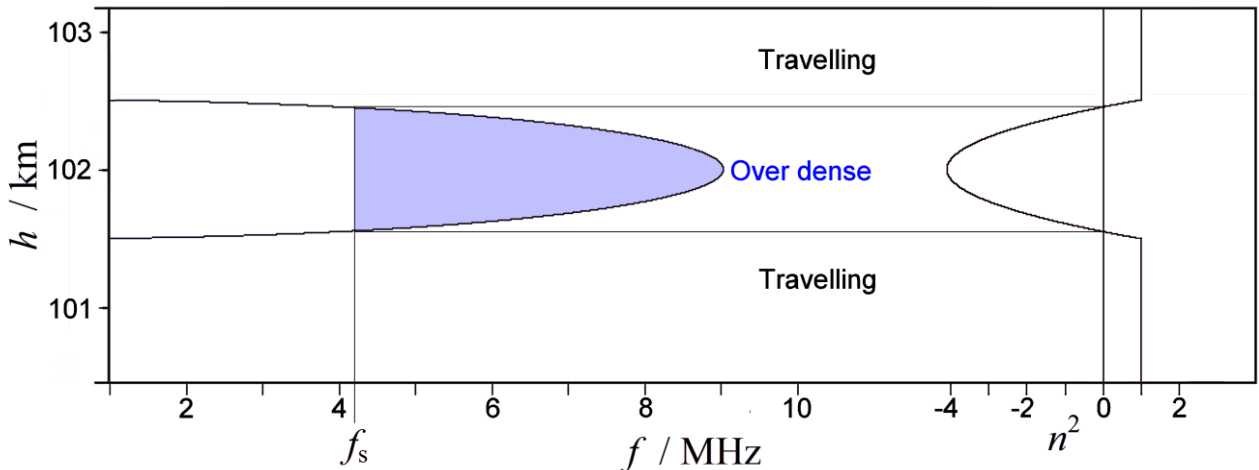

Fig. P5.1. La propagazione di un'onda evanescente attraverso un sottile strato avente $N_e$ tale da indurre $f_p > f_s$.

*Risoluzione*

Dall'equazione (3.7), risulta che un'onda radio di ampiezza $A$ e di frequneza $f_s$, nel percorrere un tratto $\Delta z$ ove $f_p > f_s$, subisce una attenuazione, diventando di ampiezza:

$$A \cdot e^{-\frac{\chi \omega}{c} \Delta x}, \qquad (5.56)$$

ove

$$\chi = \sqrt{X-1} = \sqrt{\frac{N_e(z) \cdot e^2}{m \cdot \varepsilon_0} \frac{1}{\omega} - 1} . \qquad (5.57)$$

Passando passando agli infinitesimi, avremo che la (5.56) diventa:

$$A \cdot e^{-\frac{\omega}{c} \int_{z_1}^{z_2} \chi\, dz} = A \cdot e^{-\frac{\omega}{c} \int_{z_1}^{z_2} \left[ \sqrt{\frac{N_e(z) \cdot e^2}{m \cdot \varepsilon_0} \frac{1}{\omega} - 1} \right] dz}, \qquad (5.58)$$

ove $z_1$ e $z_2$ sono le quote rispettivamente inferiori e superiori a quella del massimo della parabola, ove si ha $X=0$, per l'onda elettromagnetica della frequenza considerata.

Per conoscere poi la percentuale di energia che viene propagata da una radio onda, dovremo tenere conto che questa è proporzionale al quadrato dell'ampiezza dei campi associati all'onda stessa. Per cui, se l'energia che va a incidere su una certa superficie $S$ dello strato considerato è $A^2$, l'energia che si propaga attraverso la stessa superficie, al di sopra di esso, sarà, in base alla (5.58):

$$A^2 \cdot e^{-\frac{2\omega}{c} \int_{z_1}^{z_2} \left[ \sqrt{\frac{N_e(z) \cdot e^2}{m \cdot \varepsilon_0} \frac{1}{\omega} - 1} \right] dz} . \qquad (5.59)$$

Il termine:

$$P = e^{-F}, \qquad (5.60)$$

con

$$F = \frac{2\omega}{c} \int_{z_1}^{z_2} \left[ \sqrt{\frac{N_e(z) \cdot e^2}{m \cdot \varepsilon_0} \frac{1}{\omega} - 1} \right] dz \, , \tag{5.61}$$

rappresenta la parte di energia trasportata dall'onda evanescente da $z_1$ a $z_2$.
La tabella 5.1 riporta i risultati ottenibili numericamente per lo strato proposto.

Tab. 5.1 - I risultati ottenuti numericamente per alcuni $f_s < f_{p[max]}$.

| $f_s$ (MHz) | $z_1$ (km) | $z_2$ (km) | $F = \frac{2\omega}{c} \int_{z_1}^{z_2} \left[ \sqrt{\frac{N_e(z) \cdot e^2}{m \cdot \varepsilon_0} \frac{1}{\omega} - 1} \right] dz$ | exp(-F) |
|---|---|---|---|---|
| 1.0 | 101.50 | 102.50 | 0.14 | 0.75 |
| 2.0 | 101.51 | 102.49 | 0.14 | 0.76 |
| 3.0 | 101.53 | 102.47 | 0.13 | 0.77 |
| 4.0 | 101.55 | 102.45 | 0.12 | 0.79 |
| 5.0 | 101.59 | 102.42 | 0.10 | 0.82 |
| 6.0 | 101.63 | 102.37 | $8.22 \cdot 10^{-2}$ | 0.85 |
| 7.0 | 101.69 | 102.32 | $5.84 \cdot 10^{-2}$ | 0.89 |
| 8.0 | 101.77 | 102.23 | $3.10 \cdot 10^{-2}$ | 0.94 |

**Applicazione informatica 5.2**

Abbiamo visto nella trattazione teorica che l'avvenire la propagazione in condizioni QT o QL, dipende dal fatto che sia verificata la condizione:

$$\frac{\omega_C^2}{\omega^2} >> \left[ (1-X)^2 + Z^2 \right], \tag{5.62}$$

per l'approssimazione QT o

$$\frac{\omega_C^2}{\omega^2} << \left[ (1-X)^2 + Z^2 \right], \tag{5.63}$$

per l'approssimazione QL.

Essendo:

$$\omega_C = \omega_B f(\theta) = \omega_B \cdot \frac{1}{2} \frac{\sin(\theta)^2}{\cos(\theta)}, \tag{5.64}$$

con $\omega_B = e \cdot B/m$.
Ora possiamo pensare di trascurare le collisioni, supponendo che queste siano scarse, trasformando le relazioni appena citate nelle seguenti:

$$\frac{\omega_C^2}{\omega^2} >> \left[ (1-X)^2 \right], \tag{5.65}$$

per l'approssimazione QT, o

$$\frac{\omega_C^2}{\omega^2} << \left[(1-X)^2\right], \tag{5.66}$$

per l'approssimazione QL.

Si osservi però che così facendo, mentre le (5.62) e le (5.63) rappresentano delle condizioni necessarie e sufficienti, le (5.65) e (5.66) assumono un significato diverso:
1) la (5.65) diventa cioè una condizione necessaria nel senso che è necessario che essa sia soddisfatta affinché sia applicabile l'approssimazione QT; il vero campo di validità dell'approssimazione QT, quello che si otterrebbe applicando la (5.62), sarà in realtà più piccolo.
2) la (5.63) diventa cioè una condizione sufficiente nel senso che è sufficiente che essa sia soddisfatta affinché sia applicabile l'approssimazione QL; il vero campo di validità dell'approssimazione QT, quello che si otterrebbe applicando la (5.63), sarà in realtà più grande.

Si consideri una ionosfera ove la frequenza delle collisioni varia come:

$\nu = 10^{(a \cdot h + b)}$ s$^{-1}$,

ove

$a = -5.5 / 50$ km$^{-1}$,
$b = 8 + 7 / 50 \cdot 60$,

mentre la densità elettronica varia in modo tale che la frequenza di plasma varia con la quota:

$\omega_p = 1.2 \cdot 10^5 \cdot 10^{(h-60)/40}$ rad s$^{-1}$.

Si assuma il valore di 45000 nT per il modulo del vettore induzione magnetica della Terra.
Si studi la propagazione delle radio onde aventi pulsazione $\omega$ nel range $10^6$-$10^9$ rad$\cdot$ s$^-$, in un intervallo di quote da 60 a 140 km, per diversi valori dell'angolo $\theta$ individuato dal vettore d'onda con il vettore induzione magnetica della Terra **B**.

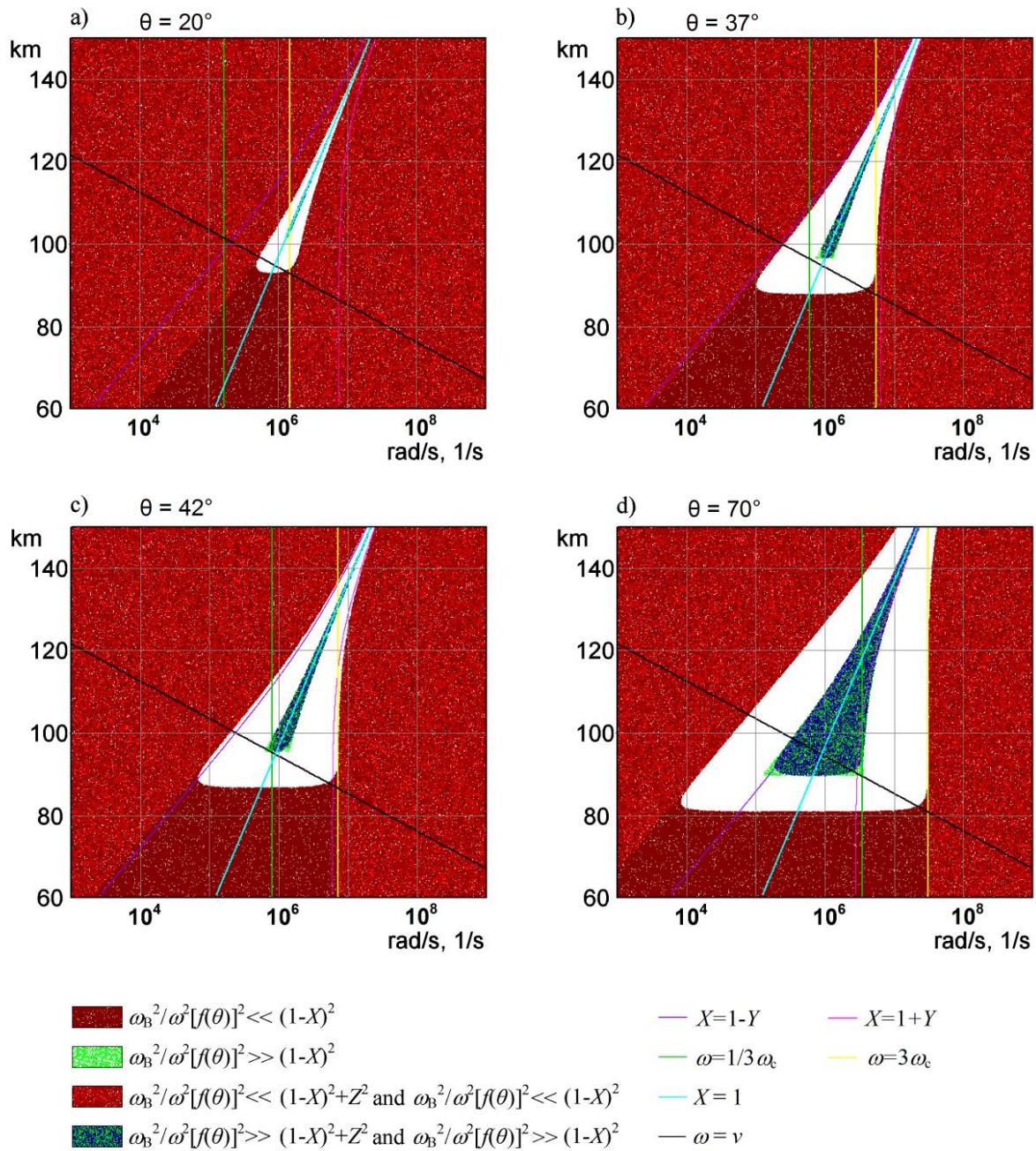

Fig. P5.2. (a)-(d) I domini di validità delle approssimazioni QT o QL, per una ionosfera modellata. Sono riportati i significati delle diverse linee colorate usate.

*Suggerimenti*

Per discutere le condizioni sotto le quali le approssimazioni QT o QL si possono applicare si deve cercare di ricavare, con una applicazione informatica, dei grafici analoghi a quelli che vengono presentati nella Fig. P5.2 (a)-(d). Una figura di questo tipo, molto esplicativa, è presentata nello storico libro di Ratcliffe (1959).

Innanzitutto si deve avere in mente che nella ionosfera sia $X$ che $Z$ variano con la quota. Infatti, secondo le definizioni date al Capitolo 2, per una certa frequenza $\omega$ fissata, $X$ è funzione di $\omega_p[N_e(h)]$ mentre $Z$ è funzione di $\nu(h)$. Nella P5.2 (a)-(d) l'informazione circa $N_e(h)$ è desumibile dalla linea celeste che rappresenta $\omega_p$ associato ad $h$ attraverso a $N_e$. Questa linea rappresenta anche l'altezza alla quale $X=1$ per una radio onda di frequenza angolare $\omega$. La frequenza delle collisioni è rappresentata dalla linea nera obliqua che rappresenta anche il luogo dei punti per il quale si ha $Z$=1. È naturale che la densità elettronica aumenti con la quota, mentre la frequenza delle collisioni diminuisca. Per quanto riguarda la quantità $\omega_C$, questa dipende da $\omega_B$ la quale dipende da **B**. Qui è stato supposto $B$=45.000 nT, secondo il testo del problema proposto, corrispondente all'intensità del campo magnetico alle medie latitudini. Per quanto riguarda l'angolo $\theta$ formato fra la direzione di propagazione e **B**, invece, sono state fatte varie assunzioni: $\theta$=20° [Fig.5.2(a)], $\theta$=37° [Fig.5.2(b)], $\theta$=42° [Fig.5.2(c)], $\theta$=70° [Fig.5.2(d)]. Nei vari casi, la linea verde rappresenta $\omega=\omega_C/3$, mentre quella gialla $\omega=3\cdot\omega_C$. Osservando le intersezioni di queste linee verticali con la retta nera, che descrive la frequenza delle collisioni, si vede che a quote superiori a quella dove la linea verde intercetta la linea nera, siamo in approssimazione QT, per $X$=1. Viceversa a quote inferiori a quella dove la linea gialla intercetta quella nera, siamo sempre in approssimazione QL. Facendo ancora riferimento alle Fig. 5.2 (a)-(d), le regioni della ionosfera modellata dove vale l'approssimazione QL sono state ricavate applicando la (5.62) e sono state disegnate di rosso scuro; quelle dove vale l'approssimazione QT, sono state ricavate applicando la (5.63) e sono state disegnate di blu scuro. La loro posizioni, relativamente alla retta $X$=1, sono in accordo con quanto visto nella teoria. In rosso e in blu brillante, invece, sono riportate le regioni corrispondenti, ottenute però tramite la (5.64) e la (5.65).

**Applicazione informatica 5.3**

Si consideri l'espressione (5.10), che esprime l'indice di rifrazione per le onde radio in assenza di campo magnetico, e in presenza di collisioni:

$$n_f^2 = 1 - \frac{X}{1-jZ} = 1 - \frac{X\cdot(1+jZ)}{1+Z^2}. \tag{5.67}$$

Se esprima $n_f$, come si è soliti fare, in dipendenza delle sue parti reali ed immaginaria ($n_f=\mu-j\cdot\chi$), e si dimostri che per $X$=1 vale la relazione:

$$\chi = \mu \approx \sqrt{\frac{1}{2}Z}, \tag{5.68}$$

essendo l'approssimazione valida se $Z^2$ può essere trascurato nei confronti di 1 (cioè se $Z$<0.1). Si progetti un programma, dove plottando la curva $\mu(X)$ si metta in evidenza la validità della (5.68) fin tanto che $Z$<0.1. Si calcoli anche la derivata

$$\left|\frac{d\mu(X)}{dX}\right|_{X=1}, \tag{5.69}$$

mostrando che questa è ~1 per $Z$~0.1 ed è crescente al decrescere di $Z$.
Un esempio dei risultati che si possono ottenere è riportato nella Fig. P5.3 (a)-(c).

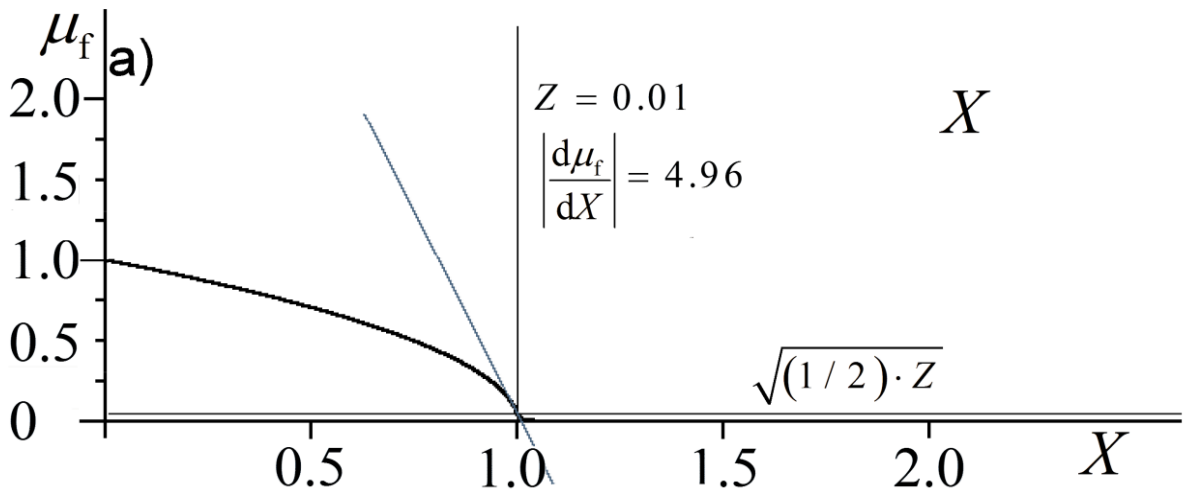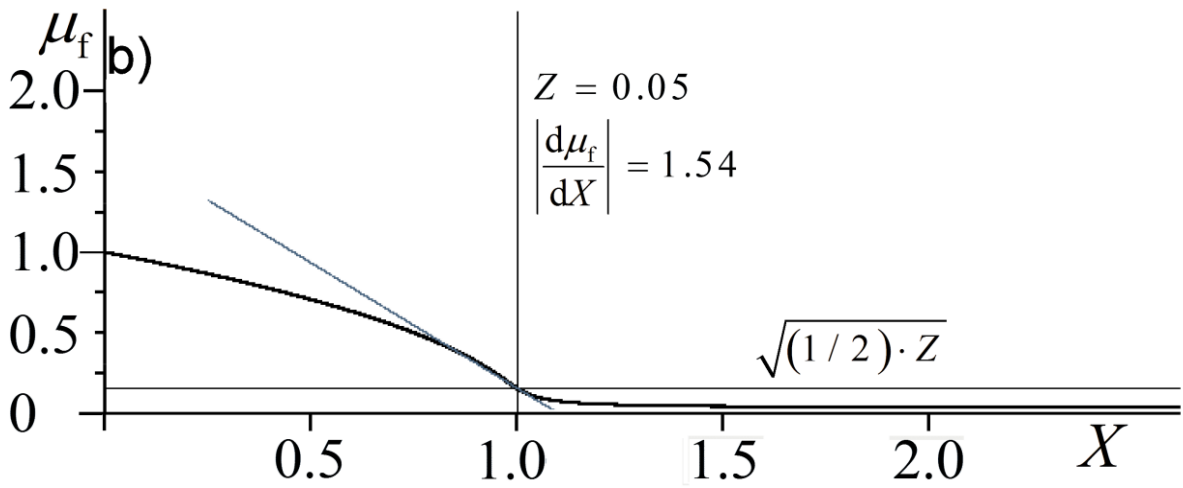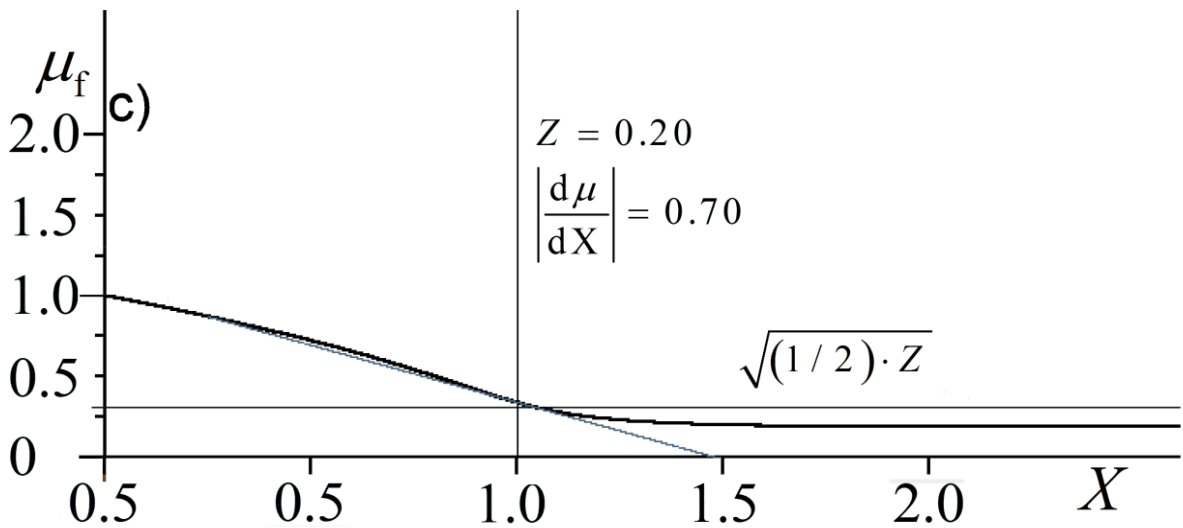

Fig. P5.3 (a)-(c). Gli andamenti di $\mu_f(X)$, per **B**=0, ottenuti dalla (5.67), secondo il Problema 5.3. Si vede che per Z<0.1, la (5.68) è verificata.

**Applicazione informatica 5.4**

Al Paragrafo 4.6, abbiamo detto, fra le altre cose, che a partire dalla relazione (2.34) è possibile ricavare $n^2=[\mu_f(\omega) - j\cdot\chi(\omega)]^2$ e, da questo, la parte reale $\mu_f$ di $n$. Per fare questo si può fare ricorso a uno dei software in commercio, capaci di svolgere il calcolo simbolico. Facendo poi l'opportuna scelta del segno nella (2.34), $\mu_f$ potrà essere identificato con $\mu_{f[ord]}$ e con $\mu_{f[ext]}$.
Si ricavino $\mu_{f[ord]}$ e $\mu_{f[ext]}$ e si riportino gli andamenti su dei grafici in funzione di $X=\omega_p^2/\omega^2$, considerando diverse condizioni radiopropagative.

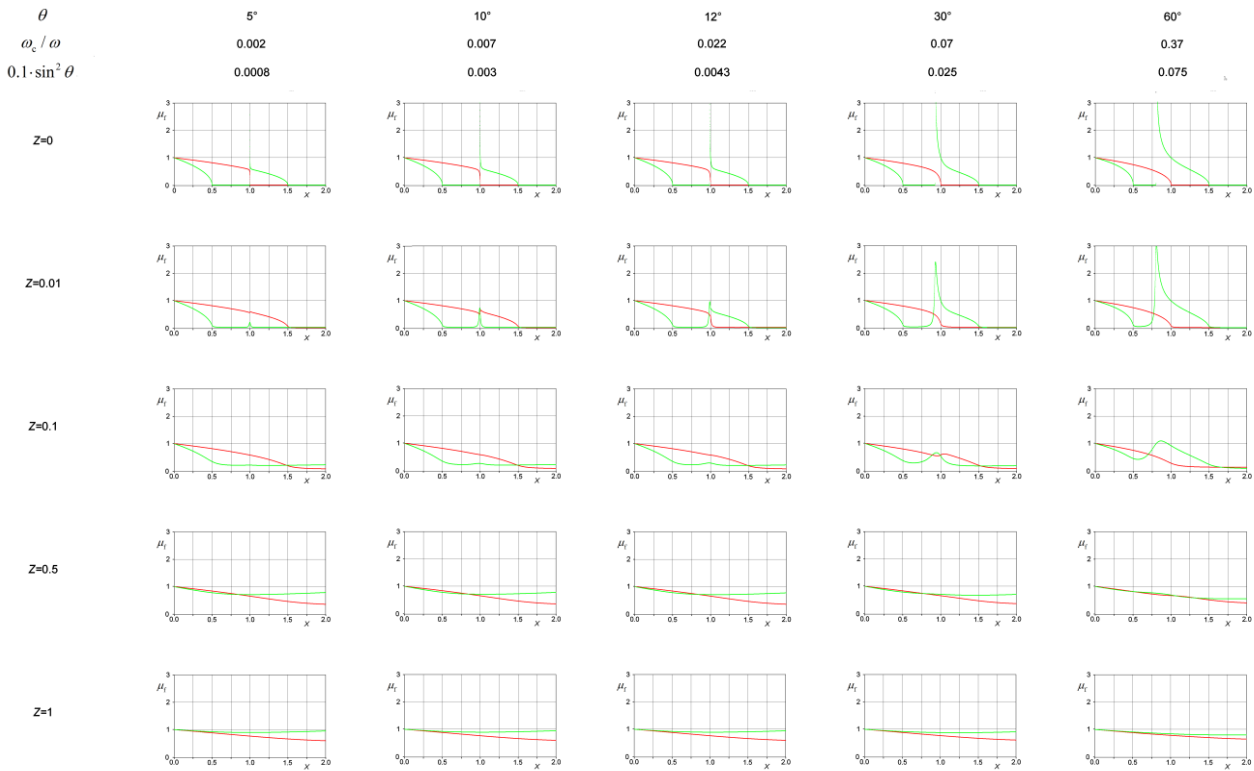

Fig. P5.4 Gli andamenti di $\mu_{f[ord]}(X)$ e $\mu_{f[ext]}(X)$, in presenza di collisioni e di campo magnetico, per diverse condizioni radiopropagative.

**Applicazione informatica 5.5**

Abbiamo visto, nei paragrafi 3.3 e 4.1, che la divergenza dell'indice di rifrazione di gruppo, garantisce la riflessione della radio onda, secondo la teoria della radio propagazione per raggi singoli. Al Paragrafo 4.6 abbiamo poi indicato come, a partire dalla (2.34), oltre a $\mu_{f[ord]}$ e $\mu_{f[ext]}$, sia possibile ricavare numericamente anche $\mu_{g[ord]}$ e con $\mu_{g[ext]}$. Infine, per quanto detto in questo Capitolo, secondo la teoria della propagazione a raggi multipli di Fresnel, fissando l'attenzione sul raggio ordinario, in caso di $dN_e(z)/dz$ non eccessivamente modesti, se $Z < 0.1\cdot\sin^2(\theta)$, e se $\omega_c>3\cdot\nu$, si ha la riflessione.
Per come è costruita, la teoria di Fresnel deve prevedere la possibilità di riflessioni che non sono spiegabili con la teoria del della propagazione a raggio singolo. Si scriva un programma che grafichi $\mu_{g[ord]}$ in funzione di $X$. Si considerino diverse condizioni radiopropagative e si identifichi almeno un caso ove la teoria di Fresnel prevede la riflessione che non è attesa usando la teoria a raggio singolo.

**Applicazione informatica 5.6**

Si utilizzi il programma sviluppato nell'applicazione informatica precedente, per graficare $\mu_{g[ord]}$ in funzione di $X$, studiando le condizioni radiopropagative per piccoli valori dell'angolo $\theta$, ricavando grafici analoghi a quelli della Fig. 5.6. Si discutano i risultati ottenuti.

**Bibliografia del Capitolo 5**

# CAPITOLO 6

## La polarizzazione delle radio onde

**Riassunto**

Viene illustrata la polarizzazione delle due componentie magnetoioniche delle onde radio nella ionosfera. La regola di Booker viene spiegata come conseguenza della continuità della relazione di polarizzazione in $X=1$.

### 6.1 Possibili stati di polarizzazione di un'onda elettromagnetica piana

Esaminiamo la rappresentazione matematica dei vari possibili stati di polarizzazione di un'onda elettromagnetica piana. Assumiamo il consueto sistema di riferimento cartesiano, facendo coincidere l'asse z con la direzione di propagazione. Utilizzando il formalismo dei numeri complessi scriviamo le componenti $x$ ed $y$ del campo elettrico **E** associato all'onda. Esse risultano:

$$E_x = A_x e^{i(\omega t - kz + \varphi_x)}, \tag{6.1}$$

$$E_y = A_y e^{i(\omega t - kz + \varphi_y)}. \tag{6.2}$$

È utile introdurre il rapporto $R = E_x/E_y$, per studiare lo stato di polarizzazione dell'onda:

$$R = \frac{E_x}{E_y} = \frac{A_x}{A_y} e^{i(\varphi_x - \varphi_y)} = \rho \cdot e^{i\delta}, \tag{6.3}$$

ove $\rho = A_x/A_y$ e $\delta = \varphi_x - \varphi_y$.
È sempre possibile scegliere l'origine dei tempi in modo tale che risulti $\varphi_y = 0$, per cui possiamo porre $\delta = \varphi_x$. Prendendo le parti reali delle (6.1) e (6.2), risulta:

$$E_x(t) = A_x \cdot \cos(\omega t + \delta) \tag{6.4}$$

e

$$E_y(t) = A_y \cdot \cos(\omega t). \tag{6.5}$$

La Fig. 6.1 riporta gli ellissi di polarizzazione per qualche valore di $\rho$ e di $\delta$.

### 6.2 Proprietà dell'equazione di polarizzazione

Nel caso delle onde che si propagano nella ionosfera, il fattore di polarizzazione $R$, risulta espresso dalla (2.27) che qui riportiamo, riscritta in modo lievemente diverso:

$$R_{+,-} = -\frac{i}{2Y_L}\left[\frac{Y_T^{\,2}}{(1-X)} \mp \sqrt{\frac{Y_T^{\,4}}{(1-X)^2} + 4Y_L^{\,2}}\right]. \tag{6.6}$$

Se teniamo conto delle collisioni fra elettroni e molecole neutre, si può dimostrare che la (6.6), diventa:

$$R_{+,-} = -\frac{i}{2Y_L}\left[\frac{Y_T^{\,2}}{(1-X-jZ)} \mp \sqrt{\frac{Y_T^{\,4}}{(1-X-jZ)^2} + 4Y_L^{\,2}}\right]. \tag{6.7}$$

Vedremo nei passaggi successivi che è importante considerare il segno di $Y_L = e \cdot B \cdot \cos(\theta)/m\omega$. Avendo fissato l'attenzione sugli elettroni, abbiamo che $e = -1.6 \cdot 10^{-19}$ C, pertanto il segno di $Y_L$ dipende dall'angolo $\theta$ che, come abbiamo già detto, rappresenta l'angolo formato dal vettore d'onda **k** con il campo magnetico della Terra. La quantità

$$\rho_{+,-} = -i\left[\frac{Y_T^{\,2}}{2Y_L(1-X-jZ)} \mp \sqrt{\frac{Y_T^{\,4}}{4Y_L^{\,2}(1-X-jZ)^2} + 1}\right], \text{ per } Y_L > 0, \tag{6.8}$$

e

$$\rho_{+,-} = -i\left[\frac{Y_T^{\,2}}{2Y_L(1-X-jZ)} \pm \sqrt{\frac{Y_T^{\,4}}{4Y_L^{\,2}(1-X-jZ)^2} + 1}\right], \text{ per } Y_L < 0. \tag{6.9}$$

Si consideri poi la quantità che appare nel primo termine del radicando:

$$\frac{Y_T^{\,2}}{2Y_L} = \frac{\omega_B^{\,2}\sin^2\theta}{\omega^2}\frac{\omega}{\omega_B\cos\theta} = \frac{1}{\omega}\omega_B\left[\frac{1}{2}\frac{\sin^2\theta}{\cos\theta}\right]. \tag{6.10}$$

Essa può essere scritta

$$\frac{Y_T^{\,2}}{2Y_L} = \frac{\omega_c}{\omega}, \tag{6.11}$$

avendo definito

$$\omega_c = \omega_B f(\theta), \tag{6.12}$$

Con

$$f(\theta) = \frac{1}{2}\frac{\sin^2\theta}{\cos\theta}. \tag{6.13}$$

Le (6.8) e (6.9) possono essere scritte nel modo seguente:

$$R_{1,2} = -i\left\{\frac{1}{\frac{\omega}{\omega_c}(1-X)-i\frac{\nu}{\omega_c}} \mp \left[\left(\frac{1}{\frac{\omega}{\omega_c}(1-X)-i\frac{\nu}{\omega_c}}\right)^2+1\right]^{\frac{1}{2}}\right\}, \text{ per } \omega_c > 0 \qquad (6.14)$$

e

$$R_{+,-} = -i\left\{\frac{1}{\frac{\omega}{\omega_c}(1-X)-i\frac{\nu}{\omega_c}} \pm \left[\left(\frac{1}{\frac{\omega}{\omega_c}(1-X)-i\frac{\nu}{\omega_c}}\right)^2+1\right]^{\frac{1}{2}}\right\}, \text{ per } \omega_c < 0. \qquad (6.15)$$

La grandezza $R$ è in generale complessa. Questo vuol dire che il vettore **E** dell'onda elettromagnetica descrive una ellisse inclinata rispetto agli assi $x$ e $y$. Si ricorda che, mentre l'asse $z$ è scelto lungo la direzione di propagazine dell'onda, l'asse $y$ è scelto in modo tale che il campo magnetico terrestre si trovi sul piano $yz$. L'asse $x$ è poi scelto di conseguenza, in modo da rendere la terna destrorsa (si veda Fig. 2.1). È poi importante osservare che dalla (6.15) si ricava:

$$R_+ R_- = 1. \qquad (6.16)$$

Questa relazione descrive una caratteristica interessante degli ellissi di polarizzazione delle due componenti, ordinaria e straordinaria.

$$\begin{cases} E'_x = E_x \cos\psi + E_y \sin\psi = E_x(\cos\psi + R_+ \sin\psi) \\ E'_y = E_y \cos\psi - E_x \sin\psi = E_x(R_+ \cos\psi - \sin\psi) \end{cases}$$

Da queste si ricava:

$$R_+' = \frac{E'_x}{E'_y} = \frac{(1+R_+ \tan\psi)}{(R_+ - \tan\psi)}; \qquad (6.18)$$

e similmente:

$$R_-' = \frac{E'_x}{E'_y} = \frac{(1+R_- \tan\psi)}{(R_- - \tan\psi)}. \qquad (6.19)$$

Dovrà essere:

$$R_+' R_-' = 1. \qquad (6.20)$$

Se consideriamo una rotazione del sistema di riferimento di $\psi = 45°$, ed usiamo le (6.18), (6.19) e (NC20), si ricava:

$$R_+' = -R_-'. \qquad (6.21)$$

Le due ellissi di polarizzazione si possono dunque ottenere come immagini riflesse l'una dall'altra, secondo un piano posto a 45° rispetto agli assi originari.

Inoltre, essendo $\delta>0$ associato alla polarizzazione sinistrorsa e $\delta<0$ a quella destrorsa, le direzioni di rotazione dei due ellisi saranno l'una l'inversa dell'altra.

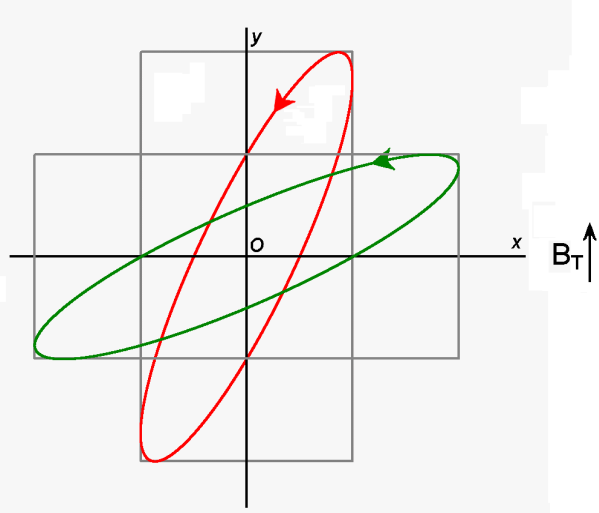

Fig. 6.1. Le polarizzazioni dei campi elettrici nel piano del fronte d'onda. Ox ed Oy sono le direzioni principali, mentre il campo magnetico della Terra ha una componente diretta lungo z e l'altra componente lungo y. Il vettore di propagazione dell'onda è diretto perpendicolarmente al piano xy nel verso entrante al foglio. L'ellisse di polarizzazione della componente ordinaria (in rosso) e quella della componente straordinaria (in verde) sono in relazione fra loro come indicato (Ratcliffe, 1959).

### 6.3 Proprietà dell'equazione di polarizzazione per $X=0$

Riprendendo le equazioni (6.14) e (6.15) e ponendo per $X=0$, si ottiene:

$$R_{+,-} = -i\left\{\frac{1}{-i\frac{v}{\omega_c}} \mp \left[\left(\frac{1}{-i\frac{v}{\omega_c}}\right)^2 + 1\right]^{\frac{1}{2}}\right\} = \left(\frac{\omega_c}{v}\right) \pm i\sqrt{-\left(\frac{\omega_c}{v}\right)^2 + 1} \text{, per } \omega_c > 0 \qquad (6.22)$$

e

$$R_{+,-} = -i\left\{\frac{1}{-i\frac{v}{\omega_c}} \pm \left[\left(\frac{1}{-i\frac{v}{\omega_c}}\right)^2 + 1\right]^{\frac{1}{2}}\right\} = \left(\frac{\omega_c}{v}\right) \mp i\sqrt{-\left(\frac{\omega_c}{v}\right)^2 + 1} \text{, per } \omega_c < 0. \qquad (6.23)$$

La (6.22), può essere scritta in modo diverso, a seconda se il termine sotto radice è positivo o negativo. Quindi per $\omega_c > 0$ si ha:

$$\begin{cases} R_{+,-} = \left(\dfrac{\omega_c}{v}\right) \pm i\sqrt{1-\left(\dfrac{\omega_c}{v}\right)^2} & \text{if } \left|\dfrac{\omega_c}{v}\right| < 1 \end{cases} \qquad (6.24a)$$

$$\begin{cases} R_{+,-} = \left(\dfrac{\omega_c}{v}\right) \mp \sqrt{-1+\left(\dfrac{\omega_c}{v}\right)^2} & \text{if } \left|\dfrac{\omega_c}{v}\right| \geq 1 \end{cases} \qquad (6.24b)$$

Allo stesso modo si può operare a partire dalla (6.23). Dunque, per $\omega_c < 0$, si ha:

$$\begin{cases} R_{+,-} = \left(\dfrac{\omega_c}{v}\right) \mp i\left[1-\left(\dfrac{\omega_c}{v}\right)^2\right]^{\frac{1}{2}} & \text{if } \left|\dfrac{\omega_c}{v}\right| < 1, \end{cases} \qquad (6.25a)$$

$$\begin{cases} R_{+,-} = \left(\dfrac{\omega_c}{v}\right) \pm \left[\left(\dfrac{\omega_c}{v}\right)^2 - 1\right]^{\frac{1}{2}} & \text{if } \left|\dfrac{\omega_c}{v}\right| \geq 1. \end{cases} \qquad (6.25b)$$

Per fissare le idee, consideriamo quest'ultima coppia di equazioni, (6.25a) e (6.25b), valide per $\omega_c < 0$. Questo caso, corrisponde alla propagazione di una radio onda dall'alto verso il basso, nell'emisfero Nord. Si osservi che dobbiamo considerare il segno positivo nella (6.25a) e quello negativo nella (NC25b) – o viceversa – affinché il modulo e l'argomento di $R$ siano continui al passaggio per $|\omega_c / v|=1$. Nella Fig. NC3 si vedono gli andamenti di $\rho$, e $\delta$ considerando il segno negativo nella (6.25a) e quello positivo nella (6.25b).

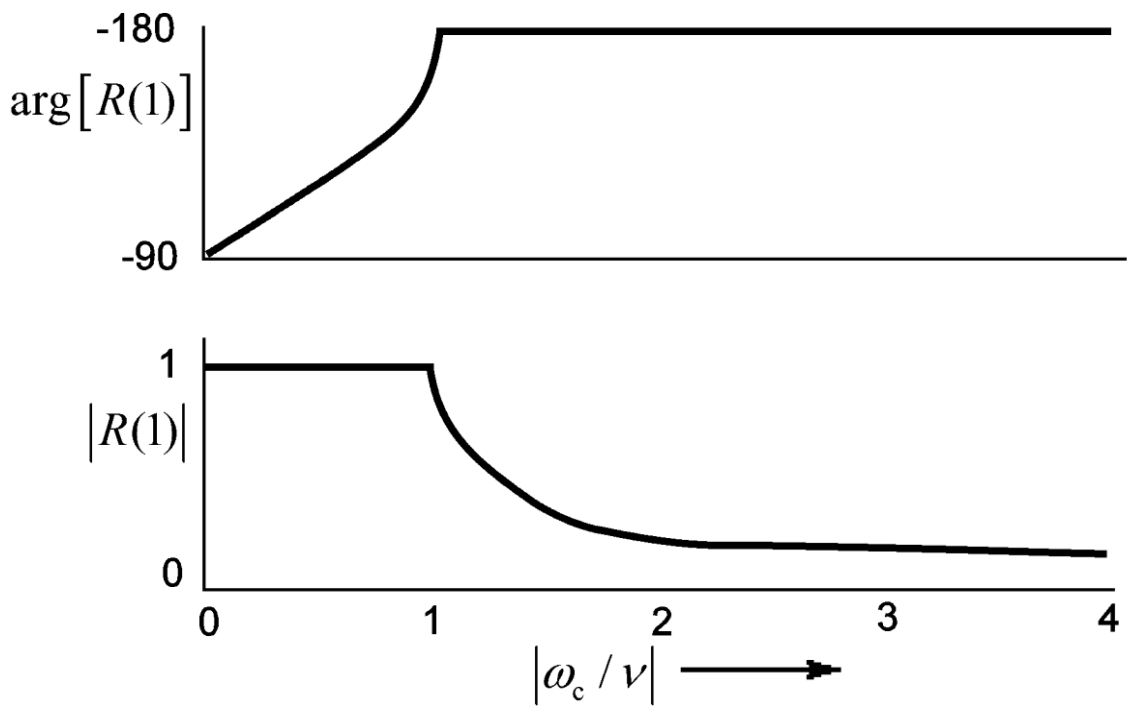

Fig. 6.2. Gli andamenti di $\rho$, e $\delta$ considerando il segno negativo nella (6.25a) e quello positivo nella (6.25b). Si vede la continuità sia di $\rho$ che di $\delta$ in $|\omega_c / v|=1$ (rfr6-Ratcliffe, 1959).

# Applicazioni Informatiche del Capitolo 6

## Applicazione Informatica 6.1

Si realizzi una applicazione informatica che utilizzando le relazioni (6.4) ed (6.5) che permettono di disegnare le ellissi di polarizzazione. Si considerino diversi casi di $R=\rho \cdot e^{i\delta}$, osservando in particolare il caso $\delta=0$, corrispondente a polarizzazione lineare e il caso $\rho=1$ e $\delta=0$, corrispondente al caso di polarizzazione circolare. Con un programma di questo tipo, si possono ottenere dei grafici simili a quelli riportati nella Fig. 6.1.

## Applicazione Informatica 6.2

Si realizzi una applicazione informatica per studiare le (6.24a) e (6.24b), valide per $\omega_c > 0$ e le (6.25a) e (6.25b), valide per $\omega_c < 0$. Si studi la continuità del modulo e l'argomento di $R$ siano al passaggio per $|\omega_c / \nu|=1$. Si verifichi la necessità di scambiare il segno superiore con quello inferiore al passaggio da $|\omega_c / \nu|<1$ a $|\omega_c / \nu|>1$. Con un programma di questo tipo, si possono ottenere dei grafici simili a quelli riportati nella Fig. P6.1.

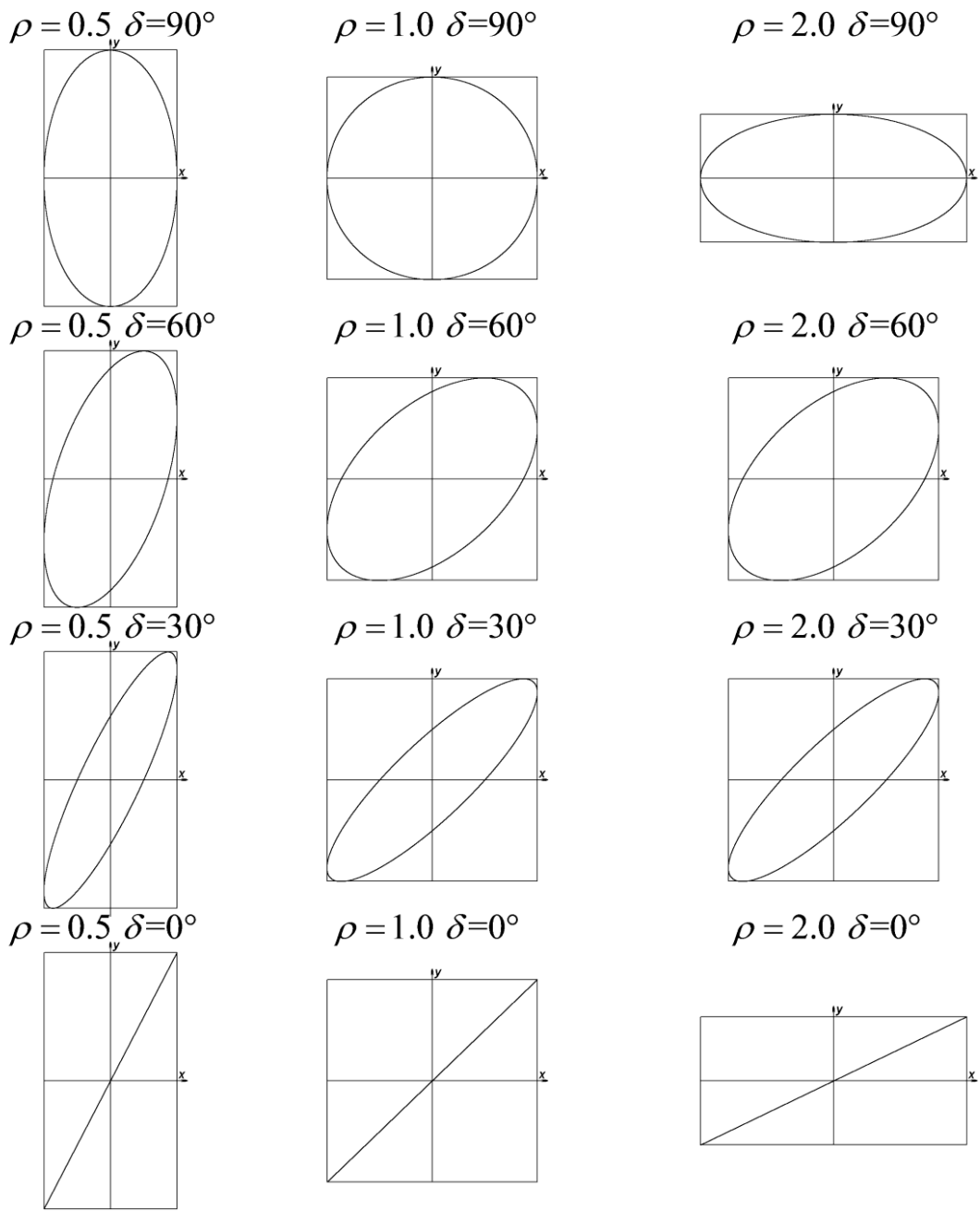

Fig. P6.1. Le ellissi di polarizzazione per diversi valori $R=\rho \cdot e^{i\delta}$. Queste ellissi sono percorse in senso antiorario. Le ellissi per $\delta<0$, sono identiche, ma percorse in senso orario.

**Applicazione Informatica 6.3**

Si consideri la relazione di polarizzazione nella sua forma generale (6.7). Attraverso uno dei software matematici disponibili sul mercato, si calcolino le parti reale ed immaginaria di $R$. Si realizzi una applicazione informatica che disegni le ellissi di polarizzazione corrispondenti a diverse condizioni radio propagative.

## Bibliografia del Capitolo 6

# CAPITOLO 7

## Propagazione ionosferica obliqua

**Riassunto**

La propagazione ionosferica viene studiata attraverso l'equazione dell'ikonale. Vengono illustrati i principali teoremi riguardanti la propagazione obliqua. Sebbene l'eqazione dell'ikonale è valida solo nel caso di mezzo isotropo, i riusultati che si possono ottenere dànno conto delle principali caratteristiche osservabili dei radiocollegamenti ionosferici.

### 7.1 L'equazione ikonale

Abbiamo visto con le (1.33a) ed (1.33b) che, nel caso di mezzo omogeneo isotropo, l'equazione delle onde può essere scritta separatamente per ciascuna componente di **B** ed **E**. Ricordando che $c^2 = (\varepsilon_0\mu_0)^{-1}$ e $v^2=(\varepsilon_r\mu_r\varepsilon_0\mu_0)^{-1}$, si ha $n^2= c^2/v^2= \varepsilon_r\mu_r$; indicando poi, la generica componente di uno di tali campi con $\psi(\mathbf{r},t)$, dalle (1.20a) e (1.20b) si può ottenere:

$$\nabla^2\Psi(\mathbf{r},t) - \frac{n^2}{c^2}\frac{\partial^2\Psi(\mathbf{r},t)}{\partial t^2} = 0. \tag{7.1}$$

Se supponiamo che la soluzione di questa equazione possa essere fattorizzata nella forma:

$$\Psi(\mathbf{r},t) = \phi(\mathbf{r})e^{-i\omega t}, \tag{7.2}$$

si ottiene:

$$\nabla^2\left[\phi(\mathbf{r})e^{-i\omega t}\right] - \frac{n^2}{c^2}\frac{\partial^2\left[\phi(\mathbf{r})e^{-i\omega t}\right]}{\partial t^2} = 0, \tag{7.3}$$

da cui:

$$e^{-i\omega t}\nabla^2\left[\phi(\mathbf{r})\right] + \frac{n^2}{c^2}\omega^2\phi(\mathbf{r})e^{-i\omega t} = 0, \tag{7.4}$$

e quindi:

$$\nabla^2\phi(\mathbf{r}) + n^2k^2\phi(\mathbf{r}) = 0, \tag{7.5}$$

Per risolvere questa equazione possiamo porre:

$$\phi(\mathbf{r}) = A(\mathbf{r})e^{ikS(\mathbf{r})}, \tag{7.6}$$

che implica:

$$\nabla \phi(\mathbf{r}) = \nabla A(\mathbf{r}) e^{ikS(\mathbf{r})} + A(\mathbf{r}) ik e^{ikS(\mathbf{r})} \nabla S(\mathbf{r}) =$$

$$= \left[ \nabla A(\mathbf{r}) + ik A(\mathbf{r}) \nabla S(\mathbf{r}) \right] e^{ikS(\mathbf{r})}. \tag{7.7}$$

Riapplicando l'operatore gradiente risulta:

$$\nabla \phi^2(\mathbf{r}) = \left[ \nabla^2 A(\mathbf{r}) + ik \nabla A(\mathbf{r}) \nabla S(\mathbf{r}) + ik A(\mathbf{r}) \nabla^2 S(\mathbf{r}) \right] \cdot e^{ikS(\mathbf{r})} +$$
$$+ \left[ \nabla A(\mathbf{r}) + ik A(\mathbf{r}) \nabla S(\mathbf{r}) \right] \cdot e^{ikS(\mathbf{r})} ik \nabla S(\mathbf{r}) =$$
$$= \left\{ \nabla^2 A(\mathbf{r}) + ik \nabla A(\mathbf{r}) \nabla S(\mathbf{r}) + ik A(\mathbf{r}) \nabla^2 S(\mathbf{r}) + \right.$$
$$\left. + ik \nabla S(\mathbf{r}) \nabla A(\mathbf{r}) - k^2 A(\mathbf{r}) \left[ \nabla S(\mathbf{r}) \right]^2 \right\} e^{ikS(\mathbf{r})}. \tag{7.8}$$

Introducendo la (7.8) e la (7.6), nella (7.5), si ha:

$$\left\{ \nabla^2 A(\mathbf{r}) + ik \nabla A(\mathbf{r}) \nabla S(\mathbf{r}) + ik A(\mathbf{r}) \nabla^2 S(\mathbf{r}) + ik \nabla S(\mathbf{r}) \nabla A(\mathbf{r}) - k^2 A(\mathbf{r}) \left[ \nabla S(\mathbf{r}) \right]^2 \right\} \cdot e^{ikS(\mathbf{r})} +$$
$$+ n^2 k^2 A(\mathbf{r}) \cdot e^{ikS(\mathbf{r})} = 0, \tag{7.9}$$

che corrisponde a:

$$\left\{ \nabla^2 A(\mathbf{r}) + 2ik \nabla A(\mathbf{r}) \nabla S(\mathbf{r}) + ik A(\mathbf{r}) \nabla^2 S(\mathbf{r}) - k^2 A(\mathbf{r}) \left| \nabla S(\mathbf{r}) \right|^2 \right\} \cdot e^{ikS(\mathbf{r})} =$$
$$= -n^2 k^2 A(\mathbf{r}) \cdot e^{ikS(\mathbf{r})}. \tag{7.10}$$

Affinché questa identità sia verificata, dovranno essere identiche separatamente le parti immaginaria e reale. Se supponiamo che variazioni di ampiezza avvengano su distanze molto maggiori rispetto alla lunghezza d'onda, avremo che:

$$k^2 \rangle\rangle \nabla^2 A(\mathbf{r}), \tag{7.11}$$

per cui, in questa ipotesi, le parti reali dei due membri della (7.11), sono identiche se:

$$\left| \nabla S(\mathbf{r}) \right|^2 = n^2, \tag{7.12}$$

che è detta *equazione eikonale* (rfr7-Paris, 1969; Arnold, 2004).

### 7.3 Equazione del raggio

Osservando la (7.6), viene naturale interpretare la superficie $S(\mathbf{r})$=const come una superficie equifase. Di conseguenza $\nabla S$ rappresenta un vettore perpendicolare a detta superficie; essendo dalla (7.12):

$$\left| \nabla S(\mathbf{r}) \right| = n, \tag{7.13}$$

il versore normale alla superficie equifase, diventa:

$$\hat{\mathbf{s}} = \frac{\nabla S(\mathbf{r})}{|\nabla S(\mathbf{r})|} = \frac{\nabla S(\mathbf{r})}{n}. \tag{7.14}$$

Esso esprime anche la direzione di propagazione della fase, e può essere pensato come diretto lungo la tangente alla curva *C* che rappresenta la traiettoria del raggio considerato.
Poiché **r**(*s*) rappresenta il vettore posizione su un certo punto del percorso *C*, considerato come funzione dell'ascissa curvilinea *s*, per l'equazione del raggio avremo:

$$\hat{\mathbf{s}} = \frac{d\mathbf{r}}{ds}. \tag{7.15}$$

D'altra parte, risulta:

$$\frac{\nabla S}{n} = \hat{\mathbf{s}}, \tag{7.16}$$

per cui si ha:

$$\frac{d\mathbf{r}}{ds} = \frac{\nabla S}{n}, \tag{7.17}$$

da cui

$$n\frac{d\mathbf{r}}{ds} = \nabla S. \tag{7.18}$$

Differenziando rispetto alla ascissa curvilinea *s* entrambi i membri della relazione precedente, abbiamo:

$$\frac{d}{ds}\left(n\frac{d\mathbf{r}}{ds}\right) = \frac{d}{ds}\nabla S. \tag{7.19}$$

Siccome per qualsiasi funzione *f* si ha d/ds(*f*)= $\hat{\mathbf{s}} \cdot \nabla f$, identificando *f* con *S*, la (7.19) può essere espressa nel modo seguente:

$$\frac{d}{ds}\left(n\frac{d\mathbf{r}}{ds}\right) = \hat{\mathbf{s}} \cdot \nabla(\nabla S), \tag{7.20}$$

la quale, ricordando la (7.16), dà:

$$\frac{d}{ds}\left(n\frac{d\mathbf{r}}{ds}\right) = \frac{\nabla S}{n} \cdot \nabla(\nabla S). \tag{7.21}$$

Questa può essere scritta nella forma seguente:

$$\frac{d}{ds}\left(n\frac{d\mathbf{r}}{ds}\right) = \frac{1}{2n} \cdot \nabla(\nabla S)^2 = \frac{1}{2n} \cdot \nabla n^2 = \nabla n, \qquad (7.22)$$

ovvero

$$\frac{d}{ds}\left(n\frac{d\mathbf{r}}{ds}\right) = \frac{1}{n} \cdot \nabla n, \qquad (7.23)$$

che è detta *equazione del raggio*.

Vale la pena ricordare che questa relazione è valida solo per un mezzo isotropo, cioè un mezzo avente un indice di rifrazione che non dipende dalla direzione del vettore d'onda, anche se può varire puntualmente. Al contrario, come abbiamo visto nel Capitolo 2, per la ionosfera terrestre, alle frequenze HF, l'indice di rifrazione dipende dall'angolo formato dal vettore d'onda con il vettore induzione magnetica della Terra. Dunque, la relazione (7.23) non può in generale essere utilizzata per determinare il percorso di un'onda radio alle frequenza HF nella ionosfera.

In una trattazione semplificata, tuttavia, si può trascurare la presenza del campo magnetico della Terra ed attribuire alle onde radio HF l'indice di espresso dalla (2.35). In detta trattazione semplificata, naturalmente, l'equazione (7.23) può essere utilmente integrata determinando il perscorso delle onde radio.

Non è necessario in tutti i casi procedure all'integrazione dell' equazione (7.23), per avere le informazioni che ci interessano riguardo al percorso dei raggi delle onde radio che si propagano nella ionosfera in modo obliquo. Infatti, vedremo nei prossimi due paragrafi le proprietà salienti di questo genere di propagazione, che riveste notevole importanza pratica poiché costituisce la tipica geometria di un radiocollegamento HF per via ionosferica. Le ipotesi che vengono fatte sono quelle di ionosfera isotropa, senza cioè considerare la presenza del campo magnetico della Terra, e di trascurare le collisioni elettrone-molecole neutre. Queste ipotesi sono le stesse con le quali è stata ricavata la (7.23).

### 7.3 Variazioni dell'ampiezza

Se consideriamo la parte immaginaria della (7.10), si ha:

$$\phi(\mathbf{r})e^{-i\omega t}2\nabla A(\mathbf{r})\nabla S(\mathbf{r}) + A(\mathbf{r})\nabla^2 S(\mathbf{r}) = 0, \qquad (7.24)$$

da cui:

$$2\nabla A \cdot \nabla S = -A\nabla^2 S. \qquad (7.25)$$

Come abbiamo già detto $\nabla S / n$ è il versore normale a $S$ e rappresenta la direzione di propagazione. Dividendo i due membri della (7.25) per $n$, otteniamo:

$$2\frac{\nabla A \cdot \nabla S}{n} = -u\frac{\nabla^2 S}{n}, \qquad (7.26)$$

ove compare un termine $\nabla A \cdot \nabla S/n$, che è la derivata di $A$ lungo la direzione di propagazione $s$. Si ha perciò la relazione seguente:

$$\frac{1}{u}\frac{dA}{ds} = -\frac{1}{2}\frac{\nabla^2 S}{n}. \tag{7.27}$$

L'integrazione numerica di questa equazione differenziale porta alla determinazione dell'ampiezza di $A$ lungo $C$ e fornisce dunque una informazione aggiuntiva rispetto alla sola equazione iconale (7.12).

**7.4 La legge della secante**

Supponiamo in questo paragrafo di effetture un esperimento di propagazione obliqua e uno di propagazione verticale, con riflessioni che avvengano sopra lo stesso punto al suolo, così come viene descritto nella Fig. 7.1. Si consideri la frequenza del sondaggio obliquo $f_{ob}$ arbitraria e si consideri la più elevata altezza di penetrazione pari a $h_1$. Si supponga ora di regolare la frequenza $f_v$ del sondaggio verticale in modo tale che anche questa altezza di penetrazione sia pari a $h_1$. La legge della secante afferma che fra le due frequenze $f_{ob}$ e $f_v$ esiste la seguente relazione:

$$f_{ob} = f_v \sec \vartheta_0. \tag{7.28}$$

Anche per le onde radio, che si propagano verso l'alto, negli strati della ionosfera con $n$ decrescente, vale la nota *legge di Snell* per la rifrazione. Secondo essa, quando un raggio ottico si propaga attraverso due mezzi trasmissivi con indice di rifrazione $\mu_1$ ed $\mu_2$, in contatto fra loro e separati da una superficie $S$, vale la relazione:

$$\mu_1 \cdot \sin \varphi_1 = \mu_2 \cdot \sin \varphi_2, \tag{7.29}$$

ove $\varphi_1$, è l'*angolo di incidenza*, che è formato dalla normale a $S$ e il raggio che si propaga nel mezzo 1, mentre $\varphi_2$ è *l'angolo di rifrazione* che è formato fra la stessa normale e il raggio che si propaga nel mezzo 2. La geometria della rifrazione di un raggio ottico fra due mezzi è illustrata nella Fig. 7.2.
Nel caso della propagazione obliqua, l'atmosfera non ionizzata corrisponde al mezzo 1. Per essa possiamo assumere indice di rifrazione $\mu_1=1$, mentre l'angolo di incidenza, come illustrato nella Fig. 7.3, è $\theta_0$. La ionosfera, corrisponde invece al mezzo 2, per essa, nel punto ove il raggio diventa orizzontale, possiamo considerare $\mu_2$ generico, indicandolo con $\mu_{h1}$ mentre per l'angolo di rifrazione, possiamo assumere $\varphi_2=\pi/2$. La (7.29) diventa così:

$$1 \cdot \sin \theta_0 = \mu_{h1} \sin(\pi/2), \tag{7.30}$$

per cui si ha, facendo uso della (7.29):

$$\sin \theta_0 = \mu_{h1} = \sqrt{1-X} = \sqrt{1 - \frac{f_p(h_1)^2}{f_{ob}^2}}, \tag{7.31}$$

Da questa otteniamo:

$$\frac{f_p(h_1)^2}{f_{ob}^2} = 1 - \sin\theta_0^2 = \cos\theta_0^2, \tag{7.32}$$

e poi:

$$f_{ob} = f_p(h_1)\frac{1}{\cos\theta_0} = f_p(h_1)\sec\theta_0, \tag{7.33}$$

Siccome $f_p(h_1)$ rapprenta la fequenza di riflessione per incidenza verticale, possiamo porre $f_p(h_1)=f_v$, per cui si ricava la (7.28), che è detta *legge della secante*.

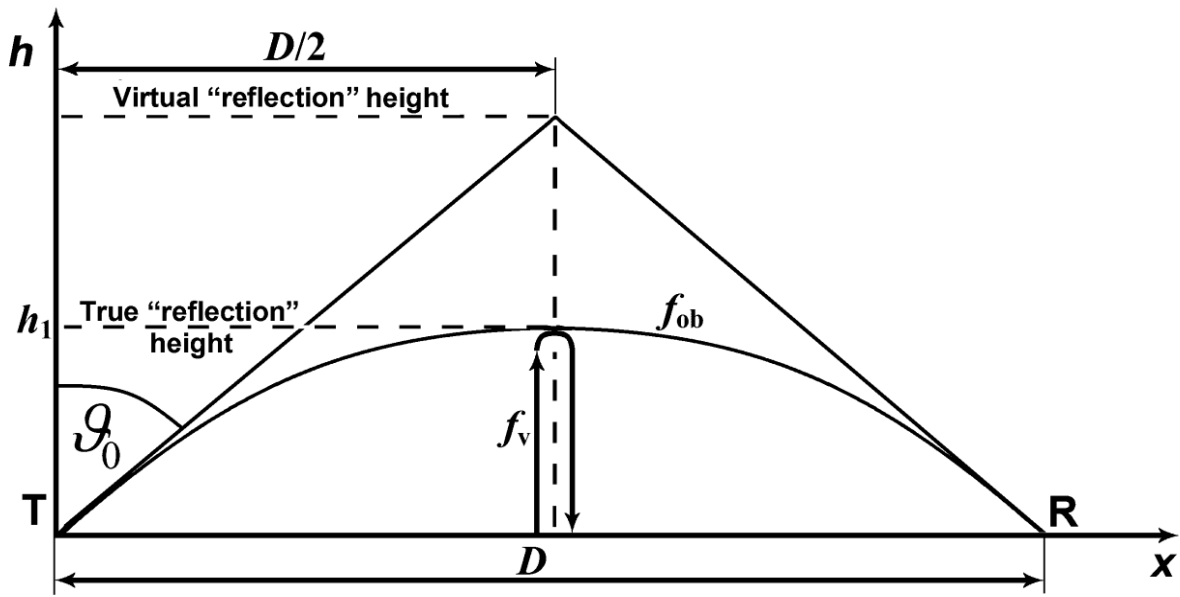

Fig. 7.1. Geometria per la propagazione di un'onda radio fra un terminale trasmittente T ed uno ricevente R. La terra e la ionosfera sono supposte piane.

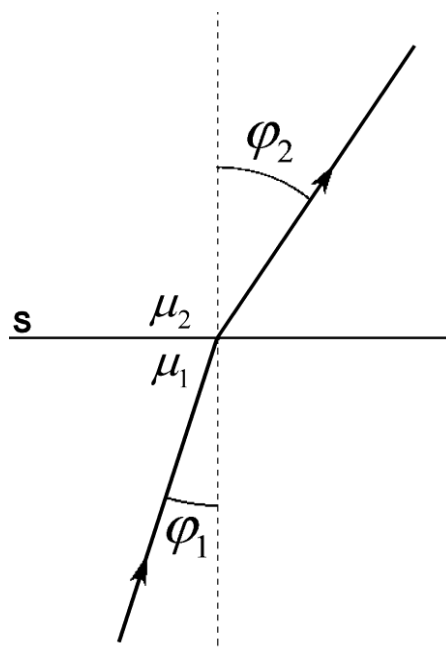

Fig. 7.2. Elementi geometrici della legge di Snell sulla rifrazione.

## 7.5 Teorema di Breit and Tuve

Con riferiemento alla Fig. 7.3, il teorema di Breit and Tuve afferma che il tempo $t$ che impiega una onda radio a raggiungere lungo il suo effettivo percorso la massima altezza $h_m$ nella ionosfera isotropa è lo stesso che impiegherebbe a raggiungere il punto P se l'onda viaggiasse nel vuoto. Questo enunciato può essere dimostrato partendo dalla seguente espressione per l'elemento di tempo d$t$ corrispondente ad un certo tratto d$s$ del percorso del raggio, che è:

$$dt = \frac{ds}{v_g(h)} = \frac{dx}{v_g(h)\sin[\vartheta(h)]}. \tag{7.34}$$

Si prosegue osservando che la velocità di gruppo può essere espressa come $v_g = c/\mu_g(h)$. Essendo, dalla (3.14), $\mu_g = \mu_f^{-1}$, si può anche scrivere $v_g = c \cdot \mu_f(h)$. La (7.34) diventa così:

$$dt = \frac{dx}{c \cdot n_f(h)\sin[\vartheta(h)]}. \tag{7.35}$$

Ricordando la (7.29), si ha:

$$dt = \frac{dx}{c \cdot n_f(h)\dfrac{\sin\vartheta_0}{n_f(h)}} = \frac{dx}{c \cdot \sin\vartheta_0}, \tag{7.34}$$

integrando la quale abbiamo:

$$t = \frac{1}{c\sin\vartheta_0}\int_0^P dx = \frac{D}{c\sin\vartheta_0} = t'. \tag{7.35}$$

quindi che il tratto PM rappresenta l'altezza alla quale dovrebbe avvenire la riflessione se questa fosse dovuta a un singolo strato riflettente posto nella ionosfera affinchè il raggio percorra il tratto TPR nel tempo $t$.

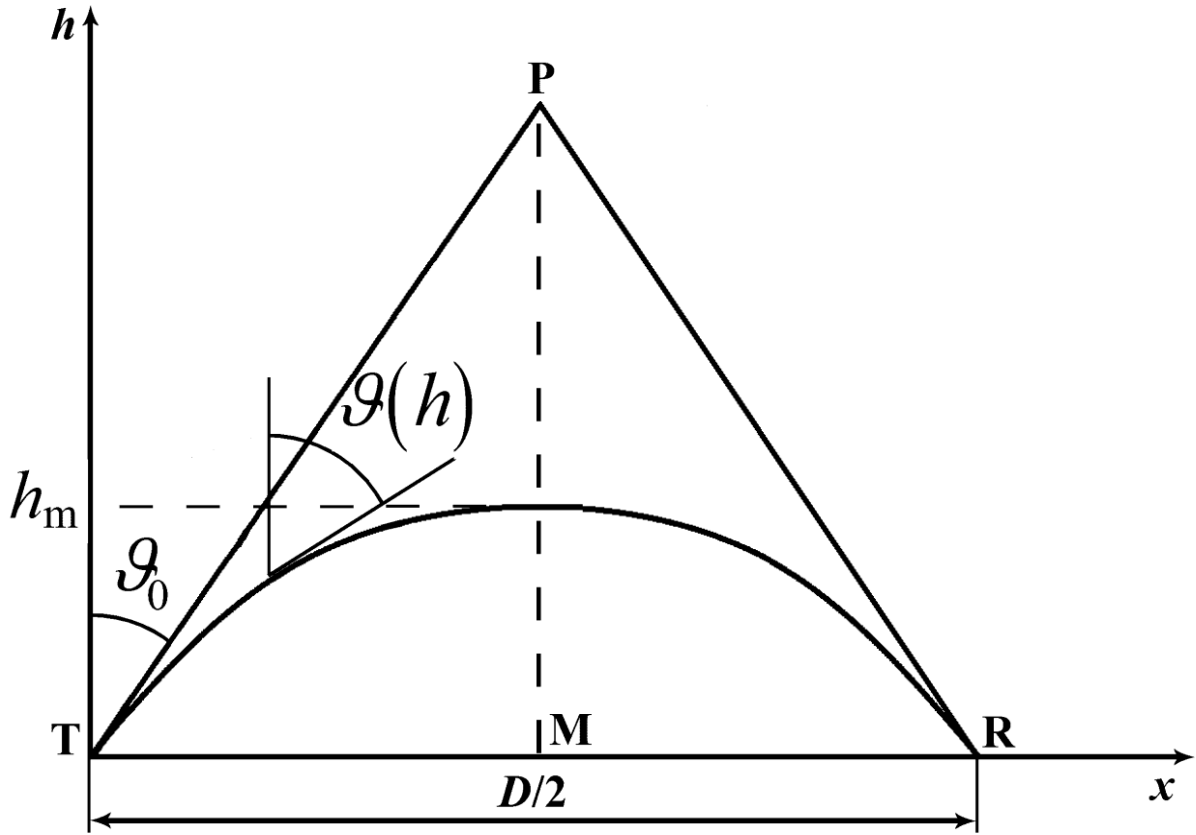

Fig. 7.3. Elementi geometrici del teorema di Breit e Tuve: il tempo che l'onda radio impiega a percorrere il tratto TR lungo il percorso curvo è lo stesso che impiegherebbe a percorrere il tratto TPR, se viaggiasse alla velocità della luce.

**7.6 Teorema di Martyn**

Esiste poi il seguente teorema di Martyn che prende in considerazione il tratto PM nella Fig. 7.3. Esso rappresenta l'altezza del percorso equivalente triangolare per il segnale di frequenza $f_{ob}$ che si propaga obliquamente. Questo tratto, secondo il teorema che stiamo enunciando, equivale all'altezza virtuale di riflessione della frequenza $f_v$ inviata verticalmente. La dimostrazione di questo teorema non viene effettuata in questa sede. Nel Problema 7.2, tuttavia, se ne propone una verifica.

**7.7 Il metodo delle curve di trasmissione**

Secondo i teoremi di Breit, Tuve e Martyn nel caso di ionosfera piana il processo di riflessione obliqua di un'onda avente frequenza $f$ su uno strato ionosferico è equivalente ad una riflessione di tipo speculare che avvenga a una altezza pari all'altezza virtuale di riflessione della frequenza $f_v$ essendo $f_v = f \cos(\vartheta_0)$, ove $\vartheta_0$ è l'angolo indicato nella Fig. 7.3. Per questo motivo, la relazione fra la frequenza verticale e quella obliqua è:

$$f_v = f_{ob} \frac{D/2}{\sqrt{\frac{D^2}{4} + h'^2}} . \tag{7.36}$$

Ora dunque ci chiediamo: se trasmettiamo una certa frequenza $f_{ob}$, e questa viene ricevuta a una certa distanza $D$, a che altezza virtuale $h'$ è avvenuta la riflessione?

A questa domanda si può rispondere considerando la (7.36) e il fatto che noi conosciamo dallo ionogramma l'andamento $h'(f_v)$. Si tratta cioè di considerare il seguente sistema:

$$\begin{cases} f_v = f_{ob} \dfrac{D/2}{\sqrt{\dfrac{D^2}{4} + h'^2}}, & (6.37a) \\ h' = h'(f_v). & (6.37b) \end{cases}$$

La soluzione di questo, nelle incognite $f_v$ e $h'$, può essere trovata per via grafica. Precisamente, si devono riportare le due funzioni (7.37a) e (7.37b) su un diagramma che abbia sulle ordinate $h'$ e sulle ascisse $f_v$, individuando il punto ove le due curve si intersecano: questo è quanto viene fatto nella Fig. 7.4. Si noti che, da questo punto di vista, $f_{ob}$ è un parametro.

Osservando la stessa Fig. 7.4, si vede che, per certi valori di $f_{ob}$, si hanno due o più soluzioni mentre per altri valori non si hanno soluzioni; questo corrisponde a dire che il radiocollegamento non è possibile sulla tratta considerata. Si vede poi che esiste una condizione limite, ove il sistema (7.37a) e (7.37b) ammette una unica soluzione, che corrisponde alla situazione ove la (7.37a) è tangente allo ionogramma. La frequenza $f_{ob}$ associata sarà dunque la massima frequenza utilizzabile su quel radiocollegamento. Essa dipende anche dalla distanza $D$ fra il terminale ricevente e quello trasmittente ed è solitamente indicata con $MUF(D)$.

Il risultato fin qui mostrato, che la *MUF* per una certa $D$ corrisponde alla condizione di tangenza fra la traccia ordinaria dello ionogramma e la curva di trasmissione per tale $D$, consentiva di determinare la $MUF(D)$ tramite il *metodo dei trasparenti fissi*. Questo metodo consisteva nel tracciare su un trasparente le curve di trasmissione standard per la distanza di interesse (usualmente 3000 km, quando si considera la riflessione sullo strato F2) e per varie frequenze. Il trasparente doveva avere le stesse scale orizzontale e verticale usate nello ionogramma. Questo veniva appoggiato sullo ionogramma avendo cura di far coincidere le scale di altezze e di frequenze per cui la $MUF(D)$ veniva desunta dal parametro della curva tangente allo ionogramma. La distanza che convenzionalmente viene tutt'oggi considerata è pari a 3000 km e il modo di propagazione è quello che implica un unico rimbalzo sulla regione F2. Si parla perciò di $MUF(3000)F2$.

Naturalmente, con l'avvento dei computers, il metodo di ricavare la MUF dagli ionogrammi verticali utilizzando i trasparenti è stato abbandonato. Oggi le curve di trasmissione vengono disegnate sugli ionogrammi digitali tramite software coi quali, con procedura semiautomatica, viene anche individuata la curva di trasmissione tangente allo ionogramma considerato.

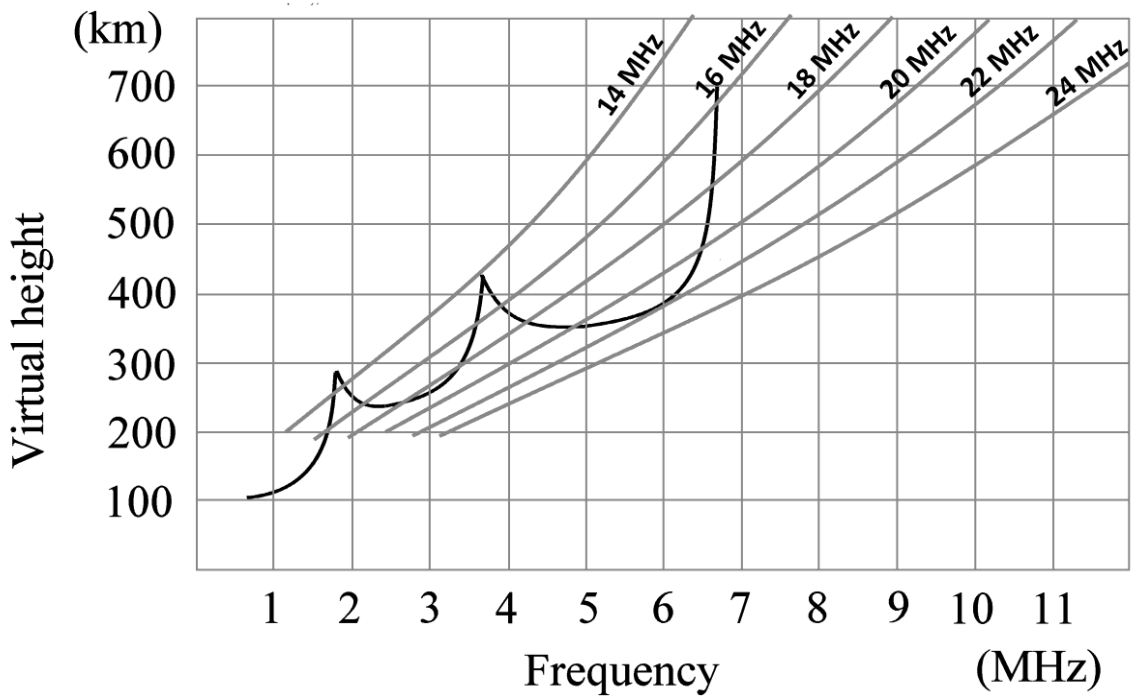

Fig. 7.4. Famiglia di curve di trasmissione $f_v = f_{ob}[(D/2)/(D^2/4+h'^2)^{1/2}]$ parametrizzate in frequenza obliqua sovrapposte alla curva $h' = h'(f)$, che rappresenta lo ionogramma. Queste curve sono disegnate per $D=2000$ km, sebbene la distanza standard usualmente considerata sia $D=3000$ km (rfr7-Davies, 1990).

### 7.8 Schema di un radiocollegamento per via ionosferica

Per fissare le idee consideriamo un radiotrasmettitore T che emette onde radio a una determinata frequenza $f_{ob}$, nel range delle HF. Supponiamo che l'antenna trasmittente emetta su tutti i possibili angoli. Questo è peraltro quanto effettivamente avviene poiché le antenne che possono essere usate nel range delle HF, non hanno mai una elevata direttività. Supponiamo inoltre che la ionosfera sia ridotta a un singolo strato a simmetria sferica, ove la frequenza di plasma è funzione soltanto della quota con un massimo pari a $f_{max} < f_{ob}$.

Per quanto abbiamo visto al paragrafo 7.4, se la frequenza $f_{ob}$ incide la ionosfera con un angolo $\vartheta$, la riflessione sarà possibile se esiste una quota dove $\cos(\vartheta) \cdot f_p(h) = f_{ob}$. Siccome $f_p(h)$ va da 0 a $f_{max}$, questa quota esiste finché $\cos(\vartheta) < f_{ob}/f_{max}$ cioè finché $\vartheta < \arccos(f_{ob}/f_{max})$. Definiamo perciò il seguente *angolo di incidenza critica*:

$$\vartheta_0^* = \arccos \frac{f_{ob}}{f_{max}}. \tag{7.38}$$

I raggi che hanno angoli di incidenza minori di $\vartheta_0^*$ non verranno riflessi, ma attraverseranno lo strato venendo al più deviati. La riflessione comincerà ad avvenire per i raggi che hanno angoli di incidenza maggiori di $\vartheta_0^*$ e prosegue sino ai raggi che vengono emessi tangenzialmente alla superficie delle Terra, detti appunto *raggi tangenti*.

Questi raggi, incideranno la ionosfera con un angolo $\vartheta_{0L}$ che chiameremo *angolo di incidenza limite*. È bene tenere presente che i raggi tangenti hanno la possibilità di propagarsi, oltre che per via ionosferica, anche lungo il suolo ove vengono guidati dalla discontinuità costituita dalla superficie terra-aria. Le onde di terra possono senz'altro superare l'orizzonte ottico ma esse comunque soffrono, almeno nella banda HF, di attenuazione elevata e la loro portata resta modesta.

In conseguenza di quanto abbiamo detto, succede che nella zona vicina a T sarà possibile ricevere il segnare irradiato, in quanto vi saranno le onde di terra. A distanze maggiori, avremo una *zona di silenzio,* in cui un utente non potrà ricevere né l'onda di terra né quella per riflessione ionosferica. Successivamente si avrà la *zona di primo salto*, in cui un utente sarà in grado di ricevere il segnale riflesso dalla ionosfera corrispondente a raggi con $\vartheta_0^* > \vartheta_0 > \vartheta_{0L}$. È bene infine tenere presente che le onde vengono riflesse al suolo, per cui, dalla zona di primo salto, possono partire altri raggi che dànno luogo a successive zone *di secondo salto*, di *terzo salto* ecc.

Abbiamo dunque visto che, ferma restando la frequenza delle radio onde, esistono valori privilegiati della distanza da T per quanto riguarda la possibilità di stabilire un radiocollegamento. Trascurando le onde terrestri, e facendo esclusivamente riferimento alla zona di primo salto, le distanze utili sono delimitate inferiormente dalla distanza coperta dai raggi con incidenza critica che è detta *distanza di skip*. Superiormente, invece, le distanze utilizzabili sono delimitate dalla cosidetta *distanza limite*, che è coperta dai raggi tangenti. La geometria di un radiocollegamento ionosferico, ove si è considerata la curvatura della Terra, è riportato nella Fig. 7.5.

Fino a questo punto abbiamo studiato lo schema di un radiocollegamento ionosferico discutendo il comportamento dei raggi di una medesima frequenza $f_{ob}$, emessi sotto diversi angoli di radiazione. Vogliamo considerare il caso di un trasmettitore T che possa irradiare radio onde di frequenza $f$ variabile, e consideriamo un ricevitore R posto a una distanza $d$ sufficientemente grande da poter trascurare le onde di terra. Vogliamo studiare quali sono le frequenze utilizzabili per detto ricevitore. Assumiamo in questo caso, per comodità, una terra piana e consideriamo che esista un singolo strato riflettente al quale ascriviamo una frequenza critica per incidenza verticale $f_{max}$. Da quanto abbiamo detto dianzi, sappiamo che a questa frequenza critica può essere associata la massima frequenza utilizzabile sul percorso considerato, risulta:

$$MUF_O(d) = f_{max} \sec \vartheta_0 . \tag{7.39}$$

In effetti nella definizione della *MUF* abbiamo sempre fatto riferimento al raggio ordinario. Questo riferimento nella (7.39), viene reso esplicito con la O posta a pedice. Tuttavia, nella pratica, è anche possibile che il radicollegamento venga a essere stabilito attraverso il raggio straordinario. In questo caso, indicando con *MUF(d)* la massima frequenza utilizzabile, considerando qualsiasi modo di propagazione si ha:

$$MUF(d) = MUF_0(d) + \frac{f_H}{2} . \tag{7.40}$$

In conclusione possiamo sintetizzare quanto abbiamo visto, affermando che, fermo restando la distanza fra T ed R, il radiocollegamento ionosferico è possibile solo se viene utilizzata una frequenza $f$ tale che:

$$f \leq MUF(d) . \tag{7.41}$$

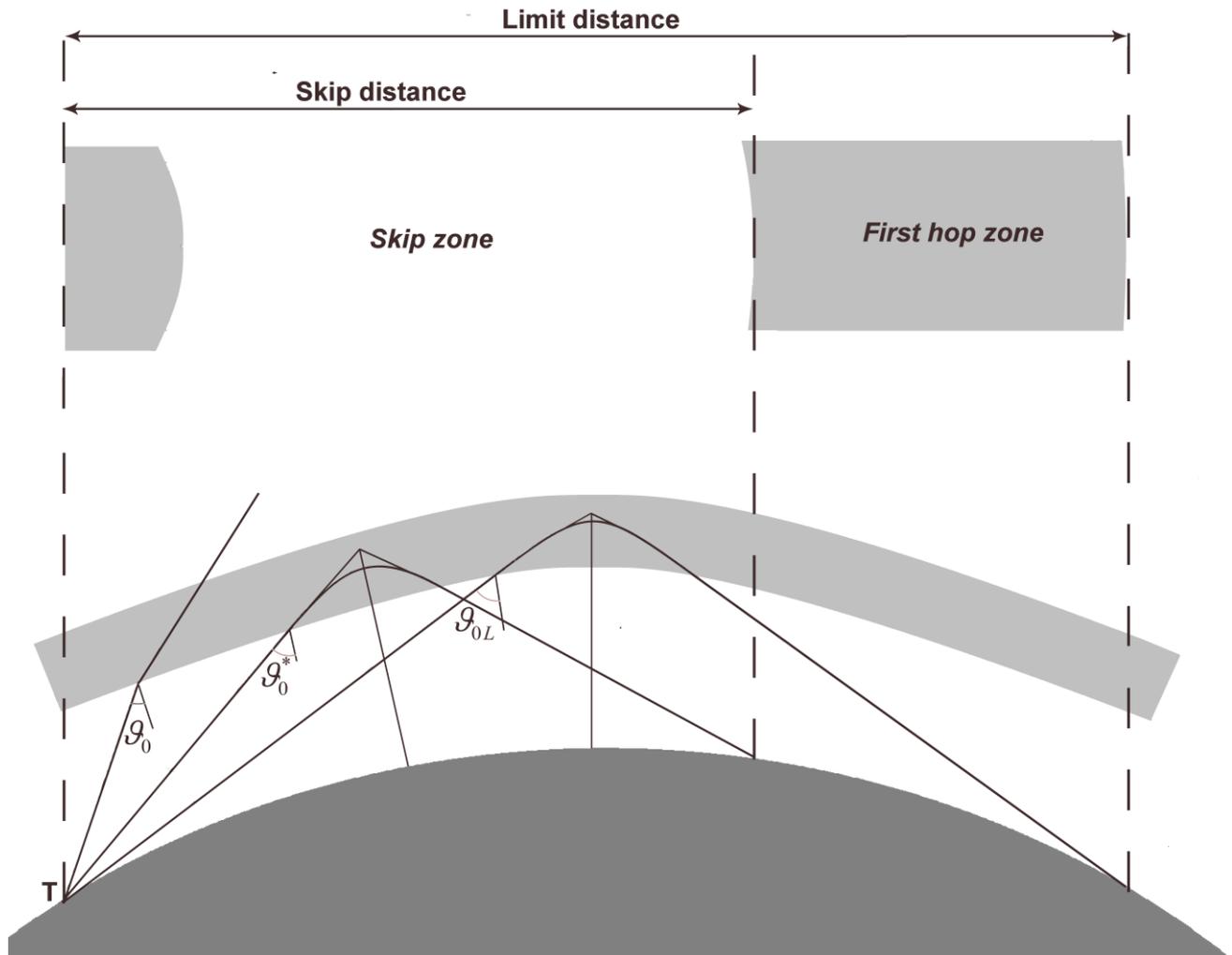

Fig. 7.5. Geometria di un collegamento radio per via ionosferica, ove è stata considerata la curvatura della Terra (rfr7-Dominici, 1971).

**7.9 Radiocollegamenti ionosferici e ray tracing**

Alcuni aspetti della geometria di un radiocollegamento ionosferico possono essere meglio studiati attraverso l'utilizzo di programmi di ray-tracing. Programmi semplici, come quelli i cui risultati vengono presentati in questo paragrafo, possono essere realizzati integrando numericamente l'equazione del raggio (7.23). Questa è stata ricavata facendo l'ipotesi di mezzo isotropo, per cui non viene considerata la presenza di **B**. Assumiamo anche l'ipotesi di assenza di collisioni per cui l'indice di rifrazione risulta $n = 1-X$, così come espresso dalla relazione (2.35). Per ricavare l'indice di rifrazione è inoltre necessario assumere un profilo di densità elettronica.

La Fig. 7.6 è stata ricavata assumendo un $N_e(h)$ con andamento parabolico. Sebbene un profilo di densità elettronica fatto in questo modo sia distante dalla realtà, esso consente di discutere le principali caratteristiche della propagazione ionosferica obliqua. In esso vengono mostrati i percorsi dei raggi emessi da una antenna sotto diversi angoli di radiazione per una certa frequenza $f$. Si vede che oltre la distanza di skip per questa frequenza $D(f)$, è possibile ricevere il segnale. Si vede, inoltre, che in realtà è possibile la ricezione lungo due percorsi: uno che si riflette a quote più basse (detto *raggio basso*) ed un altro che si riflette a quote più alte (detto *raggio alto*). Si deve però osservare che i raggi alti, riportati nella figura in amaranto, vengono emessi entro un intervallo piuttosto limitato di $\Delta$, e vanno a cadere in una regione molto estesa di territorio. L'energia che attraverso il raggio alto raggiunge il terminale ricevente sarà dunque in generale modesta.

In corrisipondenza di $\vartheta_0^*$, che abbiamo visto nel paragrafo precedente, si può definire pure un angolo di radiazione critico $\Delta_{crit}$. Per raggi che vengono irradiati con $\Delta < \Delta_{crit}$ si ha la riflessione ionosferica, mentre raggi che vengono irradiati con $\Delta > \Delta_{crit}$, penetrano la ionosfera e la loro energia si disperde nello spazio.

Nella Fig. 7.6 vengono riportati, in colore verde brillante i raggi che vengono irradiati con $\Delta$ appartenente a un intervallo estremamente piccolo intorno a $\Delta_{crit}$. Si vede che i raggi assumono traiettorie estremamente diverse, in parte disperdendosi nello spazio (raggi con $\Delta > \Delta_{crit}$), e in parte raggiungendo il suolo ma disperdendosi su grandi distanze (raggi con $\Delta < \Delta_{crit}$). Ne consegue, che, nella pratica, le radiocomunicazioni HF devono essere affidate alla propagazione tramite raggio basso.

Per vedere il comportamento della geometria della riflessione ionosferica in funzione della frequenza, cominciamo col fare riferimento alla Fig. 7.7(a). Questa riporta i raggi che si propagano in una ionosfera parabolica, tutti trasmessi ad un carta frequenza $f_1$. Si vede che la distanza $D(f_1)$, corrisponde all distanza di skip per quella frquenza. D'altra parte, per un ricevitore posto alla distanza $D(f_1)$, la frequenza $f_1$ corrisponde alla massima frequenza utilizzabile.

Questo ultimo fatto lo si capisce meglio osservando la Fig. 7.7(b) ove viene illustrata la stessa cosa, ma per una frequenza $f_2$ di poco superiore a $f_1$. Si vede che un ricevitore posto alla distanza $D(f_1)$, caso non riceve il segnale trasmesso alla frequenza $f_2$. Quindi per un utente che utilizzi un ricevitore, posto alla distanza $D(f_1)$, $f_1$ è la massima frequenza utilizzabile, in altre parole $f_1$ è la MUF($D(f_1)$). $D(f_1)$ è infatti al di sotto della distanza di skip per la frequenza $f_2$, mentre la distanza di skip per la frequenza $f_2$ è $D(f_2)$.

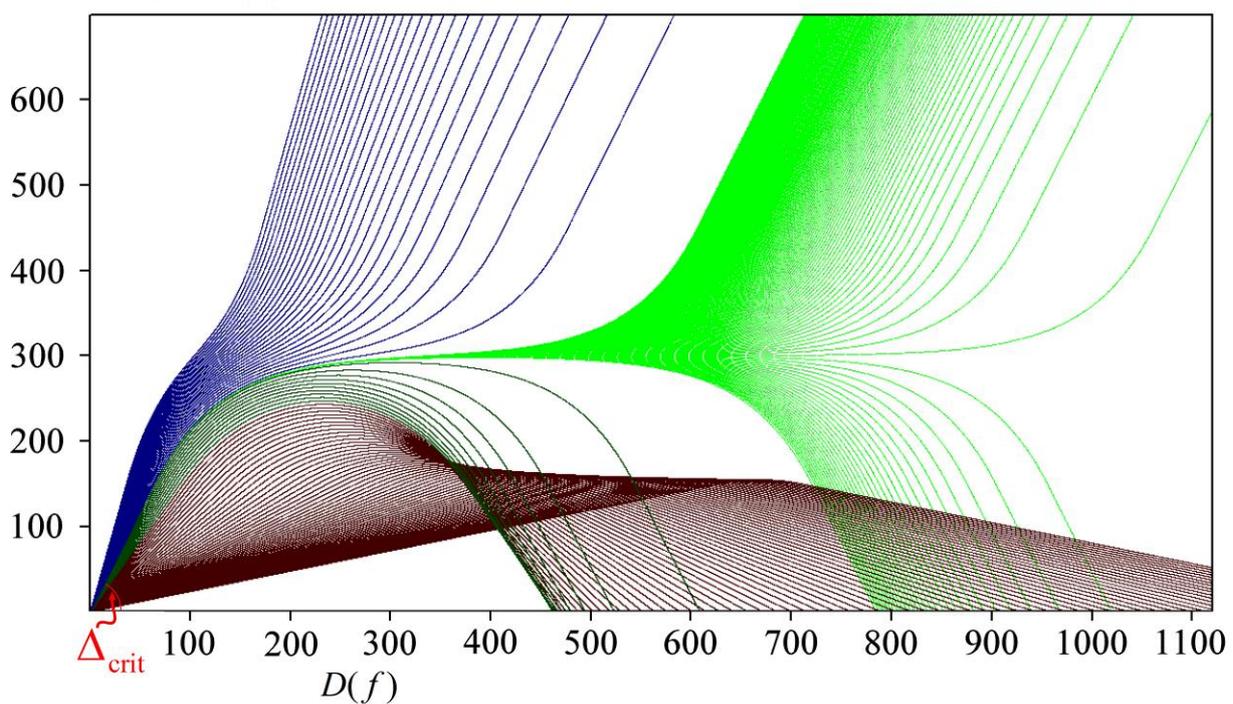

Fig. 7.6. I percorsi dei raggi associati a un'onda radio HF, provenienti da un'antenna trasmittente e sotto diversi angoli di radiazione. In blu i raggi trasmessi per $\Delta > \Delta_{crit}$, che non vengono riflessi dalla ionosfera. In verde brillante i raggi trasmessi per $\Delta_{crit}+\varepsilon > \Delta > \Delta_{crit}-\varepsilon$ con $\varepsilon = 0.1°$. I raggi aventi $\Delta_{crit}+\varepsilon > \Delta$ non vengono riflessi dalla ionosfera, quelli aventi $\Delta_{crit}-\varepsilon < \Delta$ vengono invece riflessi. Si noti che l'energia contenuta in un angolo $\Delta$ molto limitato si disperde molto nello spazio. In verde scuro ed amaranto sono riportati rispettivamente i raggi che raggiungono il suolo per riflessione ionosferica tremite il raggio alto e basso.

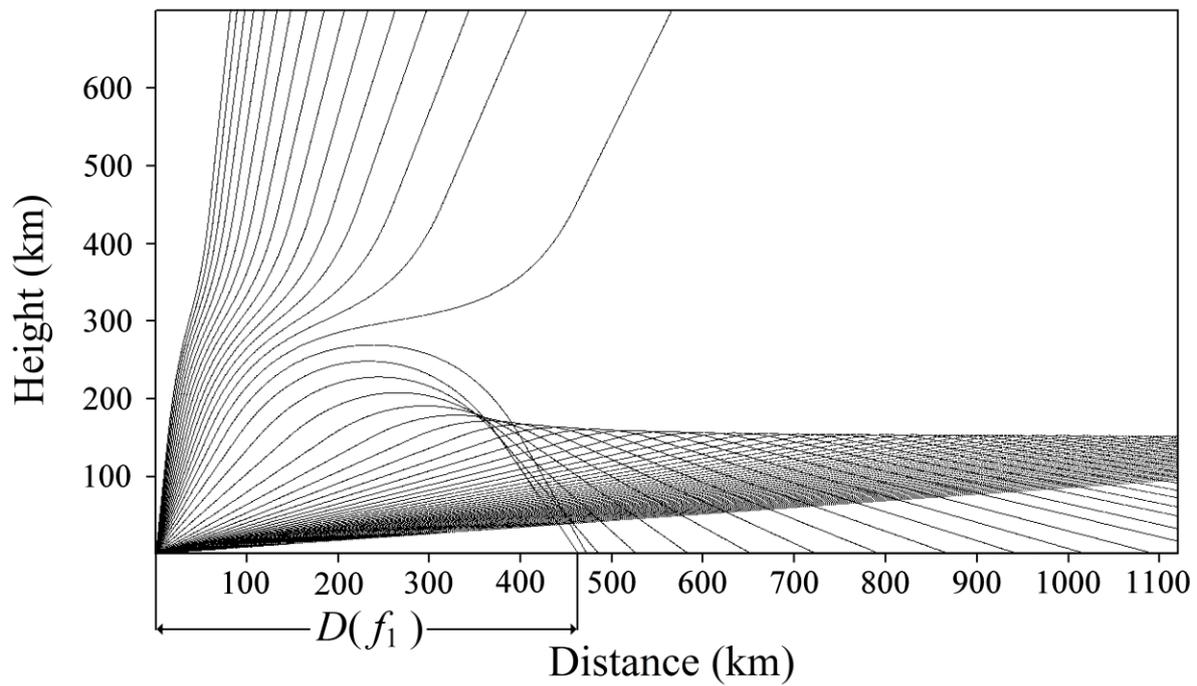

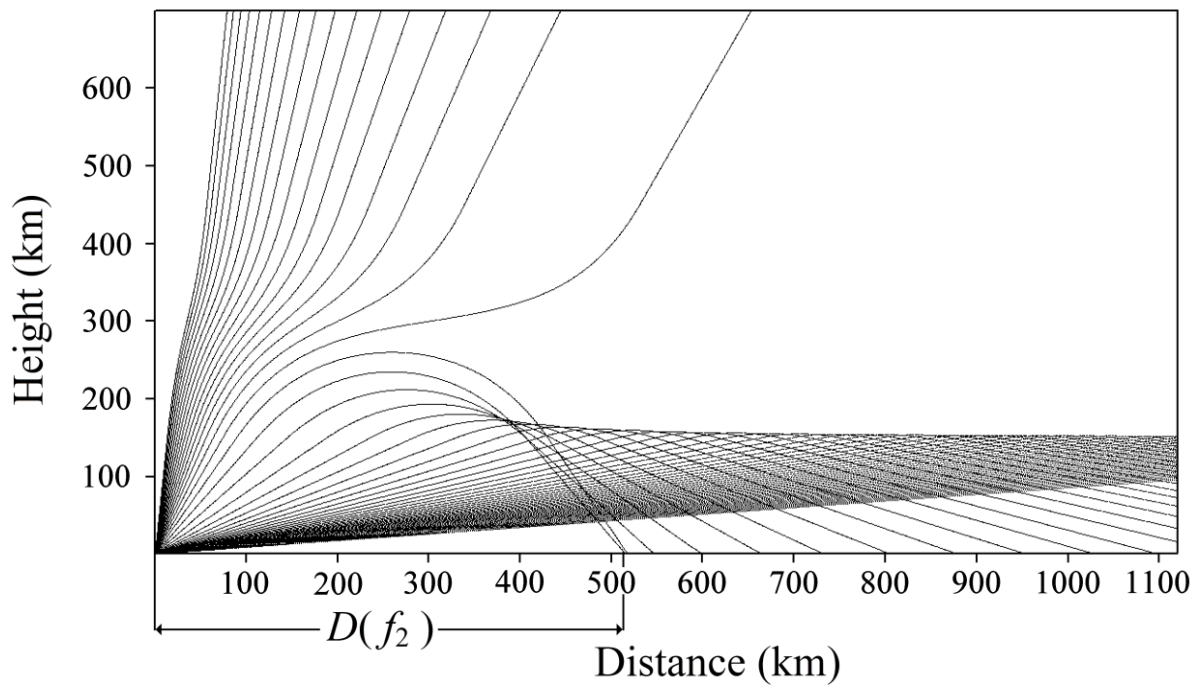

Fig. 7.7 (a)-(b). I percorsi dei raggi associati a un'onda radio HF, provenienti da un'antenna trasmittente, sotto diversi angoli di radiazione per due frequenze $f_1$ (a) e $f_2$ (b), con $f_1 < f_2$. Si vede che per le due distanze di skip $D(f_1)$ e $D(f_2)$ risulta: $D(f_1) < D(f_2)$.

# Applicazioni informatiche del capitolo 7

## Applicazione informatica 7.1

Si assuma un profilo di densità elettronico di forma parabolica, per esempio come quello proposto nell' Applicazione Informatica 4.3. Si consideri il caso di assenza di campo magnetico e di collisioni, per cui, nell'equazione i l'equazione del raggio (7.23), si può assumere *n*=1-*X*. Si scriva una applicazione informatica integrando questa equazione r determinando il percorso ottici dei raggi irradiati da un trasmettitore T, posto al suolo, per una certa frequenza fissata. Si ottengano figure analoghe alla Fig. 7.6 (a)-(b).

## Applicazione informatica 7.2

Si assuma un profilo di densità elettronico di forma parabolica, per esempio come quello proposto nel problema 4.4. Si consideri il caso di assenza di campo magnetico e di collisioni, per cui, nell'equazione del raggio (7.23), si può assumere *n*=1-*X*. Si scriva una applicazione informatica integrando questa equazione, determinando il percorso ottici del raggi irradiati da un trasmettitore T, posto al suolo, per una certa frequenza $f_{ob}$ fissata ed un certo angolo di elevazione *Δ*.
Si scelgano $f_{ob}$ e *Δ* tali che si abbia la riflessione ionosferica del raggio considerato. Si realizzi una applicazione informatica con la quale si verifica il teorema di Martyn.

**Bibliografia del Capitolo 7**

# CAPITOLO 8

## Assorbimento ionosferico

**Riassunto**

L'assorbimento delle onde radio nella ionosfera viene valutato quantitativamente per diverse condizioni propagative. Vengono proposte delle formule per la componente ordinaria e straordinaria, sia in condizioni normali di propagazione, che in prossimità della riflessione.

### 1. Introduzione

L'interesse per l'assorbimento delle onde radio nella ionosfera, è cominciato sin dal momento in cui sono stati effettuati i primi sondaggi ionosferici, quando vennero notate variazioni di ampiezza del segnale ricevuto (rfr8-Pillet, 1960). Fu chiaro da subito che l'onda attraversava un mezzo ionizzato prima di incontrare un livello dove la densità elettronica era sufficiente per far avvenire la riflessione. Dunque fu presto altrettanto evidente che prima della riflessione l'abbattimento del segnale doveva essere causato dall'assorbimento in tale mezzo ionizzato. Inizialmente si pensò che l'assorbimento avesse luogo nella regione E. Vennero dunque effettuati numerosi studi registrando l'ampiezza delle onde riflesse dalla regione F, sia in caso di incidenza verticale che in caso di incidenza obliqua.
Tuttavia, già nel 1930, rfr8Appleton e Ratcliffe () pubblicarono i risultati di misure di intensità di eco a distanze differenti dopo la riflessione nella regione E. Essi giunsero alla conclusione che l'assorbimento dovesse avvenire molto al di sotto della quota di riflessione, in una regione distinta alla quale attribuirono il nome di regione D. Dal punto di vista teorico, vi fu l'importante contributo di rfr8Booker (1935), che dimostrò che una onda radio può essere assorbita anche in una zona dove l'indice di rifrazione è poco diverso dall'unità. Questa zona corrisponde alla regione D supposta da Appleton e Ratcliffe. Successivamente rfr8Farmer e Ratcliffe (1935) effettuarono degli esperimenti che confermavano l'ipotesi dell'esistenza della regione D. Venne constatato un forte aumento del coefficiente di riflessione nelle ore della sera, aumento che venne attribuito alla diminuzione dell'assorbimento nella regione D al crepuscolo.

### 8.2 Assorbimento in assenza di campo magnetico

L'attenuazione di un segnale attraverso la ionosfera può essere descritta attraverso la diminuzione esponenziale dell'intensità di campo $E$, attraverso una relazione del tipo:

$$E(t) = E_0 \exp(-k \cdot s), \qquad (8.1)$$

essendo $s$ l'ascissa curvilinea lungo la traiettoria del raggio dell'onda.
Come abbiamo visto al Paragrafo 3.2, il coefficiente di assorbimento è espresso da $k = \omega \cdot \chi / c$ essendo $\chi$ la parte immaginaria dell'indice di rifrazione $n = \mu - i \cdot \chi$.
Per ricavare $n$, riprendiamo la relazione (5.10) che corrisponde all'equazione di Appleton, ove si è trascurata la presenza del campo magnetico. Si ha:

$$n^2 = 1 - \frac{X}{1 - jZ}, \qquad (8.2)$$

che, allo scopo di ricavare le parti reale ed immaginaria, può essere scritta come:

$$n^2 = 1 - \frac{X(1+iZ)}{(1+Z^2)} = 1 - \frac{X}{(1+Z^2)} - \frac{iXZ}{(1+Z^2)}. \tag{8.3}$$

Da questa, esplicitando $n^2 = (\mu + i\cdot\chi)^2 = (\mu^2 + \chi^2) - i\, 2\cdot\mu\cdot\chi$, si ricava:

$$\begin{cases} \mu^2 + \chi^2 = 1 - \dfrac{X}{(1+Z^2)}, & (7.4a) \\ -2\mu\chi = -\dfrac{XZ}{(1+Z^2)}. & (7.4b) \end{cases}$$

La soluzione di questo sistema dà:

$$\chi = \sqrt{\frac{1 - \dfrac{X}{1+Z^2} + \dfrac{\sqrt{(-1+X-XZ-Z^2)(-1+X+XZ-Z^2)}}{1+Z^2}}{2}}, \tag{8.5}$$

dalla quale si può ricavare il coefficiente di assorbimento $k$ attraverso la relazione $k = \omega\cdot\chi/c$. Nella Fig. 8.1 è riportato il grafico di $k$ in funzione di $X$ in assenza di **B**, nell'intervallo $0 < X < 1$.

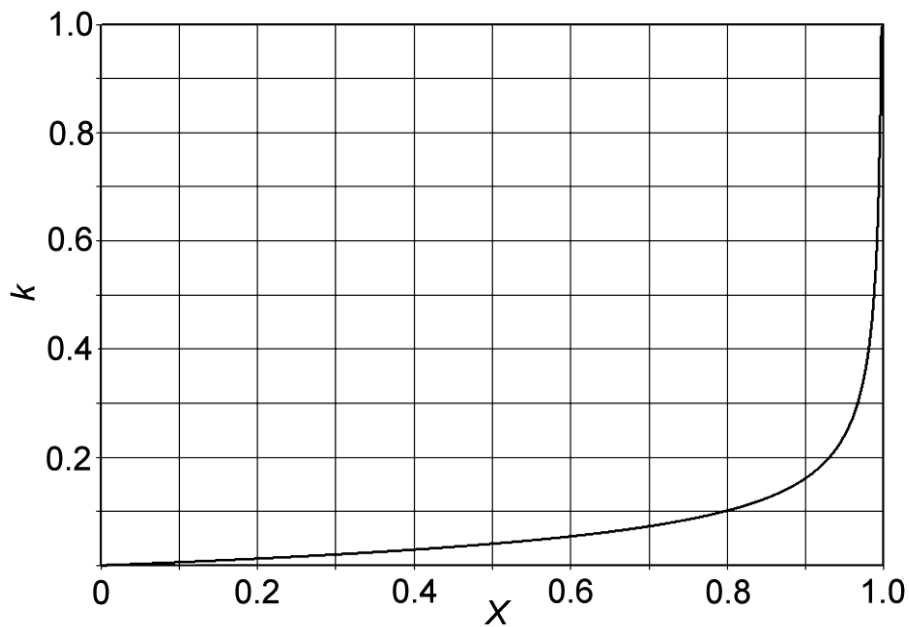

Fig. 8.1. Il grafico del coefficiente di assorbimento $k = \omega\cdot\chi/c$ in funzione di $X$ in assenza di **B**, nell'intervallo $0 < X < 1$.

### 8. 3 Assorbimento ionosferico in presenza di campo magnetico

Con una procedura simile a quella appena vista è in realtà possibile calcolare il coefficiente di assorbimento anche in presenza di **B**. Si dovrà partire dalla nota relazione (2.34) di Appleton che qui riportiamo per comodità:

$$n^2 = 1 - \frac{X}{1 - \frac{Y_T^2}{2(1-X-jZ)} \pm \sqrt{\frac{Y_T^4}{4(1-X-jZ)^2} + Y_L^2}}. \qquad (8.6)$$

In questa relazione dobbiamo esprimere *n* nelle sue parti reale ed immaginaria $n=\mu-i\cdot\chi$ per identificare $\mu$ e $\chi$ dalla relazione precedente. Le espressioni che si ricavano sono complicate e sono di difficile interpretazione. Con l'ausilio di uno dei software matematici in commercio che eseguono il calcolo simbolico, possono essere ricavate espressioni analitiche che successivamente possono essere graficate. Ad esempio nella Fig. 8.2 sono riportati gli esempi dei grafici di $\chi$ ricavati in questo modo, per diverse condizioni radiopropagative. Da $\chi$, facendo di nuovo ricorso alla relazione $k=\omega\cdot\chi/c$ è possibile ricavare il coefficiente *k*.

Per valutare l'assorbimento da una radio onda nel corso della sua propagazione all'interno del mezzo ionosferico, è necessario conoscere *k*, lungo o tutto il percorso di propagazione. Per effettuare questa operazione si rende di conseguenza necessaria la conoscenza, in ogni punto del percorso, di *N*, *v*, e **B**. Prima dell'avvento dei calcolatori elettronici si rendeva necessario suddividere il problema in due parti per le quali valgono con buona approssimazione delle formule diverse. Precisamente si è soliti suddividere il problema in:

a) assorbimento non deviativo, che avviene in regioni dove $\mu$ è prossimo all'unità ed il percorso può essere supposto rettilineo.

b) assorbimento deviativo, che avviene in regioni dove $\mu$ è variabile, e dove si ha la riflessione per incidenza obliqua.

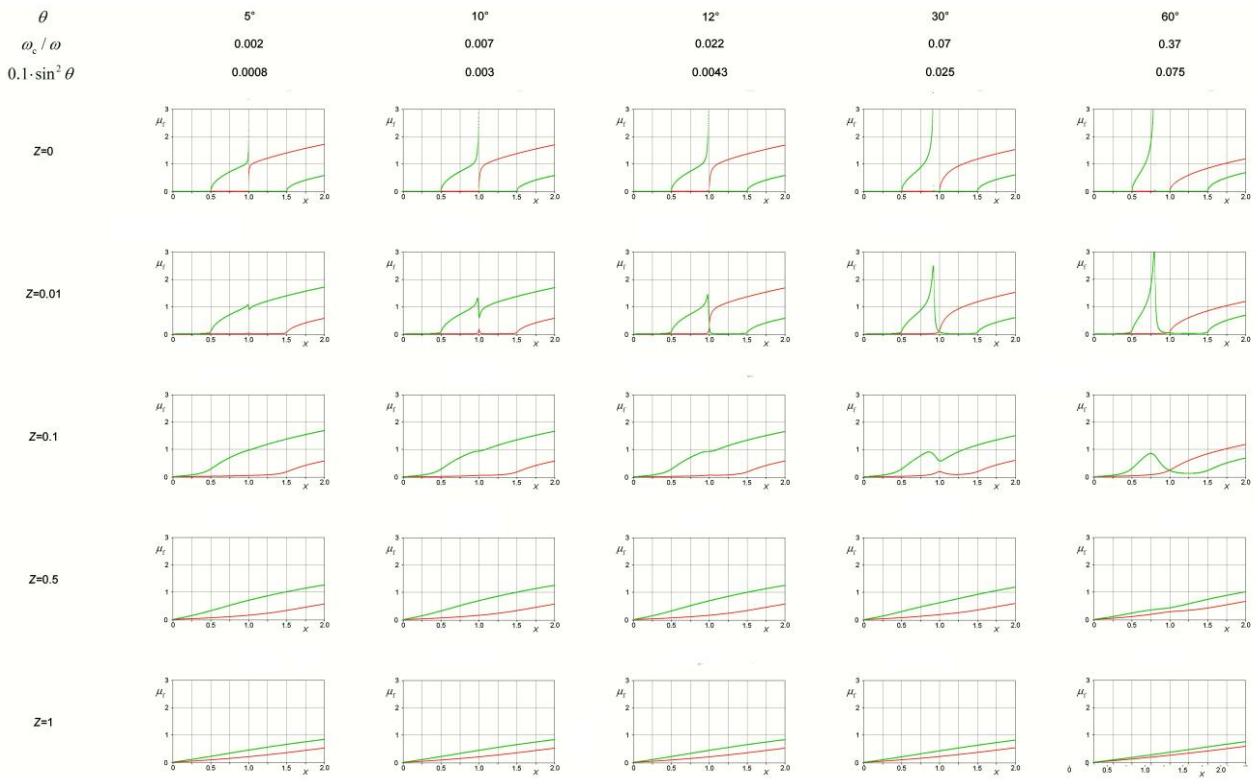

Fig. 8.2 Gli andamenti di $\chi_{f[ord]}(X)$ e $\chi_{f[ext]}(X)$, in presenza di collisioni e di campo magnetico, per diverse condizioni radiopropagative.

## *Assorbimento non deviativo*

Abbiamo visto nel Paragrafo 5.5 e nel Applicazione Informatica 5.2, che, nelle zone lontane dalla riflessione ionosferica, ($X=1$ per l'ordinario e $X=1\pm Y_L$ per lo straordinario) la propagazione avviene in condizioni QL. In queste zone, sarà $\mu\approx 1$ per cui si parlerà di assorbimento non deviativo. Per quanto riguarda l'indice di rifrazione $n=\mu-i\cdot\chi$ in base alla (5.22), varrà la relazione:

$$(\mu - i\chi)^2 = 1 - \frac{X}{\{1 \pm Y_L - jZ\}}. \tag{8.7}$$

Da questa si ricava il seguente sistema di due equazioni nelle due incognite $\mu$ e $\chi$:

$$\begin{cases} \mu^2 - \chi^2 = 1 - \dfrac{X\cdot\left[1\pm |Y_L|\right]}{\left(1\pm Y_L\right)^2 + Z^2}, & (7.8a) \\ -2\mu\chi = -\dfrac{Z\cdot X}{\left(1\pm Y_L\right)^2 + Z^2}. & (7.8b) \end{cases}$$

Dalla (8.8b) si ottiene:

$$\chi = \frac{1}{2\mu} \frac{Z \cdot X}{\left(1 \pm Y_L\right)^2 + Z^2}. \tag{8.9}$$

Sostituendo in essa i valori di X, Z e Y, secondo le definizioni date nel capitolo 2, si ha:

$$\chi = \frac{1}{2\mu} \frac{\frac{v}{\omega} \cdot \frac{\omega_p^2}{\omega^2}}{\left(1 \pm \frac{\omega_L}{\omega}\right)^2 + \frac{v^2}{\omega^2}} = \frac{1}{2\mu} \frac{\frac{v}{\omega} \cdot \frac{\omega_p^2}{\omega^2}}{\frac{1}{\omega^2}\left(\omega \pm \omega_L\right)^2 + \frac{v^2}{\omega^2}} =$$

$$= \frac{1}{2\mu} \frac{\omega_p^2 \frac{v}{\omega}}{\left(\omega \pm \omega_B\right)^2 + v^2} = \frac{1}{2\mu} \frac{Ne^2}{m\varepsilon_0} \frac{1}{\omega} \frac{v}{\left(\omega \pm \omega_L\right)^2 + v^2}. \tag{8.10}$$

Essendo poi $k=\omega/(c\cdot\chi)$, risulta:

$$k = \frac{1}{2\mu c} \frac{Ne^2}{m\varepsilon_0} \frac{v}{\left(\omega \pm \omega_L\right)^2 + v^2}. \tag{8.11}$$

Prendendo il segno positivo per il raggio ordinario e il segno negativo per il raggio straordinario. L'utilizzo di questa relazione è stato proposto in generale (Pillet, 1960) rfr8, non limitatamente alla condizione QT, dove essa è a rigore valida.
Se $\theta<40°$, poi, è stato proposto di seguire il suggerimento di Ratcliffe (1959) rfr8, e porre $\omega_B \approx \omega_L$. Avremo così, in luogo della (8.10):

$$\chi = \frac{1}{2\mu} \frac{Ne^2}{m\varepsilon_0} \frac{1}{\omega} \frac{v}{\left(\omega \pm \omega_B\right)^2 + v^2}, \tag{8.12}$$

e, in luogo della (8.11):

$$k \approx \frac{1}{\mu} \frac{e^2}{2\varepsilon_0 me} \frac{Nv}{\left(\omega \pm \omega_B\right)^2 + v^2}. \tag{8.13}$$

*Assorbimento deviativo*

Abbiamo visto, sempre nel Paragrafo 5.5 e nel Problema 5.2, che si ha la riflessione ionosferica per lo straordinario per $X=1\pm Y_L$ se $\theta<40°$ e questa riflessione avviene in consizioni QL. Questo implica che, in tali condizioni, la precedente relazione (8.11) può essere utilizzate anche per calcolare l'assorbimento deviativo del raggio straordinario.
Per quanto riguarda invece l'assorbimento deviativo del raggio ordinario, si deve ricordare che la riflessione avviene per $X=1$, in condizioni QT. In questo caso, sarà valida le relazione (8.2), dalla quale abbiamo ricavato la soluzione 8.5). Tuttavia, osservando la simmetria del sistema (8.4a) e (8.4b), si poteva pure ricavare subito $\mu=\chi$ ed una soluzione formalmente simile alla (8.9):

$$\chi = \frac{1}{\mu} \frac{XZ}{(1+Z^2)} \quad , \tag{8.14}$$

dalla quale di ottiene:

$$k = \frac{1}{2\mu c} \frac{Ne^2}{m\varepsilon_0} \frac{\nu}{\omega^2 + \nu^2} . \tag{8.15}$$

## 8.4 Sommario

In generale, il problema di determinare il coefficiente di assorbimento $k$, si risolve come abbiamo detto all'inizio del Paragrafo precedente, ricavando $\chi$, a partire dall'espressione (8.6) per l'indice di rifrazione ed usando poi la relazione $k = \chi \cdot \omega/c$.

Per casi particolari, invece, si possono usare le seguenti relazioni semplificate che abbiamo visto dianzi e che qui ripetiamo sinteticamente:

$$k \approx \frac{1}{\mu} \frac{e^2}{2\varepsilon_0 me} \frac{N\nu}{\omega^2 + \nu^2} , \tag{8.16}$$

per l'assorbimento deviativo del raggio ordinario, che avviene per $X=1$ in condizioni QT;

$$k \approx \frac{1}{\mu} \frac{e^2}{2\varepsilon_0 me} \frac{N\nu}{(\omega \pm \omega_B)^2 + \nu^2} , \tag{8.17}$$

per l'assorbimento deviativo del raggio straordianario, che per $\theta < 40°$, avviene in condizioni QL;

$$k \approx \frac{1}{\mu} \frac{e^2}{2\varepsilon_0 me} \frac{N\nu}{(\omega \pm \omega_L)^2 + \nu^2} , \tag{8.18}$$

ove $\omega_L = \omega_B \cos(\theta)$ per l'assorbimento non deviativo dell'ordinario (segno positivo) e dello straordinario (segno negativo); questa formula non è a rigore vera per qualsiasi disrezione di propagazione.

# Applicazioni informatiche del Capitolo 8

### Applicazione informatica 8.1 (assorbimento in assenza di campo magnetico)

Si usi l'espressione (8.5) per calcolare $\chi$ in funzione di $X$. Da questa si ricavi il coefficiente di assorbimento $k$ attraverso la relazione $k = \omega \cdot \chi / c$.
Si realizzi un'applicazione informatica con la quale si disegni il grafico di $k$ in funzione di $X$, nell'intervallo $0 < X < 1$, ottenendo un grafico come quello della Fig. 8.1.

### Applicazione informatica 8.2 (assorbimento non deviativo per il raggio ordianario)

Si parta dall'equazione (8.6) ed usando uno dei sofware matematici in commercio, si ricavino la parte reale $\mu_{ord}$ e quella immaginaria $\chi_{ord}$ dell'indice di rifrazione $n_{ord}=\mu_{ord}$-i$\cdot$ $\chi_{ord}$. Si ponga l'attenzione sulla parte immaginaria $\chi_{ord}$, si ricavi il coefficiente di assorbimento $k_{ord}$ attraverso la relazione $k_{ord} = \omega \cdot \chi_{ord}/c$.
Considerando diverse condizioni radioprogative, si ricavino dei grafici di $k_{ord}(\omega)$. Si riporti sugli stessi grafici $k_{ord}(\omega)$ ottenuti attraverso la (8.12).

### Apllicazione informatica 8.3 (assorbimento deviativo e non deviativo per il raggio strordianario)

Analogamente a quanto fatto nell'esercizio precedente, si ricavino dei grafici $k_{ext}(\omega)$ e si confrontino coi corrispondenti che si possono ricavare attraverso la (8.12).

### Applicazione informatica 8.4 (assorbimento deviativo per il raggio ordianario)

Si parta dall'equazione (8.6) ed usando uno dei sofware matematici in commercio, si ricavino la parte reale $\mu_{ord}$ e quella immaginaria $\chi_{ord}$ dell'indice di rifrazione $n_{ord}=\mu_{ord}$-i$\cdot$ $\chi_{ord}$. Si ponga l'attenzione sulla parte immaginaria $\chi_{ord}$, si ricavi il coefficiente di assorbimento $k_{ord}$ attraverso la relazione $k_{ord} = \omega \cdot \chi_{ord}/c$.
Considerando diverse condizioni radioprogative, si ricavino dei grafici di $k_{ord}(\omega)$. Si riporti sugli stessi grafici, $k_{ord}(\omega)$ ottenuti attraverso la (8.15).

## Bibliografia del Capitolo 8

# APPENDICE 1

## Onde Evanescenti

Consideriamo un conduttore sottoposto ad un campo elettrico oscillante, che possiamo scrivere nella forma E=$E_0$·exp (-$i\omega t$). Considereremo poi, per semplicità, il caso unidimensionale, nell'approssimazione di *campi deboli*, così da poter trascurare gli effetti magnetici nel conduttore. Si noti che questa ipotesi è la stessa nella quale è stata fatta tutta la nostra trattazione. Se facciamo pure l'ipotesi che le cariche siano soggette a una forza che tende a smorzarne il moto, la seconda legge di Newton, per gli elettroni assume la forma:

$$m(\ddot{\mathbf{x}} + \gamma \dot{\mathbf{x}}) = -eE_0 e^{i\omega t}. \tag{A1.1}$$

La cui soluzione, per la velocità è:

$$\dot{\mathbf{x}}(t) = \mathbf{v}(t) = A_0 \cdot e^{i\omega t}$$

e, per l'accelerazione: (A1.2)

$$\ddot{\mathbf{x}}(t) = \dot{\mathbf{v}}(t) = i\omega \mathbf{A_0} \cdot e^{i\omega t}. \tag{A1.3}$$

Da questa:

$$m(i\omega \mathbf{A_0} e^{i\omega t} + \gamma \mathbf{A_0} e^{i\omega t}) = -e\mathbf{E_0} e^{i\omega t}, \tag{A1.4}$$

e

$$\mathbf{A}_0 = -\frac{e\mathbf{E}_0}{m(i\omega + \gamma)}, \tag{A1.5}$$

e

$$\mathbf{v} = -\frac{e\mathbf{E}}{m(i\omega + \gamma)}. \tag{A1.6}$$

Ricordando che, per definizione di densità di corrente **j**, si ha:

$$\mathbf{j} = Ne\mathbf{v}, \tag{A1.7}$$

dove *N* si indica il numero di elettroni per unità di volume, si ottiene:

$$\mathbf{J} = \frac{Ne^2 \mathbf{E}}{m(i\omega + \gamma)}. \tag{A1.8}$$

In questa relazione può essere introdotta la relazione (1.6) per la frequenza di plasma $\omega_p^2 = Ne^2/m\varepsilon_0$. Si ottiene così:

$$\mathbf{J} = \frac{\omega_p^2}{(i\omega + \gamma)} \varepsilon_0 \mathbf{E}. \tag{A1.9}$$

Ricordando la relazione $\mathbf{J} = \sigma \mathbf{E}$, è possibile introdurre la conducibilità generalizzata che risulta:

$$\sigma = \frac{\omega_p^2}{(i\omega + \gamma)} \varepsilon_0. \tag{A1.10}$$

Questa è in generale una grandezza complessa. Nel caso di basse frequenze $\omega \ll \gamma$, essa è puramente reale e si riconduce alla legge di Ohm.
Se consideriamo la relazione (P1.47), che è detta *equazione del telegrafo*, ove poniamo $\mu = 1$ e $\sigma$ in base alla (A1.10), abbiamo:

$$\nabla^2 \mathbf{E} - \sigma \frac{\partial \mathbf{E}}{\partial t} - \frac{\partial^2 \mathbf{E}}{\partial t^2} = 0. \tag{A1.11}$$

Questa ha soluzione del tipo $\mathbf{E} = \mathbf{E}_0 \exp(i \cdot k \cdot x - \omega t)$. Essa può essere sostituita, nell'equazione d'onda, ricavando la seguente relazione di dispersione:

$$\omega^2 = k^2 c^2 \frac{i\omega}{i\omega - \gamma} \omega_p^2. \tag{A1.12}$$

Che, nel limite di alte frequenze, $\omega \gg \gamma$, porta a:

$$\omega^2 = k^2 c^2 + \omega_p^2, \tag{A1.13}$$

dalla quale:

$$k^2 = \frac{\omega^2 - \omega_p^2}{c^2}. \tag{A1.14}$$

Nel caso $\omega^2 < \omega_p^2$ il numero d'onda $k$ è puramente immaginario ed il campo elettrico nel conduttore assume la forma:

$$\mathbf{E} = \mathbf{E}_0 e^{-\frac{x}{l_p}} e^{-i\omega t}. \tag{A1.15}$$

ovvero, il campo elettrico non si propaga all'interno del conduttore, ma penetra solo sino ad una distanza $l_p$, che è detta *distanza di pelle inerziale*:

$$l_p = \sqrt{\frac{c^2}{\omega_p^2 - \omega^2}}. \tag{A1.15}$$

In virtù di questa caratteristica, questo genere di onde vengono definite *evanescenti*.

# APPENDICE 2

## Velocità di fase e velocità di gruppo

Ricordiamo che si definisce *onda* una perturbazione viaggiante che dà luogo a trasporto di energia senza trasporto di materia.

Un'*onda progressiva* lungo una direzione **r** è un'onda che dà luogo a trasporto di energia in quella direzione. Inoltre, un' *onda armonica* è un'onda che può venire espressa tramite relazioni sinusoidali. Se consideriamo la consueta terna di riferimento, con assi $x$, $y$, $z$ possiamo esprimere un'*onda armonica* progressiva diretta lungo l'asse $x$, nella forma:

$$E = a \cdot \sin b(x-ct). \tag{A2.1}$$

dove $a$ e $b$ sono opportune costanti. Imponendo all'argomento b·($x$-$c$·$t$) la periodicità rispetto al tempo, con periodo $T$ o con frequenza $f=T^{-1}$, l'espressione assume la forma seguente:

$$E = A\sin\left(\omega t - kx + \varphi\right). \tag{A2.2}$$

con $\omega = 2\pi f = 2\pi/T$, $k = \omega/c = 2\pi/\lambda$. In queste relazioni $A$ è l'*ampiezza*, $\omega$ è la *frequenza angolare* della radio onda, $\lambda$ è la *lunghezza d'onda* cioè la distanza percorsa dall'onda in un periodo, $k$ è il *numero d'onda*, $\varphi$ è la *fase iniziale all'origine*, cioè la fase all'istante $t=0$ e a distanza $x=0$; naturalmente, scegliendo opportunamente l'istante iniziale, si può sempre rendere $\varphi=0$.

Se l'onda non è armonica, cioè non è esprimibile attraverso una singola espressione sinusoidale, attraverso il teorema di Fourier essa può essere espressa come somma di funzioni sinusoidali (componenti armoniche) ciascuna avente la propria ampiezza, frequenza, fase iniziale.

Consideriamo ora il caso semplice di un'onda che possa essere espressa come composizione di due onde armoniche, di uguale ampiezza, esprimibili come:

$$\begin{cases} E_1 = A\sin\left(\omega_1 t - k_1 x\right), & \text{(A2.3a)} \\ E_2 = A\sin\left(\omega_2 t - k_2 x\right). & \text{(A2.3b)} \end{cases}$$

Applicando relazioni trigonometriche note, per l'onda risultante si ottiene:

$$E = E_1 + E_2 = \left[2A\cos\left(\frac{\omega_1 - \omega_2}{2}t - \frac{k_1 - k_2}{2}x\right)\right] \cdot$$

$$\sin\left(\frac{\omega_1 + \omega_2}{2}t - \frac{k_1 + k_2}{2}x\right). \tag{A2.4}$$

La velocità di fase $v$ con la quale si propaga la fase [($\omega_1+\omega_2$)·$t$ / 2-($k_1+k_2$)·$x$ / 2], è:

$$v = \frac{\omega_1 + \omega_2}{k_1 + k_2}, \tag{A2.5}$$

mentre la velocità $v'$ con cui si propaga l'onda di ampiezza, cioè la velocità con cui si propaga il "gruppo di onde", inviluppato da tale onda, detto *velocità di gruppo*, è:

$$v_g = \frac{\omega_1 - \omega_2}{k_1 - k_2}. \tag{A2.6}$$

Se vogliamo generalizzare, basta supporre che le due onde differiscano di una quantità infinitesima, cioè porre $\omega_1-\omega_2=\mathrm{d}\omega$ e $k_1-k_2=\mathrm{d}k$. Si ha così, dalla relazione precedente:

$$v_g = \frac{\mathrm{d}\omega}{\mathrm{d}k}. \tag{A2.7}$$